\documentclass[twocolumn,english,floatfix,prb,eqsecnum]{revtex4-1}
\usepackage{amsmath}
\usepackage{epsfig,color}
\usepackage{graphicx}
\usepackage{dcolumn}
\usepackage{bm}
 \usepackage{multirow}

\newcommand{\ve}{\varepsilon}

\newcommand{\bk}{{\bf k}}
\newcommand{\bp}{{\bf p}}
\newcommand{\bq}{{\bf q}}

\newcommand{\nn}{\nonumber}
\newcommand{\beq}{\begin{equation}}
\newcommand{\eeq}{\end{equation}}
\newcommand{\bea}{\begin{eqnarray}}
\newcommand{\eea}{\end{eqnarray}}
\newcommand{\bse}{\begin{subequations}}
\newcommand{\ese}{\end{subequations}}
\newcommand{\bwt}{\begin{widetext}}
\newcommand{\ewt}{\end{widetext}}

\newcommand{\bkp}{{\bf k}'}
\newcommand{\bpp}{{\bf p}'}

\newcommand{\bv}{{\bf v}}

\newcommand{\I}{\mathrm{Im}}

\newcommand{\bsu}{\begin{subequations}}
\newcommand{\esu}{\end{subequations}}

\newcommand{\lr}{\left(}
\newcommand{\rr}{\right)}
\newcommand{\ls}{\left[}
\newcommand{\rs}{\right]}
\newcommand{\tk}{{\bar k}}
\newcommand{\bo}{{\bar\Omega}}
\newcommand{\bg}{{\bar g}}
\newcommand{\sse}{\sigma'_\Sigma(\Omega)}

\begin{document}

\title{
Optical conductivity of a two-dimensional metal\\ at the onset of 
spin-density-wave order}
\author{Andrey V. Chubukov$^{a)}$, Dmitrii L. Maslov$^{b)}$, and Vladimir I. Yudson$^{c)}$}
\date{\today}
\affiliation{
$^{a)}$Department of Physics, University of
Wisconsin-Madison, 1150 Univ. Ave., Madison, WI 53706-1390\\
$^{b)}$Department of Physics, University of
Florida, P. O. Box 118440, Gainesville, FL
32611-8440\\
 $^{c)}$Institute for
Spectroscopy, Russian Academy of Sciences, Troitsk, Moscow region,
142190, Russia
}
\date{\today}

\begin{abstract}
We consider
the
optical conductivity
of a clean
two-dimensional metal
 near
 a
 quantum
spin-density-wave
 transition.
 Critical magnetic fluctuations are known to
  destroy
 fermionic coherence
   at
    \lq\lq hot
  spots\rq\rq\/ of the Fermi surface
 but
     coherent quasiparticles survive
     in
      the  rest of the Fermi surface.
 A large part of the Fermi surface is not really \lq\lq cold\rq\rq\/
 but rather \lq\lq lukewarm\rq\rq\/ in a sense that
  coherent quasiparticles in that part
   survive
  
   but are strongly renormalized
 compared to the non-interacting case.
 We discuss
 the self-energy
 of lukewarm fermions and their contribution to the optical conductivity,
 $\sigma(\Omega)$,
  focusing specifically on scattering off composite
 bosons made
 of
 two critical magnetic fluctuations.
 Recent study [S.A. Hartnoll et al., Phys. Rev. B {\bf 84}, 125115 (2011)]
  found  that
 composite scattering
   gives the strongest contribution to
    the
    self-energy 
    of
     lukewarm fermions
     and
 suggested that this  may
  give rise
  to
 a
  non-Fermi liquid behavior of
 the optical conductivity
  at
  the lowest frequencies.
   We show  that
  the
  most
   singular term in the conductivity coming from self-energy
   insertions into the conductivity bubble,
   $\sigma'(\Omega)\propto \ln^3\Omega/\Omega^{1/3}$,
    is canceled  out by the vertex-correction and Aslamazov-Larkin diagrams.
However, the cancelation does not hold
 beyond logarithmic accuracy,
and the
remaining conductivity
still
 diverges as $1/\Omega^{1/3}$.
 We further argue that
 the
 $1/\Omega^{1/3}$ behavior holds only
  at
  asymptotically low
  frequencies, well inside the frequency range affected by superconductivity.
   At larger $\Omega$,
   up to frequencies above the Fermi energy, $\sigma'
    (\Omega)$ scales as
  $1/\Omega$, which is reminiscent of the behavior
  observed in the superconducting cuprates.
\end{abstract}

\maketitle
\protect
\section{Introduction}
\label{sec:intro}
Understanding the behavior
of fermions near a quantum-critical point (QCP) remains one of the most challenging problems
 in the physics of strongly correlated
 materials.
As one possible manifestation of quantum criticality, the resistivity $\rho(T)$
 of optimally-doped
cuprates, Fe-pnictides, heavy-fermion compounds, and other materials
exhibits a linear-in-$T$ behavior over a wide range of temperatures\cite{hussey:2011,bruin:2013,hartnoll:2013}
  instead of the $T^2$ behavior, expected
  for
  a
  Fermi liquid (FL) with umklapp scattering
  \cite{baym}.
  Another
  type
  of the
   non-FL (NFL)
  behavior, $\rho (T) \propto T^b$ with $b \approx 3/2$,
   has been
  observed near the end point of
  the superconducting phase in
  the
  hole-
  and electron-doped cuprates,\cite{t32,comm_a}
  whereas
  $\rho(T)\propto T^c$ with $c\approx 5/3$
 has been
 observed
near ferromagnetic criticality in a number of three-dimensional itinerant ferromagnets.~\cite{fm}

  In addition to the {\em dc} resistivity, the optical conductivity provides
  useful information about the energy dependences of the scattering rate and
  effective mass.   The real part of
  of the conductivity $\sigma'(\Omega)$, measured at $\Omega \gg T$,
    can be described by a \lq\lq generalized Drude formula\rq\rq\/ \cite{basov:11}
    \beq
  \sigma'(\Omega)
 = \frac{\Omega_{\mathrm{p}}^2}{4\pi}
  \frac{1/\tau_{\mathrm{tr}} (\Omega)}{\left[\Omega \frac{m^*_{\mathrm{tr}}(\Omega)}{m} \right]^2 + \left(\frac{1}{\tau_{\mathrm{tr}} (\Omega)}\right)^2},
    \label{ch_1}
    \eeq
    where $\Omega_{\mathrm{p}}$ is the effective plasma frequency, $\tau_{\mathrm{tr}}(\Omega)$ is the transport
    scattering
    time, 
    and $
    m^*_{\mathrm{tr}} (\Omega)$ is the 
   \lq\lq transport effective mass\rq\rq\/ ($m$ is
    bare electron mass).
    If
    the
    fermionic self-energy $\Sigma = \Sigma'+i\Sigma''$ has
     a stronger dependence on
     the
     frequency
     than on the momentum
     across
     the Fermi surface (FS),  $m^*_{\mathrm{tr}}(\Omega)/m$ is
     equal
      to
      $1/Z (\Omega)$, where
    $Z=\left(1+\frac{\partial\Sigma'}{\partial\Omega}\right)^{-1}$ is the quasiparticle residue.  The transport
    scattering rate
    $1/\tau_{\mathrm{tr}} (\Omega)$
  is proportional to
    $\Sigma'' (\Omega)$, but
    may be much smaller than the latter if the dominant scattering mechanism
    involves
    small momentum transfers.
  For an ordinary FL with
 interactions roughly the same at all momentum transfers,
   $\Sigma''(\Omega, T) \sim 1/\tau_{\mathrm{tr}} (\Omega, T) \propto \max\{\Omega^2,T^2\}$
and $Z=\mathrm{const}$ (Ref.~\onlinecite
{comm^*}).
    Equation (\ref{ch_1})
     then
     predicts that $\sigma'(\Omega)=\mathrm{const}$ at
   low frequencies, when $Z\Sigma''(\Omega) \ll \Omega$.
   Instead,
  the
  measured
  $\sigma'(\Omega)$ of many strongly-correlated materials
  depends strongly on
  the
  frequency, often as a power-law  $\sigma'(\Omega) \propto 1/\Omega^{d}$ with positive exponent $d$,
  meaning
  that
  $\sigma'(\Omega)$ increases as $\Omega$ gets smaller.
  For example,
  $\sigma'(\Omega)$
 of
  several underdoped and optimally doped
  cuprates
   in
   the
 $(x,\Omega)$
  domain
  outside the pseudogap
   phase ($x$ stands for doping)
 was
 described
  by a power-law form
  with
  either
  $d\approx 1$ (Ref.~\onlinecite{basov}) in a wide frequency range, roughly from $
 100$ meV to about $1$ eV,
 or
 with
  $d\approx 0.7$ (Ref.~\onlinecite{azrak:1994}) and $d =0.65$ (Ref.\onlinecite{dirk,mike})
 in the intermediate frequency range
$\Omega \sim
100-500$ meV.
   Likewise, $\sigma'(\Omega)$ of the ruthenates SrRuO$_3$ (Ref.~\onlinecite{dodge:2000}) and CaRuO$_3$ (Refs.~\onlinecite{lee:2004,dodge:2006}),
   as well of the helimagnet MnSi (at ambient pressure, Ref.~\onlinecite{marel:2003}), follows
 a
   power-law form with $
   d \approx 1/2$.
   \bwt
   \begin{table*}[ht]
\caption{List of notations}
\centering
\begin{tabular}{|c| c |c|}
\hline
Notation & Meaning &Relation to other parameters   \\ [0.5ex] 
\hline
$\bar g$ & coupling constant of the spin-fermion model (in units of energy)&$-''-$  \\ 
\hline
$\gamma$ & Landau damping coefficient &$ 
\gamma=
4 {\bar g}/\pi v^2_F$\\
\hline
$\bk_F$ & arbitrary point on the Fermi surface &\\
\hline
$\bk_{\mathrm{h.s.}}$ & location of the hot spot &\\
\hline
$\bq_{\pi}$ & SDW wavevector& 
$\bq_\pi=
(\pi,\pi)$\\
\hline
$k_\perp$ & component of $\bk$ along the normal to the Fermi surface &\\
\hline
$\delta k$ & distance from the hot spot along the Fermi surface &\\
\hline
$m^*$ & effective mass defined by Eq.~(\ref{disp})&\\
\hline
$E_F^*$ & effective Fermi energy &$E_F=m^*v_F^2/2$\\
\hline
$K$&$(2+1)$ momentum & $K=(\bk,\Omega)$\\
\hline
$K_F$ &$(2+1)$ momentum on the Fermi surface & $K_F=(\bk_F,\Omega)$\\
\hline
$z$ & dynamic scaling exponent &\\
\hline
$Z_{\bk_F}=Z_{\delta k}$& quasiparticle residue & Eq.~(\ref{sa2})\\
\hline
$\Gamma(P,K;P',K')
\equiv \Gamma(P,K;Q)$& composite vertex & Eq.~(\ref{1.7})\\
\hline
$\Sigma_{\bq_{\pi}}$ & self-energy due to scattering by a single SDW fluctuation & Eq.~(\ref{1.3})\\
\hline
$ \Sigma_{\mathrm{comp}_1}$ & one-loop self-energy due to scattering by composite bosons & Eqs.~(\ref{1.8}) and
(\ref{1.81})\\
\hline
\multirow{2} {*}
{$\Sigma_{\mathrm{comp}_2}$} & two-loop self-energy in the 2D regime & Eqs.~(\ref{2.1}) and
(\ref{2.1.1})\\
& two-loop self-energy in the 1D regime & Eqs.~(\ref{se2_10}) and
(\ref{se2_5})\\
\hline
$ \Sigma_{\mathrm{comp}_3}$ & three-loop self-energy & \\
\hline
$\Omega_{\min}$ & lower boundary of the $1/\Omega$ behavior of the conductivity &$\bar g^2/E_F$ \\
\hline
$\Omega_{\max}$ & upper boundary 
of
 the $1/\Omega$ behavior of the conductivity &$E^2_F/{\bar g}$  \\
\hline
$\Omega_b$ & crossover between FL and NFL forms of the self-energy & $ (v_F \delta k)^2/\bar g$\\
\hline
{$\tk$} &\raisebox{1ex}{ dimensionless distance from the hot spot along the Fermi surface} & $\tk=v_F\delta k/\bar g$\\
\hline
$\bar\Omega$ & \raisebox{1ex}{dimensionless frequency} &$ \bar\Omega=\Omega/\bar g$\\
\hline
$\bo_{\min}$ &\raisebox{1ex}{ dimensionless form of $\Omega_{\min}$}&$\bo_{\min}/\bg=\bar g/E_F$ \\
\hline
$\bo_{\max}$ &\raisebox{1ex}{ dimensionless form of $\Omega_{\max}$}&$\bo_{\max}/\bg=\left(\bar g/E_F\right)^2$ \\
\hline
$\bo_b$ &\raisebox{1ex}{ dimensionless form of $\Omega_b$}  & $\Omega_b/\bg= \tk^2$\\
\hline
$\sigma'(\Omega)$ & real part of the optical conductivity at $T=0$ &\\
\hline
$\sigma'_{\Sigma}(\Omega)$ & $\sigma'(\Omega)$ obtained by taking into account self-energy insertions only& \\
[1ex]
\hline
\end{tabular}
\label{table:notations}
\end{table*}
\ewt
   Among the various deviations from the FL scenario, the linear scaling of $\rho(T)$ with $T$ and
   concomitant
   $1/\Omega$ scaling
   of $\sigma'(\Omega)$ are considered as the most ubiquitous and universal ones.\cite{bruin:2013,hartnoll:2013}
   As the temperature and frequency dependences of the conductivity
   are
    likely to
    originate from the same scattering mechanism,
  the    combination
    of these two scalings
    imposes some important
  constraints on the form of the fermionic self-energy.

 These constraints form the basis of the
   phenomenological
   \lq\lq marginal FL\rq\rq\/ (MFL) theory,~\cite{mfl} which stipulates
    that
 $\Sigma''(\Omega, T)$  scales as $\Omega$ or
     $T$ (whichever is larger)
     at any point on the FS,
      and
      also that  $\Sigma''(\Omega, T)$
      is comparable to
       $1/\tau_{\mathrm{tr}} (\Omega, T)$.
     However,
attempts to
   derive the
     MFL
      form of $1/\tau_{\mathrm{tr}} (\Omega, T)$ microscopically,
      in some model for
      interacting electrons
   near a QCP
   in 2D,
   have been largely unsuccessful.
  Problems
  arise both
  for
   Pomeranchuk ($q=0$) and
   density-wave (finite $q$) types of a QCP, in either charge or spin channel.
For
a
$q=0$  QCP,
 the fact that
 critical scattering involves small momentum transfers
implies that
$1/\tau_{\mathrm{tr}}$
 is smaller than $\Sigma''(\Omega, T)$, and
not only differs
from
 it  in magnitude but also scales differently with $\Omega$ and $T$,
 so that a MFL behavior of the self-energy does not translate into that of the conductivity.
  For
 a
  finite-$q$
  QCP, only
  a
  subset of points on the FS
  around
  hot spots
  is
  affected by criticality, while fermions
  on
  the rest of the FS preserve a regular FL behavior.
  Because
  both
  resistivity and optical conductivity are obtained by averaging over the Fermi surface,
  a
 NFL
  contribution
  from hot fermions
  is
  short-circuited by
   the contributions from other parts of the FS, i.e., the largest contributions to $\rho (T)$ and $\sigma'(\Omega)$ come from
   outside
   the
   hot regions
        (the \lq\lq Hlubina-Rice conundrum\rq\rq\/~\cite{hlubina_rice}).

 In
 the
 MFL phenomenology, this problem is by-passed by assuming that the
 critical
 bosonic field
 is purely local, i.e., that the corresponding
 susceptibility does not depend on $q$ (Ref.~\onlinecite{varma_MFL})
  and diverges at 
the
QCP for all
   momenta.
  Then, on one hand, typical $1/\tau_{\mathrm{tr}}$ 
  is
  of the same order as
$\Sigma''(\Omega)$,
on the other, every point on the FS is hot.  However,
 a scenario
in which
a
bosonic susceptibility softens simultaneously at all momenta is very special and not likely
to be applicable to
all systems
in which
 a
linear-in-$T$ resistivity
 and $1/\Omega$
 scaling
  of the optical conductivity
have been observed.

An alternative
 route,
  which we will follow in this paper,
is to
revisit
 the
\lq\lq conventional\rq\rq\/ theory of a density-wave instability
with soft fluctuations
 peaked near a particular $
 \bq$,
and to
 analyze in more detail  contributions to
 the
 resistivity and optical conductivity coming from fermions outside
 the
 hot regions.

 An
 important
  step in this direction has recently been made by  Hartnoll,
   Hofman, Metlitski, and Sachdev (HHMS) in Ref.~\onlinecite{max_last}.
   They considered the optical conductivity of a 2D metal near a spin-density-wave
   (SDW)
    instability with
    ordering wavevector
    $ \bq_\pi = (\pi,\pi)$
     and focused primarily on the
   contribution to $\sigma(\Omega)$
    coming from fermions in \lq\lq lukewarm\rq\rq\/ regions, located just outside the hot regions.
    Lukewarm fermions
    behave as FL quasiparticles
     even
     right
     at
     the
     QCP,
     but their
     residue
 is small and
     depends on the distance to the hot spot
     in a singular way.
     The leading contribution to $\Sigma''(\Omega)$
      of lukewarm fermions comes not from
     direct scattering by $\bq_\pi$,
      as the initial and final states of this process
       cannot be
      simultaneously
      near the FS, but from
      a composite scattering process which involves two critical bosonic fields.
   Scattering by one field takes
a  fermion out of the FS, while scattering by another
  brings it back to the FS and, furthermore, to the vicinity
  of its original location.
  The self-energy $\Sigma''(\Omega)$
    from  
    composite scattering
  has a FL form
  but
 depends in a singular way
 on
 the distance to
 the nearest hot spot.
 Substituting
this self-energy into the conductivity bubble,
  HHMS
 obtained
  a NFL form of the
 optical conductivity at the smallest $\Omega$:
 $\sigma'(\Omega) \propto
 1/\Omega^{1/3}$
 to two loop-order.
 (Here and in the rest of the paper, $\Omega$ is assumed to be positive so that all non-analytic functions of $\Omega$ are to be understood as real.)

 HHMS further argued that the self-energy
  from
  composite scattering
 comes predominantly from $2k_F$
  processes  (two incoming particles have nearly
 opposite momenta),
 in which case
vertex corrections
do not cancel the self-energy contribution,
 hence the final result for $\sigma'(\Omega)$
  should be
  the same as obtained simply by replacing $1/\tau_{\mathrm{tr}}$ by $\Sigma''$.

   In this paper, we report two
   results. First, we analyzed the interplay between
   the
   self-energy,
   vertex-correction, and
   Aslamazov-Larkin diagrams
   and found that
   the leading contribution to $\sigma'(\Omega)$
   is canceled  between different diagrams.
   In this respect our
      result
     differs
     from
     that by HHMS.  We found, however, that the cancelation does not
    hold
    beyond the
 logarithmic accuracy, and
 even after cancelations
 $\sigma' (\Omega)$
 still diverges at
 $\Omega\to 0$
 as
  $1/\Omega^{1/3}$.

  Second,
 we found that, if the ratio of the spin-fermion coupling to the Fermi energy is treated as a small
 parameter of the theory,
the $1/\Omega^{1/3}$ behavior holds only at asymptotically
low frequencies,
 below some scale $\Omega_{\min}$
which is parametrically smaller than the scale of
 the
 $d-$wave superconducting transition temperature $T_c$.
 At
 higher
frequencies,
the optical conductivity behaves in a MFL way:
$\sigma'(\Omega) \propto 1/\Omega$.
This last behavior holds up to frequencies
  on the order of the fermionic bandwidth.

In the next 
subsection,
we present a brief summary
of the theoretical results for the optical conductivity
 near 
 both
 $q =0$ and finite $q$ critical points, 
 obtained without taking into account
 composite scattering.
Then, in Sec.~\ref{sec:summary}, 
we summarize our results  and describe
their relation to
 those
 by HHMS.

\subsection{Optical conductivity near a QCP: summary of
prior results}
\label{sec:prior}
\subsubsection{
Pomeranchuk QCP ($q=0$)}

A Pomeranchuk-like QCP separates
 two spatially uniform phases, e.g., a paramagnet and ferromagnet.
 This is  a continuous phase transition, and the correlation length of
 long-wavelength order-parameter fluctuations
 diverges at the critical point.
 Scattering of fermions by these fluctuations is strong, but typical momentum transfers ${\tilde q}$ are small compared to $k_F$.
 For a generic FS,\cite{comment_1} the optical conductivity is finite even in the absence of umklapp scattering and disorder,
 and is described by  Eq.~(\ref{ch_1}) with $1/\tau_{\mathrm{tr}}(\Omega)$
 that
 differs from  $\Sigma''(\Omega)$  by
 the
  \lq\lq transport factor\rq\rq\/ $({\tilde q}/k_F)^2$. Critical scaling implies that ${\tilde q}\propto \Omega^{2/z}$,  where $z=3$ is the dynamical exponent for a Pomeranchuk transition.\cite{schofield}
 As a result,
 scaling of the conductivity is different from that of $\Sigma''$, both in 3D and 2D.\cite{schofield,rosch:2005,maslov:2011,comment_1}
 In 3D,
 both
$\Sigma''(\Omega)\propto \Omega^{D/z}$
and
$m^*_{\mathrm{tr}} /m = 1/Z = 1/|\ln \Omega|$
fit the MFL scheme,
 while
the transport scattering rate $1/\tau_{\mathrm{tr}}\sim \Sigma''(\Omega)(\tilde q/k_F)^2\propto \Sigma''(\Omega)\Omega^{2/z}$
 scales as $\Omega^{5/3}$.
Then the conductivity $\sigma'(\Omega)\sim  \Sigma''(\Omega) Z^2 \Omega^{\frac{2}{z} -2}$
 scales as $1/\Omega^{1/3}$ (modulo a logarithm),
which
is very different from
the
MFL, $1/\Omega$ form.
    In 2D,  both $\Sigma'(\Omega)$ and $\Sigma''(\Omega)$ scale as $\Omega^{2/3}$
    (modulo
    logarithmic
    renormalizations
    by
    higher-order processes~\cite{ss_lee,ms}). In this situation, the quasiparticle $Z$ factor
    is frequency dependent and scales as $Z (\Omega) \approx (\partial \Sigma'/\partial \Omega)^{-1}  \propto
    \Omega^{1/3}$
    for frequencies below some characteristic scale.
      As a result,
   $\sigma' (\Omega) \sim
  \Sigma''(\Omega) Z^2 (\Omega) \Omega^{\frac{2}{z} -2}$ tends to a constant value in the low-frequency limit,
  as in an ordinary FL. At higher frequencies, when the $Z$ factor is almost equal to unity,
    $\sigma' (\Omega)$ behaves as
    $\Sigma''(\Omega)\Omega^{2/z}/\Omega^2\propto \Omega^{-2/3}$.
         At even higher frequencies, where $\tilde q\sim k_F$,  $\sigma'(\Omega)$ scales as $\Omega^{-4/3}$.

\subsubsection{Density-wave QCP (finite $q$)}
\label{sec:neq}

A density-wave QCP
separates
a uniform disordered phase and a spatially modulated ordered phase,  e.g.,
a paramagnet
and
 spin-density wave (SDW).
For definiteness,
we
 consider an
SDW
with ordering momentum
$\bq_\pi=(\pi,\pi, \pi)$ in 3D and $
\bq_\pi=
(\pi,\pi)$ in 2D (the lattice constant is set
 to unity).
Because  only a subset of points on the FS
 is
connected by
  $\bq_\pi$,
   critical fluctuations
    affect mostly the fermions
  near such \lq\lq hot
  lines\rq\rq\/ (in 3D) or hot spots (in 2D);
 see Fig.~\ref{fig:hotspot} for the  2D case.
 On
 the rest of the FS,
    the interaction mediated by
    critical fluctuations
    transfers
    a fermion from
    a
    FS point $\bk_F$ to
    $\bk_F+\bq_\pi$, which is away from the FS.
     The  energy
 of the final state
  (measured from the Fermi level)
     $|\ve_{{\bf k}_F + \bq_\pi}|$
     is finite,
     hence
      at frequencies smaller than this scale quantum criticality does not play a role,
 and the self-energy retains the same FL form as away from 
 the
 QCP.

 In 3D, the effective interaction between hot fermions,
 mediated by critical fluctuations,
 yields $\Sigma''(\Omega) \propto \Omega$  and $m^*_{\mathrm{tr}} /m = 1/Z = 1/|\ln \Omega|$,
  same as for 
  a
  $q=0$ QCP.
     Because $q_{\pi}$ is a large momentum transfer, $1/\tau_{\mathrm{tr}} (\Omega)$ is the same  as $\Sigma'' (\Omega)$.
 However the width of the hot region (the distance from the hot line where $\Sigma''(\Omega) \propto \Omega$)
 by itself scales as $\Omega^{1/2}$, hence the conductivity $\sigma'(\Omega)
\propto  \Sigma''(\Omega) Z^2 (\Omega) \Omega^{1/2}/\Omega^2
 $ scales as $1/\Omega^{1/2}$, up to logarithmic
  corrections.

 In 2D, the effective interaction mediated by critical fluctuations
    destroys FL quasiparticles in hot regions,
     which is manifested by
     a NFL form of the
        self-energy. 
        At one-loop level,
          $\Sigma''(\Omega) \sim \Sigma' (\Omega) \propto \sqrt{\Omega}$, and $Z (\Omega) \propto \sqrt{\Omega}$.
   The width of the hot region scales as $\Omega^{1/2}$,
   as in 3D, and the contribution to the conductivity from hot fermions
  tends to a constant value at  vanishing $\Omega$. Beyond one-loop level,
      this
  contribution
  is further reduced by vertex corrections~\cite{max_last}  and
         scales as $\Omega^{a}$
         with $a >0$, i.e., it
          vanishes at $\Omega =0$.

    For $\Omega$ larger than the
      maximum
      value of
      $|\ve_{{\bf k}_F + \bq_\pi}|$
      the whole FS is hot,
      i.e.,
     $\Sigma''(\Omega)$ is approximately the same
    at
    all
     points
     on
    the FS.
    Parametrically,
    this  holds only at rather high energies, larger than the bandwidth (see Sec.\ref{sec:sigma_se} below),
     but the scale may be reduced by a small numerical prefactor.
    If
    the self-energy
still
obeys
the
    quantum-critical
    form
     in this regime,
        $\sigma'(\Omega)$ scales as $1/\Omega$ in 3D,
     and
     either
     as $1/\sqrt{\Omega}$ or
     $1/\Omega^{3/2}$ in 2D, depending on
     whether
     $Z$  in this regime still
     scales as $\sqrt{\Omega}$
     or has already saturated at $Z=1$.~\cite{acs}
\vspace{2 cm}
\subsection{Summary of the results of this paper}
\label{sec:summary}

 Following earlier work, \cite{acs,ms,max_last}
 we adopt
 the
 spin-fermion model as a microscopic low-energy
 theory for a system of interacting fermions at an
 SDW
 instability.
 This model has two characteristic energy scales
 -- the Fermi energy $E_F \sim v_F k_F$ and the effective spin-mediated 
 four-fermion interaction ${\bar g}$.
 To decouple the low-
  and high-energy sectors of the theory, we choose
 the ratio ${\bar g}/E_F$ to be
 small.
   We found that
   in this case the whole FS
   becomes hot only
   at
  $\Omega > \Omega_{\max}  \equiv
   E^2_F/{\bar g} > E_F$.
    At such high energies,
   results
 of the low-energy theory  can hardly  be valid.
  To obtain
  a
  true low-energy behavior, one then has to consider the situation when
  only
   some parts of the FS
   are hot
   while the rest  of it is cold.
   In this case,
   $\sigma'(\Omega)$ is
     given by an average over the FS:
        \beq
        \sigma'(\Omega) \propto \oint d {\bk_F}
        \frac{1/\tau_{\mathrm{tr}}
        ({\bk_F},\Omega)}
       {\lr\Omega /Z_{\bk_F}\rr^2 + \ls 1/\tau_{\mathrm{tr}}
       ({\bk_F},\Omega)
\rs^2},\label{cond}
        \eeq
        where
         ${\bk_F}$ indicates a point on the FS and $Z_{\bk_F} = m/
         m^*_{\mathrm{tr}} ({\bk_F},\Omega)
         $ depends on the
          position of  ${\bk_F}$ relative to the nearest hot spot.

  Hot fermions have the largest self-energy but the smallest $Z_{\bk_F}$, and also the
    width of the hot region
       shrinks as $\Omega$ decreases.
         To two-loop order,
         the
         contribution
          from hot fermions
        to the conductivity $\sigma' (\Omega \to 0)$
   is
         a frequency-independent
         constant,
         which simply adds up to FL-like, constant contributions from the cold regions.
         Beyond two-loop order this contribution
         is further reduced by vertex corrections.\cite{max_last}
\begin{figure}
\includegraphics[scale=0.35]{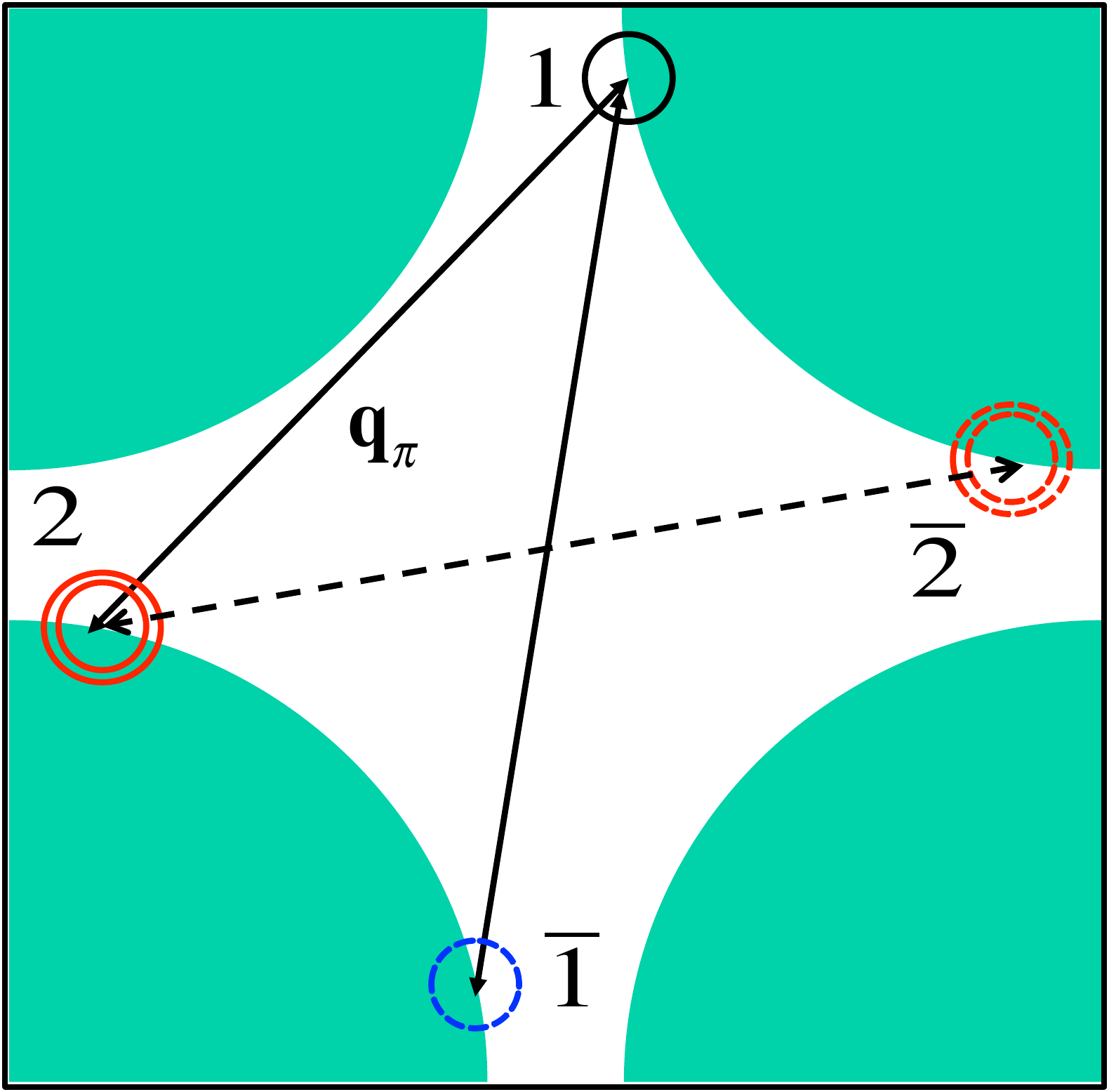}
\caption{A two-dimensional Fermi surface with hot spots denoted by $1$, $\bar 1$, $2$, and $\bar 2$. Hot spots $1$ and $2$ are connected by the spin-density-waver ordering vector $\bq_\pi=(\pi,\pi)$. Hot spots $\bar 1$ and $\bar 2$ are the mirror images of hot spots $1$ and $2$, correspondingly.}
\label{fig:hotspot}
\end{figure}

The issue we
considered, following HHMS,
is whether
 fermions,
 located
away
from the hot regions,
  can
  give rise to a NFL  behavior of  the optical conductivity
   at
an SDW instability.
 Phenomenological models that take into account the interplay between the hot and cold regions in various transport phenomena
          have been considered by 
               several
                 many authors.\cite{previous}
  We considered  this interplay within a microscopic theory.

 At first glance,
 the interaction
  between
 fermions
 located
 away from the hot regions
  is
 unable
  to
 give rise to
  a
   NFL behavior
of $\sigma' (\Omega)$.
Indeed,
    the interaction peaked at
     $\bq_\pi$ moves a
  fermion away from the FS, into a region where its energy
  (measured from the Fermi level)
  is finite. One could then expect
 quantum criticality
 to be
 irrelevant,
  and the
  corresponding
   contribution to $\sigma'(\Omega)$
 to approach a constant value at $T=0$,
   as
  in an ordinary FL.   This reasoning is, however, oversimplified because,
   besides processes with momentum transfer $
\bq_\pi$, there
  also exist
   composite processes, which involve
   an
    even number of
    critical bosonic fields.
  These composite processes have been
  introduced in Ref.~\onlinecite{ms} and considered in detail
   by HHMS.  (For more recent work,
   see Ref.~\onlinecite{peter}.)
 HHMS
 introduced
 a  composite boson,
 with a propagator
  made from
  two
 critical
 propagators of
 the
 primary bosonic fields and two Green's functions of intermediate fermions
(see Fig.~\ref{fig:vertex}).
They found (and we confirmed their result) that the imaginary part of the
 fermionic
 self-energy
  from
  \lq\lq one-loop\rq\rq\/ composite scattering
  (diagram
  $a$ in Fig.~\ref{fig:se})
  scales as
  $\Sigma''_{\text {comp}_1}  (\Omega)\propto \Omega^{3/2}$
  for any point
  on  the FS.
  This singular self-energy, however, does not
 crucially
 affect $\sigma'(\Omega)$ because
   the $\Omega^{3/2}$ term
   comes from small-momentum scattering, and $1/\tau_{\mathrm{tr}} (\Omega)$ is smaller
   than $\Sigma''(\Omega)$ by power of $\Omega$,
   which makes this contribution smaller  than
   a
   regular
   FL term.

A more interesting contribution to the
   self-energy  comes from
  \lq\lq two-loop\rq\rq\/
  composite scattering
 of lukewarm fermions (Fig.~\ref{fig:se_2loop}).
 To be specific, a fermion located   away from
 a
  hot spot by distance $\delta k$ along the FS is
   classified as \lq\lq lukewarm\rq\rq\/,
   if
      $\delta k$ is large enough for the self-energy to assume a FL form with $\Sigma\sim\mathcal{O}(\Omega)+i\mathcal{O}(\Omega^2)$,
       yet small
 enough such that $\partial \Sigma'(\delta k,\Omega) /\partial \Omega >1$.
   The
      characteristic frequency separating
       the hot and lukewarm
       regimes is
       $\Omega_{b}
       \equiv (v_F \delta k)^2/{\bar g}$,
       and the boundary between lukewarm and cold regimes is $|\delta k| \sim {\bar g}/v_F$;
        the lukewarm  behavior
      holds for
          $k_F (\Omega {\bar g}/E^2_F)^{1/2}<|\delta k| < {\bar g}/v_F$.

HHMS put a special emphasis on a particular
two-loop composite process (Fig. \ref{fig:se_2loop}) in which
    intermediate
    fermions belong to
 \lq\lq lukewarm\rq\rq\/ regions near
 {\em opposite}
  hot spots located at $\bk_{\mathrm{hs}}$ and $-\bk_{\mathrm{hs}}$,
  e.g., spots $1$ and $\bar 1$ in Fig.~\ref{fig:hotspot}.
 For a system with a circular FS, such a process
 is often
 called a \lq\lq $2k_F$ process\rq\rq\/
 and we will use this terminology
 here.\cite{comment_2kF}

  Our result for the
   self-energy of
   a lukewarm fermion
    from 
       both
   two-loop
   forward and
   $2k_F$ composite processes
    is
  FL-like at the smallest $\Omega$, i.e.,
  $\Sigma''$ scales as $\Omega^2$; however,
 the
 prefactor of the $\Omega^2$ term
 depends strongly on $\delta k$:
   $\Sigma''_{\text {comp}_2}  (\Omega)
      \propto
       \left[{\bar g}^2 \Omega^2  E_F/(v_F \delta k)^4
       \right]
        \ln^3{\frac{\Omega_b}{\Omega}}$.
        This is parametrically
        (by a factor of
        $k_F/|\delta k|$)
       larger  than $\Sigma''_{\bq_\pi}  \propto {\bar g}^2 \Omega^2 /(v_F \delta k)^3$
      from
      direct scattering by ${\bf q}_{\pi}$.\cite{acs,ms,senthil}
         The HHMS result for $\Sigma''_{\text {comp}_2}  (\Omega)$  differs from ours
         -- they found
         that
         $\Sigma''_{\text {comp}_2}  (\Omega)$
        is
        the same as $\Sigma^{''}_{q_\pi}$, up to a logarithmic
         factor.
      The difference is due to the fact that we considered
      a
       2D FS with
       finite curvature, of order
       $
       1/k_F$, while HHMS assumed that the curvature is zero at the bare level
       but generated
       dynamically
       by the interaction.
       The $\Omega^2$  behavior
       of $\Sigma''_{\mathrm{comp}2}$
        holds up to
       a characteristic scale
       $
       \Omega_{b1}
      \equiv
      (v_F \delta k)^3/({\bar g} E_F) \sim \Omega_b (\delta k/k_F)
      <\Omega_b
      $ (modulo logarithms).
     In the frequency interval $\Omega_{b1} < \Omega < \Omega_b$,
       the curvature of the FS becomes irrelevant, and
         composite scattering becomes effectively
       a
        1D process. In this regime,
       we confirmed
             the HHMS result that the self-energy acquires
       a MFL-like
           form
           with $\Sigma''_{\text {comp}_2}  (\Omega)
      \propto {\bar g} \Omega/(v_F \delta k)$.

      We extended the analysis to the region of larger $\Omega > {\bar g}$ and $v_F |\delta k| > {\bar g}$, not considered by HHMS,
       and obtained the full forms of the
     self-energy due to one-loop and two-loop composite scattering (see Figs. \ref{fig:sigma_vs_omega} and \ref{fig:sigma_vs_k_a}).

A substitution
of
 these full forms
 into the current bubble
 gives
    the \lq\lq self-energy\rq\rq\/  contribution to the optical conductivity,  $\sigma'_{\Sigma} (\Omega)$,
    which does not take into account vertex corrections.
         We found that
     $\sigma'_{\Sigma}(\Omega)\propto
     \ln^3\Omega
     /\Omega^{1/3}$ at
     frequencies below the lowest energy scale of the model, i.e., for
     $ \Omega < \Omega_{\min}
     \equiv
     {\bar g}^2/E_F
     < {\bar g}$.
     The exponent $1/3$
     coincides with the HHMS result
     to two-loop order.
      At higher frequencies,
      $\Omega_{\min} < \Omega < {\bar g}$, we
      found
      that
       two-loop composite scattering of lukewarm fermions
        give rise to
        a
        MFL-like
        conductivity:
$ \sigma'_{\Sigma}(\Omega)\propto 1/\Omega$.
 The
 $1/\Omega$
 behavior actually extends to
  even higher frequencies,
      up to
      $
      \Omega_{\max} \equiv E^2_F/{\bar g}
      >E_F $, at which scale the whole FS becomes hot.
       In the range ${\bar g} < \Omega < \Omega_{\max}$, the dominant contribution to
        conductivity comes from
        direct ${\bf q}_\pi$ scattering.

 Whether
   $\sigma'_{\Sigma} (\Omega)$
gives a good approximation for
 the actual optical conductivity
 depends on
 the interplay between self-energy and vertex-correction insertions into the conductivity bubble.
  HHMS
       argued
        that
  the self-energy and vertex-correction
  diagrams for $2k_F$ scattering
  add up rather than cancel each other
   because
   the current vertices in the self-energy diagram are near the same hot spot,
   while the current vertices
   in the vertex-correction diagram are near the opposite hot spots.
  We obtained a somewhat different result.
Namely, we found that
  that
  the $\ln^3\Omega/\Omega^{1/3}$
 contributions  to $\sigma'(\Omega)$
 are
 canceled
within each of the
 two groups of diagrams.  The first group contains the self-energy and vertex-correction insertions (diagrams $A$ and $B$ in Fig.~\ref{fig:AB}),
while the second one contains
two Aslamazov-Larkin--type diagrams (diagrams $C$ and $D$ in Fig.~\ref{fig:CD}).
 HHMS
 considered only one diagram in each group, and,
 consequently,
 did not find the cancelation.

We found, however, that the cancelation does not
hold
 beyond the
 logarithmic accuracy:
 after cancelations,
 $\sigma' (\Omega
 )$
 still
 diverges at vanishing frequency as
  $\sigma' (\Omega) \propto 1/\Omega^{1/3}$.
 We also found that
   at higher frequencies,
    $\Omega > \Omega_{\min}$,
 the
  vertex corrections change
the
numerical
prefactor but not the functional form of the
$1/\Omega$ scaling,
i.e., the
the final result for the conductivity in this range is
$\sigma'(\Omega) \sim \sigma'_{\Sigma}(\Omega)\propto 1/\Omega$.

The outcome of our analysis is that
 composite scattering
of
lukewarm fermions does
give rise to a NFL behavior of the optical conductivity at an SDW
instability,
namely
\bea
\sigma'(\Omega)\propto\left\{
\begin{array}{l}
\Omega^{-1/3},\;\mathrm{for}\;\Omega<\Omega_{\min};\\
\Omega^{-1},\;\mathrm{for}\;\Omega_{\min}<\Omega<
 {\bar g}.
\end{array}
\right.
\label{sigma_final}
\eea
 The $1/\Omega$ behavior furthermore extends to even higher frequencies, up to $\Omega_{\max}$.
At  ${\bar g} < \Omega < \Omega_{\max}$ it comes from hot fermions.
 These are the key results of this paper.

  The
  rest of the
  paper is organized as follows.
 Section~\ref{sec:sfm} is devoted to the fermionic self-energy. We briefly review
 the
 spin-fermion model near an
 SDW transition in Sec.~\ref{sec:sfm_intro} and discuss
 the
 fermionic self-energy
 for hot, lukewarm, and cold fermions due to
  large-$q$ scattering by a primary bosonic field in Sec.~\ref{sec:1loop}.
In Sec.~\ref{sec:comp}, we consider
 small-$q$ scattering by
    a composite field made
 from
 two
 primary fields. In Sec.~\ref{sec:se_summary}, we summarize the results for the self-energy
 to two-loop order
 and  present the hierarchies
 of crossovers 
in
 $\Sigma$ as a function of $\Omega$ and $\delta k$.  In Sec.~\ref{sec:higher_loops}, we analyze the effect of higher loop corrections.
 Section \ref{sec:3} is devoted to the optical conductivity.
 In Sec.~\ref{sec:sigma_se}, we consider
the contribution to
the
 conductivity
 obtained by inserting the
 fermionic self-energy into the  conductivity bubble.
In Sec.~\ref{sec:cancel},
we
show
the self-energy and vertex-correction diagrams mutually cancel each other {\em if} one neglects the
variations of the quasiparticle residue over the FS.
In Sec.~\ref{sec:no cancel}, 
we show, however,
 that
if this variation is taken account, the NFL power-law singularities in the conductivity [Eq.~(\ref{sigma_final})]
survive after cancelations between the self-energy and vertex-correction diagrams.
In Sec.~\ref{sec:qual}, we explain how this result can be understood in the framework of the semiclassical Boltzmann
equation.
     Our conclusions are presented
     Sec.~\ref{sec:4}.
     For 
     the readers
     convenience 
      the list of notations is given in Table~\ref{table:notations}.

 \section{Spin-fermion model and fermionic self-energy}
 \label{sec:sfm}

\subsection{Spin-fermion model}
\label{sec:sfm_intro}
The spin-fermion model has been discussed  several times in recent literature,\cite{acs,ms,max_last}
so we will be brief.  The model assumes that
the
low-energy physics
  near
 a
   SDW instability
  in a 2D metal
  can be described via
  approximating
  the
  fully renormalized fermion-fermion interaction
   by
an
    effective  interaction in
   the
    spin channel.
    This interaction is mediated by
    nearly-gapless
    antiferromagnetic spin fluctuations:
   \beq
   H_{\mathrm{int}} =
   \sum_{\bk\dots\bpp}
    V(
   \bk-\bp) c^{\dag}_{
   \bk, \alpha} c^{\dag}_{
   \bkp\beta}c^{\phantom{\dagger}}_{
   \bk+\bkp-\bp,\gamma}
 c^{\phantom{\dagger}}_{
 \bp,\delta} \vec{\sigma}_{\alpha\delta}\cdot\vec{\sigma}_{\beta\gamma},
 \label{wed1}
 \eeq
 with
\beq
V(\bk-\bp) = {\bar g} \chi (\bk-\bp),
\label{sa1}
\eeq
where ${\bar g}$ is the effective coupling
(in units of energy), and
 \beq
\chi(\bq)=
\chi (\bq, \Omega =0)
 =
\frac{1}{\xi^{-2} + |{\bf q} - {\bf q}_\pi|^2},
\label{wed2}
\eeq
 is
 proportional to
 the static spin susceptibility
peaked near
$\bq_\pi$.

       The input parameters
       of
        the model are
     ${\bar g}$,
     the
      spin correlation length $\xi$,
       and the Fermi velocity ${\bf v}_F$ which,
       in general,
       depends on the location
       along the FS.
           The coupling ${\bar g}$ is assumed to be smaller than the fermionic bandwidth, otherwise
        the
        low-
        and high-energy sectors
        of the theory do not
       decouple.
       Landau damping of spin fluctuations
       is generated dynamically,
      as
       the
        bosonic self-energy,
        and is
        due to the same spin-fermion coupling (\ref{wed1})
        that
         gives rise to
         the
          fermionic self-energy.

As
 in previous work,
 we consider
 a
  FS
  that
  crosses the magnetic Brillouin zone boundary at eight points --
  the
  hot spots (see Fig.~\ref{fig:hotspot}). There are two hot spots in each quadrant of the Brillouin zone, and four
  out of the eight hot spots are
  the mirror images of
 the other four.

  The Fermi velocities
   at the two hot spots connected by
   $\bq_\pi$
   are given by
       ${\bf v}_F (\bk_{\mathrm{hs}}) = (v_x,v_y)$ and ${\bf v}_F (\bk_{\mathrm{hs}} + \bq_\pi) = (-v_x,v_y)$,
       where the local $x$ axis is along the $(\pi,\pi)$ vector connecting the two hot spots and $y$ is orthogonal to it.
       Instead of $v_x$ and $v_y$,
        it is
        more
        convenient to use $v_F = (v^2_x+ v^2_y)^
       {1/2}
       $ and the angle $\theta$ between
       ${\bf v}_F (\bk_{\mathrm{hs}})$ and ${\bf v}_F (\bk_{\mathrm{hs}} + \bq_\pi)$: $\theta = \arccos{(v^2_x-v^2_y)/v^2_F}$.
        The dependence of the self-energy
          on $\theta$ is not crucial, as long as $\theta$ is not
          too
          small
          and, to shorten the formulas below, we
  assume that  $\theta = \pi/2$ (i.e., $v_x = v_y$).
  This
  assumption
  holds when
  the
  hot spots are located close
   to
  $(0,\pi)$ and symmetry-related points.

 For a FS
 of the type
 shown in Fig. \ref{fig:hotspot}, the fermionic bandwidth is of the same order as the Fermi energy $E_F \sim v_F k_F$, where
 $k_F$ is the
  Fermi momentum
  averaged over the FS.
  At the QCP, where $\xi^{-1} =0$, we then have only two relevant energy scales:
  $E_F$ and ${\bar g}$
  (we remind that $\bar g$ is chosen to be smaller than $E_F$). We will see that the
  frequency dependence of the conductivity
  exhibits crossovers
   at two
   energies:
  \beq
  \Omega_{\min}
  \equiv \frac{{\bar g}^2}{E_F}~~   \mathrm{and}~~ \Omega_{\max}
  \equiv
   \frac{E^2_F}{{\bar g}}.
  \label{ch_3}
  \eeq
  The hierarchy of
 energy
  scales in the model is then
  \beq
  \Omega_{\min}
  < {\bar g}
  < E_F
  < \Omega_{\max}.
  \label{ch_4}
  \eeq
Here and in the rest of the paper, we use weak inequalities ($<$ and $>$) instead of strong ones ($\ll$ and $\gg$) because the actual crossovers are determined not only by parameters but also by numbers, which we do not attempt to compute in this paper. Also, $\sim$ means \lq\lq equal in order of magnitude\rq\rq\/
 and $\approx$ means \lq\lq approximately equal\rq\rq\/.

{\subsection{Self-energy due to $
\bq_{\pi} $ scattering}
\label{sec:1loop}
\subsubsection{One-loop order}

First, we
 consider
the
fermionic self-energy due to scattering mediated by a single spin fluctuation peaked at $\bq_\pi = (\pi,\pi)$.
 A self-consistent
 treatment
 of
 the
 fermionic and bosonic self-energies shows\cite{acs,ms,senthil} that
 close to criticality,
 i.e.,
 when
  $\xi{\bar g}/
  v_F
  $ is larger than unity,
  the
 fermionic self-energy
  depends much stronger on
 the
 frequency than
 on
 the momentum transverse to the FS. The self-energy is also the
 largest
 at the hot spots,
 because
  a fermion scattered
  from one of the
   hot spots lands almost
  exactly on another
   hot
    spot.
  To one-loop order, the
   bosonic self-energy (the Landau damping term) is
  equal to
   $\gamma \Omega$, where
\beq
\gamma = \frac{4
{\bar g}}
{\pi v^2_F}
\label{1.2}
\eeq
 is
 the
 Landau damping coefficient.
  The
  fermionic
   self-energy right at
  the hot spot is
  given by
  \beq
\Sigma_{\bq_\pi} (\bk_{\mathrm{hs}}, \Omega) =
 i \frac{3 {\bar g}}{2\pi v_F \gamma} \left(\sqrt{-i\gamma \Omega + \xi^{-2}} - \xi^{-1}\right).
\label{1.1}
\eeq
The one-loop bosonic self-energy  can be absorbed into the staggered spin susceptibility.
Correspondingly, the effective interaction
 becomes
 dynamic:
  $V(\bq)\to V(\bq,\Omega)$,
where
  \beq
  V(\bq, \Omega) =\frac{\bar g}{\xi^{-2}+
 (\bq-\bq_\pi)^2  - i \gamma \Omega}.
  \label{vd}
  \eeq

As long as $\xi$ is finite,
 $\Sigma_{\bq_\pi} ({\bk_{\mathrm{hs}}}, \Omega)$
 at the lowest $\Omega$
  has a canonical FL form, with
    $\Sigma'_{\bq_\pi} ({\bk_{\mathrm{hs}}}, \Omega)
   \propto \Omega$ and
   $\Sigma''_{\bq_\pi} ({\bk_{\mathrm{hs}}}, \Omega) \propto \Omega^2$. Right at
   the
   QCP, $\xi = \infty$, and
   $\Sigma_{\bq_\pi} ({\bk_{\mathrm{hs}}}, \Omega)$
   has a NFL form:
   $\Sigma_{\bq_\pi} ({\bk_{\mathrm{hs}}}, \Omega)
    \propto \sqrt{i \Omega}$. In this
   case,
   $\Sigma'_{\bq_\pi} ({\bk_{\mathrm{hs}}}, \Omega)$ and $\Sigma''_{\bq_\pi} ({\bk_{\mathrm{hs}}}, \Omega)$
     are of comparable magnitudes, and both are larger than the bare $\Omega$ term in the fermionic propagator.

For a fermion located away from a hot spot, a FL behavior holds even at
criticality ($\xi = \infty$),
but the prefactors of
of the FL forms of
$\Sigma'_{\bq_\pi} ({\bk_F}, \Omega)$ and $\Sigma''_{\bq_\pi} ({\bk_F}, \Omega)$
  depend crucially on the
  distance
  from a hot spot
    along the FS, $\delta k$.
  At $\xi = \infty$,
\bea
&&\Sigma_{\bq_\pi} ({\bk_F}, \Omega)  =
i \frac{3 {\bar g}}{2\pi v_F \gamma} \left(\sqrt{-i\gamma \Omega + (\delta k)^{2}} - |\delta k|\right)
\nonumber \\
&&
\equiv\Omega \frac{3{\bar g}}{4\pi v_F |\delta k|}
S\left(\frac{{\bar g}\Omega}{(v_F |\delta k|)^2}\right),
\label{1.3}
\eea
 where
 \beq
 S(x)=
 \frac{i \pi}{2 x}\lr\sqrt{1-\frac{ 4 i x}{\pi}}-1\rr
 \eeq
 with
 $S(0) =1$ and
  $S(x \gg 1) \approx (i\pi/x)^{1/2}$.
  Finite $\delta k$ plays the same role as
   finite
  $\xi^{-1}$:
  both weaken a NFL behavior of the fermionic self-energy.
 Expanding Eq.~(\ref{1.3})  in $\Omega$,  we obtain
\beq
\Sigma_{\bq_\pi} ({\bk_F}, \Omega)
 =
\Omega \left(\frac{1}{Z_{\bk_F}} -1 \right) +
\frac{3
}{4\pi^2} \frac{{\bar g}^2}{(v_F |\delta k|)^3} i\Omega^2 + \dots
\label{1.4}
\eeq
where
\beq
\frac{1}{Z_{\bk_F}}
= 1 +
 \frac{3{\bar g}}{4\pi v_F |\delta k|}.
\label{sa2}
\eeq
A crossover
between
the
FL and
 NFL
regimes occurs at
the
characteristic energy
\beq
\Omega_{b}
\equiv
 \frac{(v_F |\delta k|)^2}{{\bar g}} \sim \Omega_{\max}  \left(\frac{|\delta k|}{k_F}\right)^2.
 \label{omegab}
\eeq
At $\Omega < \Omega_{b}$, the self-energy  has a FL form, Eq. (\ref{1.4}),
  at $\Omega > \Omega_{b}$,
$\Sigma_{\bq_\pi} ({\bk_F}, \Omega)$
 scales as $\sqrt{\Omega}$.

In the rest of the paper, we will be focusing on scaling dependences while discarding numerical prefactors.
\subsubsection{
Classification of fermions as \lq\lq cold\rq\rq\/, \lq\lq lukewarm\rq\rq\/, and \lq\lq hot\rq\rq\/
in the presence of
$\bq_{\pi}$-scattering}
\label{sec:3aa}
 It is convenient to measure
 energies
 in units of ${\bar g}$ and
 momenta
 in units of  ${\bar g}/v_F$.
 Accordingly, we define
 the
 dimensionless energy and momentum as
 \beq
 {\bar \Omega}
 \equiv \frac{\Omega}{\bar g}, ~~  \tk
 \equiv \frac{v_F \delta k}{{\bar g}}.
 \label{ch_6}
 \eeq
 We also introduce dimensionless
 quantities
  \bea
 {\bar \Omega}_{\max}
  &=&\frac{\Omega_{\max}}{\bar g}
  = \left(\frac{E_F}{\bar g}\right)^2,   ~~  {\bar\Omega}_{\min} =
\frac{\Omega_{\min}}{\bar g} = \frac{\bar g} {E_F},\nn\\
 {\bar \Omega}_b &=& \frac{\Omega_b}{\bar g} = \tk^2.
  \label{ch_7}
 \eea
 In these variables, a crossover
between
the
FL and
 NFL
regimes occurs at
${\bar \Omega} \sim {\bar \Omega}_b = \tk^2$.

  The behaviors of $\Sigma''_{\bq_\pi} (\tk,{\bar \Omega}) \equiv \Sigma''_{\bq_\pi}  (\bk_F,\Omega)$
   and $Z_\tk ({\bar \Omega}) \equiv Z_{\bk_F} $ are shown in Fig.~\ref{fig:qpi_sigma_vs_omega}, as a function of $\bo$ at fixed $\tk$, and in Fig.~\ref{fig:qpi_sigma_vs_k}, as a function of $\tk$ at fixed $\bo$.
 The distinction between cold, lukewarm, and hot
behaviors
 depends on
the
 energy, and is best seen
in Fig.~\ref{fig:qpi_sigma_vs_k},
where
$\Sigma''_{\bq_\pi} (\tk,{\bar \Omega})$ and $Z _\tk({\bar \Omega})$
are plotted
as a function of $\tk$.

We
define
a fermion as
 \lq\lq cold\rq\rq\/ if, at
given ${\bar \Omega}$, $\Sigma''_{\bq_\pi}( \tk,{\bar \Omega})$ has a FL,
$\bo^2$,
 form and $Z_{\tk} ({\bar \Omega}) \approx 1$.
 The cold regions are indicated in the right panels of Figs.~\ref{fig:qpi_sigma_vs_omega} and \ref{fig:qpi_sigma_vs_k}.
 With this definition, cold fermions are described by
 a
 weak-coupling FL theory, and thus contribute to
 the FL-like part of
 of the conductivity.
Next, we define a fermion as \lq\lq lukewarm\rq\rq\/ if   $\Sigma''_{\bq_\pi}(\tk,{\bar \Omega})$ still has a FL form but
$Z_{\tk}(\bo)$
is smaller than
unity
 and
 scales
 as $\tk$
 for
  $\tk<1$. The corresponding regions are shown in the left panels of Figs.~\ref{fig:qpi_sigma_vs_omega} and   \ref{fig:qpi_sigma_vs_k}.
 Finally,
 we define a fermion as \lq\lq hot\rq\rq\/
 if  $\Sigma''_{\bq_\pi}(\tk,{\bar \Omega})$ has a NFL form, i.e., $\Sigma''_{\bq_\pi}\propto\sqrt{
\bar \Omega
 }$
 to one-loop order.
  With this last definition,
 the
 hot region gradually extends with increasing frequency and,
  for ${\bar \Omega} >1$, includes
  the range where the
  quasiparticle residue
  is close to
  unity.
 \begin{figure}
\begin{centering}
\includegraphics[scale=0.35]{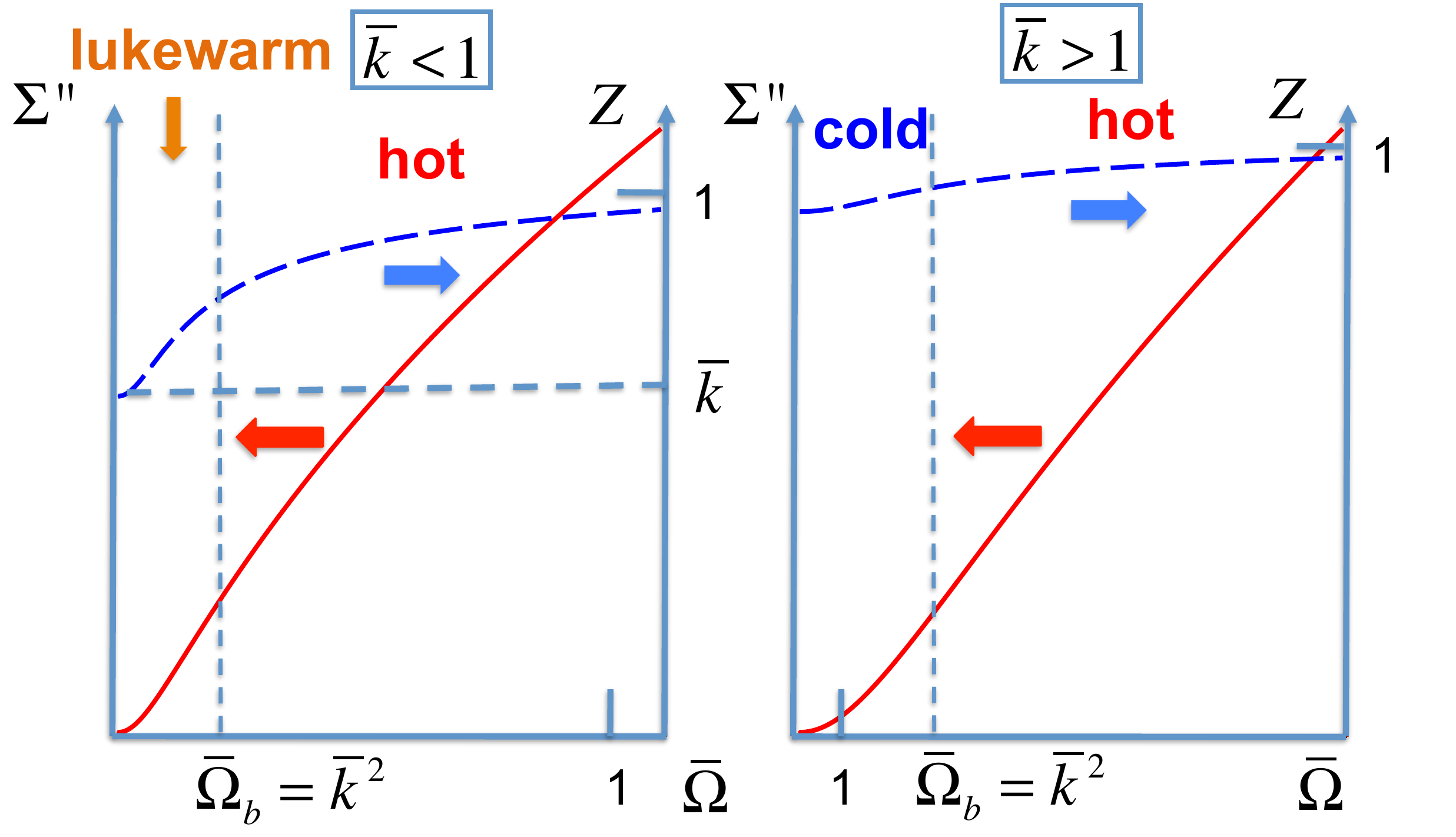}
\end{centering}
\caption{
Imaginary part of the fermionic self-energy from $\bq_\pi$ scattering, $\Sigma''_{\bq_\pi}(\tk,\bo)$ (left axis), and the quasiparticle residue, $Z_{\tk}(\bo)$ (right axis), as a function of $\bo$. Left panel: $\tk<1$; right panel: $\tk>1$. Dimensionless variables are defined according to Eqs.~(\ref{ch_6}) and (\ref{ch_7}).}
\label{fig:qpi_sigma_vs_omega}
\end{figure}
  One could, in principle, separate this range from a truly \lq\lq hot\rq\rq\/, NFL  behavior at ${\bar \Omega} <1$, where
  not only $\Sigma''_{\bq_\pi} (\tk,{\bar \Omega})$ scales as $
   \sqrt{\bar \Omega}$ but also the quasiparticle residue is smaller than unity. We will not do this, however, because our
  main goal is to distinguish between
  the FL-and
  NFL-like forms of the optical conductivity,
   which is determined primarily by $\Sigma''$.
   Besides, as we discuss in Sec.~\ref{sec:hclw}, the distinction between hot and lukewarm fermions becomes more 
   subtle
   once
   composite scattering
   is taken into account.
 \begin{flushleft}
 \begin{figure}[tbp]
\includegraphics[scale=0.25]{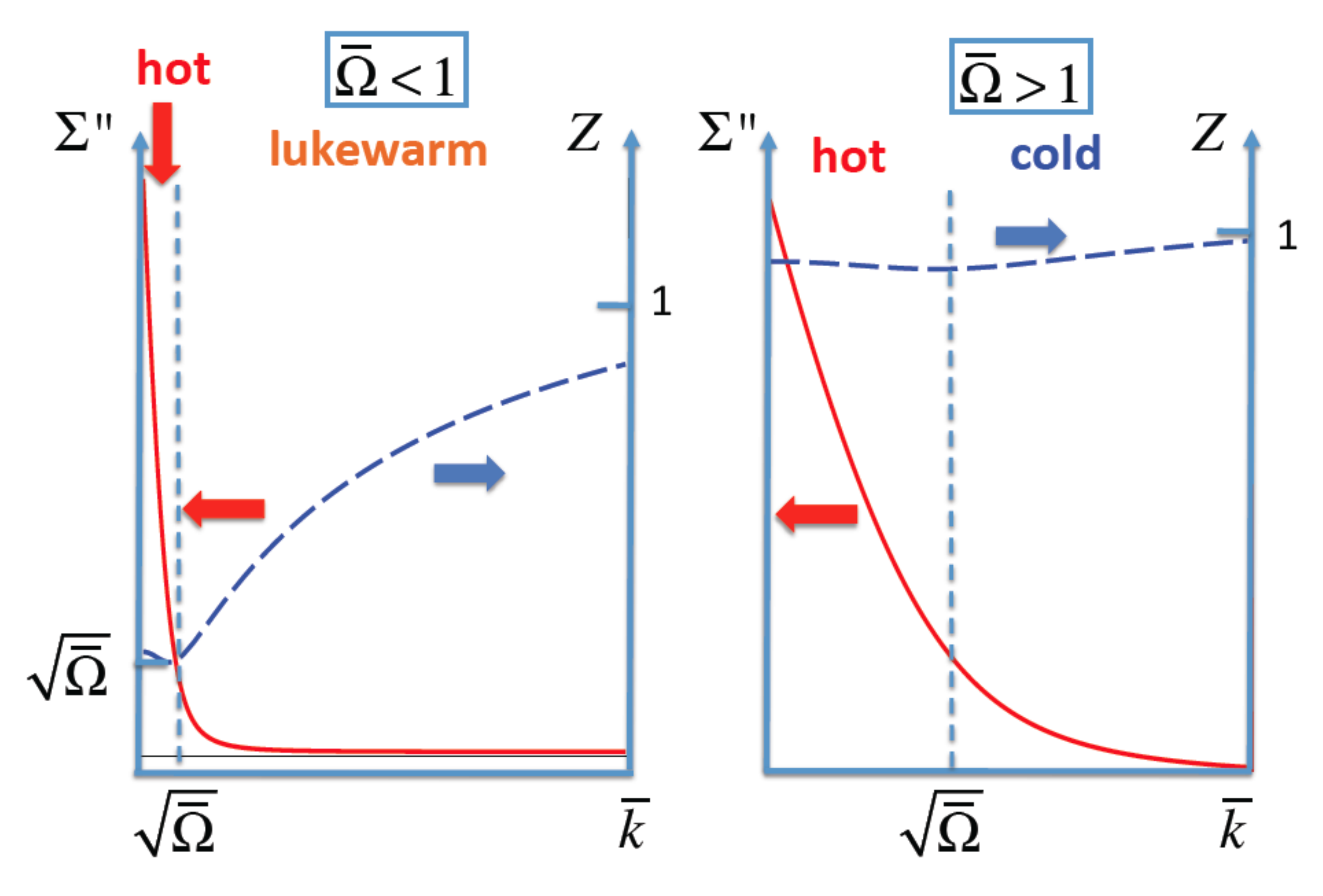}
\caption{Imaginary part of the fermionic self-energy from $\bq_\pi$ scattering, $\Sigma''_{\bq_\pi}(\tk,\bo)$ (left axis), and the quasiparticle residue, $Z_{\tk}(\bo)$ (right axis), as a function of $\tk$. Left panel: $\bo<1$; right panel: $\bo>1$. Dimensionless variables are defined according to Eqs.~(\ref{ch_6}) and (\ref{ch_7}).}
\label{fig:qpi_sigma_vs_k}
\end{figure}
\end{flushleft}

 With these definitions,
 at fixed $|\tk| <1$
 a crossover between
 the lukewarm
 and hot
  regimes
 occurs at $\bar\Omega\sim{\bar \Omega} _{b}
  \sim \tk^2 <1$.
 There is no range  for the cold behavior
 in this case.
At  $|\tk|
 >1$, on the contrary, there is
 no range for the lukewarm behavior:
 as the frequency increases,
 the crossover between the cold and hot regimes occurs at ${\bar \Omega}\sim
\bar \Omega_b\sim\tk^2>1
$.
Again,
   we will see in Sec.~\ref{sec:hclw},
 that the
 structure of
 crossovers
 changes once
 composite scattering is included.

\subsubsection{Higher-order terms and the accuracy of the perturbation theory}
\label{sec:3a}
 A
peculiar feature  of the spin-fermion model
near criticality
 is  the absence of a
 natural
 small parameter,
 even
 if
     the coupling ${\bar g}$ is
     chosen to be
     small (compared with the Fermi energy).
   Although
   the loop expansion
   goes formally
    in powers of ${\bar g}$,
   a
    dimensionless parameter
   of
   the
    perturbative expansion is not ${\bar g}/E_F$ but rather
     $\delta
     \equiv
      {\bar g} v^2_F/\gamma$, where
     $\gamma$ is the Landau damping coefficient
     [Eq.~(\ref{1.2})].
     Because $\gamma$ by itself scales as ${\bar g}/v^2_F$, the spin-fermion coupling
     drops
     out, and  $\delta
     \sim 1$, i.e., higher-order terms in the loop expansion for the self-energy are of the same order as
     the one-loop expression.
    The functional forms of the leading terms in the
      higher-loop
      fermionic and bosonic  self-energies
     are
      then
       the same as the one-loop result.
     On a more careful look, however,
     the
     two-loop
       terms contain additional
        logarithmic factors ($\ln \Omega$ or $\ln|\delta k|$, depending on the regime), and the powers of
        logarithms
         increase
        with the loop order.\cite{acs,ms,senthil}

One can try to
control
the logarithmic series
by extending the model to $N$ fermionic flavors
 and taking the limit $N
 \to\infty
 $. In this case, the Landau damping parameter becomes of order $N$ and the expansion parameter
becomes  small
as $1/N$.
     However, it
     has recently been
     found that this
     procedure
     brings the theory only under partial control because some
     perturbative terms
     from
      $n \geq 4$-loop orders
       do not contain $1/N$.
      \cite{ss_lee,ms,senthil}
      Having this in mind, we will
        keep $N=1$ in our analysis and check whether higher-order terms in the loop expansion introduce
        a
         qualitatively new behavior, not seen at lower orders.
To be more specific, in
the next section we discuss
  how higher-loop terms affect the structure of the imaginary part of the self-energy
    for a lukewarm fermion.
     At one-loop order, $\Sigma'' ({\bk_F}, \Omega) \propto {\bar g}^2 \Omega^2/(v_F |\delta k|)^3$.
    It turns out that, beyond the
    one-loop level,
     there are contributions
         that
     give parametrically larger $\Sigma'' ({\bk_F}, \Omega)$,
      with a stronger dependence either on $\Omega$ or on $\delta k$.  To analyze
      these
       terms, we follow Ref.~\onlinecite{max_last} and introduce the notion of composite scattering.

 \subsection{Self-energy due to  composite scattering}
\label{sec:comp}
\subsubsection{Composite scattering
vertex}
\label{sec:vertex}

  In a composite scattering process,
a fermion located on
 the FS undergoes an even number of scatterings by the
 the primary bosonic field
 with a
  propagator
 peaked at
 $\bq= \bq_\pi$ [Eq.~(\ref{vd})]. At intermediate stages, the fermion can move far away from the FS but
  it
   eventually comes back to
 the
  vicinity of the point of origin.

\begin{figure}
\includegraphics[scale=0.35]{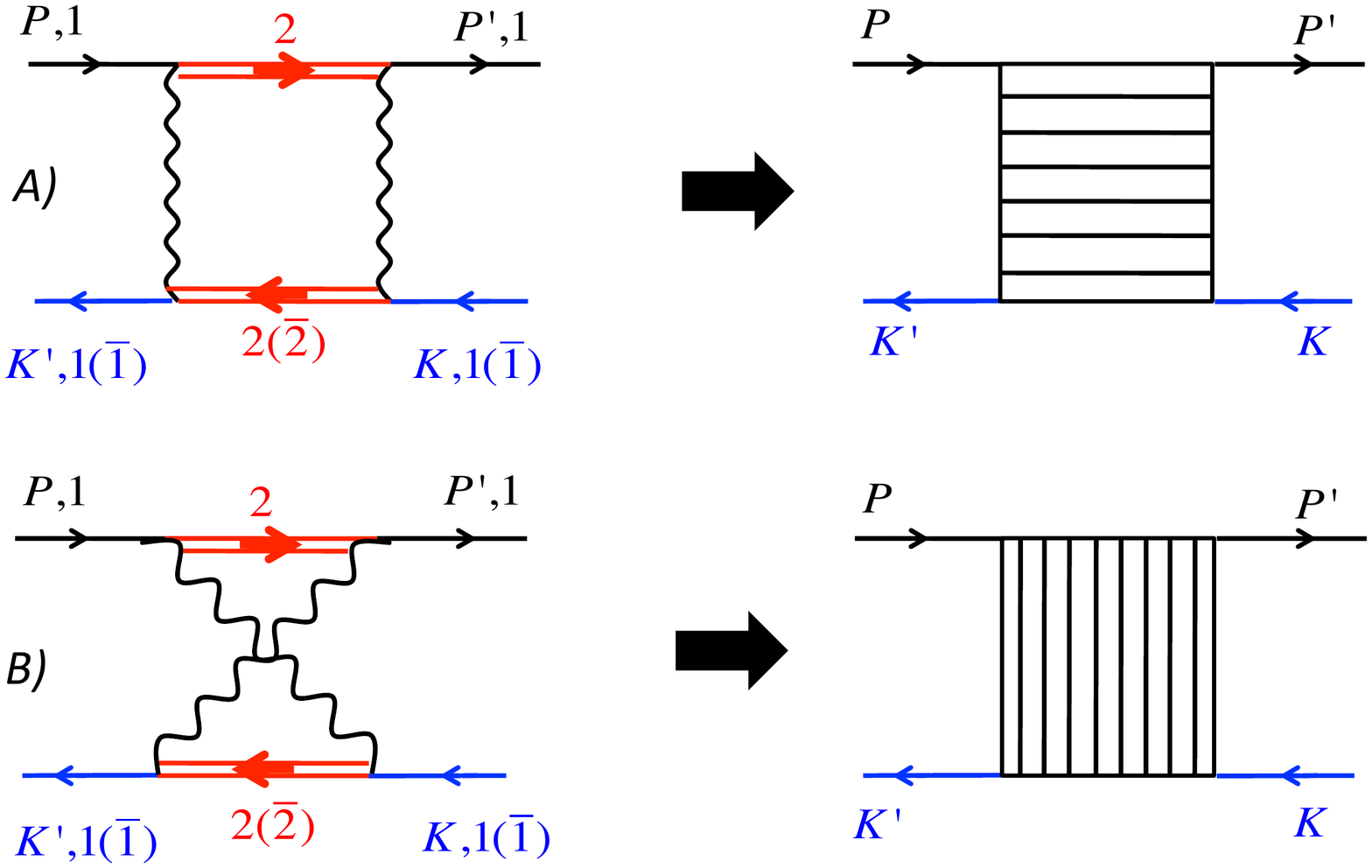}
\vspace{-1 in}
\hspace{-0.4in}
\includegraphics[scale=0.35]{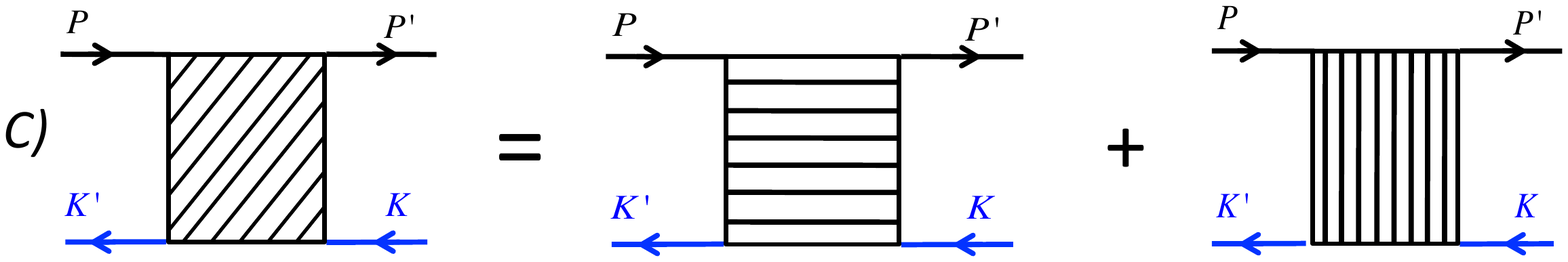}
\vspace{-6 cm}
\caption{
Composite vertex with intermediate processes
in the particle-hole ($A$) and
  particle-particle channel  ($B$).
   Labels $1$, $\bar 1$, $2$, and $\bar 2$ correspond to hot spots in Fig.~\ref{fig:hotspot}. The notation \lq\lq $P,1$\rq\rq\/ means that the 2D component of $P=(\bp,\Omega_p)$ is near hot spot $1$, and
   similarly for other $2+1$ momenta.  In a forward-scattering event, the initial and final states belong to the same hot spot, e.g., spot $1$.  A $2k_F$ event involves fermions from opposite hot spots, $1$ and $\bar 1$. Double solid lines denote off-shell fermions at hot spots $2$ (forward scattering) and $\bar 2$ ($2k_F$ scattering).
The total composite vertex ($C$) is a sum of vertices $A$ and $B$.
 The wavy line denotes the effective dynamic interaction carrying a momentum close to $\bq_\pi$ [Eq.~(\ref{vd})].}
\label{fig:vertex}
\end{figure}
Composite  scattering processes
can be viewed as
$2n$-loop processes in terms of the original
spin-fluctuation propagator.
However,
 it is more convenient to  view them as 
 separate 
 a
 subclass of processes,  which
involve
  small momentum scattering governed by new composite
 vertices.
 The
  lowest-order
 composite vertex
 involves two scatterings by momenta
 $\bq_\pi
 + {\bf q}_1$ and ${\bf q}_\pi
 + {\bf q}_2$, in which both $q_1$ and $q_2$ are small.
 One can construct two vertices of this kind, with
  intermediate
  processes in the particle-hole and particle-particle channels.
    Such two vertices are depicted in
    panels $A$ and $B$ of Fig.~\ref{fig:vertex}, correspondingly.
Each
 vertex is a
convolution of two spin-fluctuation propagators
with two propagators of intermediate fermions.

As an example, we analyze the particle-hole vertex
 (panel $A$ in Fig.~\ref{fig:vertex})
for composite scattering between fermions with
the initial (2+1) momenta
$P = ({\bf p},
\Omega_p)$ and $K = ({\bf k}, \Omega)$,
and final momenta $P'=P-Q$ and $K'=K+Q$,
with $Q = ({\bf q}, \Omega_q)$.
To simplify calculations, we will first compute the composite vertex and self-energy
 in Matsubara frequencies and then
 perform
 analytical continuation.
 Neglecting spin indices
 for a moment, we have
\bea
&&\Gamma
(P,K; Q)
={\bar g}^2 \int \frac{d^3 Q_1}{(2\pi)^3} G(P+ Q_\pi+ Q_1) \label{1.5}\\
&& \times G(K+Q+Q_\pi+Q_1) \chi (Q_\pi+Q_1) \chi (Q_\pi+Q_1+Q),\nn
\eea
where $Q_\pi=(\bq_\pi,0)$.
We choose the initial states to be  on the FS with (2+1) momenta $P=P_F\equiv (\bp_F,\Omega_p
)$ and $K=K_F\equiv (\bk_F,\Omega)$
 with small $\Omega
$ and $\Omega_p$, and at distances $\delta p$ and $\delta k$
 from the corresponding
 hot spots.
 Later,
 we will
 choose $\bp_F$ and $\bk_F$ to be either near the same hot spot, e.g., hot spot $1$, or near
 diametrically opposite hot spots, e.g, hot spots $1$ and $\bar 1$ in Fig.~\ref{fig:hotspot}.

One can make sure that the largest contribution to  $\Gamma
(P_F,K_F; Q)$ at small $Q$ comes from the range
 of
integration
 when all three components of
 $Q_1$ are small.
 Such
 a
 scattering event
 transfers fermions from
 the
 points $\bp_F$ and $\bk_F$ on the FS
 to 
 the
 intermediate states with $2$-momenta about
$\bp_F + \bq_\pi$ and $\bk_F + \bq_\pi$, while
 changing
 the
 frequencies
 only by a small amount (of order $\Omega_q$). Since the energies of 
 the
 intermediate fermions, $\ve_{\bp_F + \bq_\pi}$ and $\ve_{\bk_F + \bq_\pi}$, are large compared with their frequencies, $\Omega_p+\Omega_{q_1}$ and $\Omega
 +\Omega_q+\Omega_{q_1}$,
  the corresponding fermionic Green's functions can be approximated by
  their static limits,
 $1/v_F \delta p$ and $1/v_F
  \delta k$,
 and taken outside the integral.
   The
   remainder of
   $\Gamma
   (P_F, K_F; Q)$
    contains
    a product of
    two spin propagators
   integrated over
   the
   $2+1$ momentum $Q_1 = ({\bf q}_1, \Omega_{q_1})$
   \beq
   \int  \frac{d^2 q_1 d \Omega_{q_1}}{(2\pi)^3} \frac{1}{{\bf q_1}^2
  + \gamma |\Omega_{q_1}|} \frac{1}{({\bf q}_1 + {\bf q})^2
  + \gamma |\Omega_{q_1} + \Omega_q|}.
   \label{1.6}
   \eeq
   The integral diverges logarithmically
   at the lower limit
   and, to logarithmic
   accuracy, yields $(1/
  4\pi^2 \gamma) \ln\ls{\Lambda^2
   /({\bf q}^2
   + \gamma |\Omega_q|)}\rs$,
   where
   $\Lambda\sim\min\{
    |\delta k|,
   |\delta p|\}$.
   Using
    Eq.~(\ref{1.2}) for $\gamma$,
   one
   obtains~\cite{max_last}
   \beq
  \Gamma
  (P_F, K_F; Q)
  =
 \frac{\bar g}{16\pi}\frac{1}{\delta k \delta p} \ln\frac{\Lambda^2}{
  q^2
  +\gamma |\Omega_q|}.
  \label{1.7}
  \eeq
Notice that the vertex in Eq.~(\ref{1.7})  depends only on $\Omega_q$, although
 the
 original
  vertex in Eq.~(\ref{1.5})
  depends
  in general
  on
  all
  the three frequencies:
 $\Omega_p$, $\Omega
 $, and $\Omega_q$.
The dependence on $\Omega_p$ and $\Omega
$
was
eliminated by
  replacing
  the intermediate states' Green's functions by
    their
   static values.
  This circumstance will be crucial for cancelations between diagrams for the conductivity in Sec.~\ref{sec:cond}.

The particle-particle vertex (Fig.~\ref{fig:vertex}, panel $B$)
 differs
 from
  the particle-hole one
only in that the $(2+1)$ momentum on the double line is replaced by $K-Q_1-Q_\pi$. However, since the intermediate fermions are again away from the FS, their Green's functions can also be replaced by
their static values,
 $1/v_F\delta p$ and $1/v_F\delta k$, upon which the particle-particle vertex becomes equal to the particle-hole one.
 The total vertex (Fig.~\ref{fig:vertex}, panel $C$) is equal to the sum of the particle-hole and particle-particle ones.

It is instructive to compare the composite vertex with the
 bare
  interaction $V(\bk-\bp)$ in Eq.~(\ref{sa1}).
 First, we observe that  the composite vertex scales as ${\bar g}$ rather than ${\bar g}^2$ despite
 the fact
  that it is
  formally
  of
  second order in the spin-fluctuation propagator.
 One
 factor of
 $\bar g$
 is canceled out by the Landau damping coefficient $\gamma$ in the denominator.
 Next, for fermions in the lukewarm region, $V(\bk-\bp) \sim {\bar g}/
 \left[(\delta k)^2 + (\delta p)^2
 \right]$.
 For comparable $\delta k$ and $\delta p$,
 the original and composite vertices are
 then
 both of order ${\bar g}/(\delta k)^2$, but the composite vertex has an additional logarithmic factor
 $\ln\frac{
 \Lambda^2}{
  q^2
  +\gamma |\Omega_q|}$.
  This extra logarithm gives rise to an additional
  factor of
  $\ln {|\delta k|}$
  in the ${\mathcal O}(\Omega)$ term in the real part of the self-energy.
  In addition,
  the same logarithm
  leads to two effects in the imaginary part of the self-energy.
  The first one is
   a nonanalytic, $\Omega^{3/2}$ term
   due to one-loop composite scattering,
   discussed in Sec.~\ref{sec:1LC}.
   The second
   one
   is the
    enhancement of the
     prefactor
   in  the self-energy
   due
   to
    two-loop
    composite scattering, discussed in Sec.~\ref{sec:3b}. We will consider the one-
   and two-loop composite
   processes separately.

  \subsubsection{
  \lq\lq One-loop\rq\rq\/
  self-energy
  due to
  composite scattering}
\label{sec:1LC}
  \begin{figure}

\includegraphics[scale=0.33]{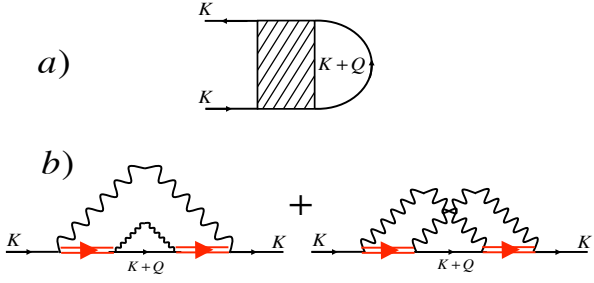}
\vspace{-3 cm}
\caption{
$a)$: One-loop self-energy due to composite scattering. The hatched box is the composite vertex in Fig.~\ref{fig:vertex}$C$.
$b)$: Equivalent representations of diagram $a)$ in terms of the original interaction in Eq.~(\ref{vd}).
As in Fig.~\ref{fig:vertex}, a double
line denotes an off-shell fermion.}
\label{fig:se}
\end{figure}
  The
  lowest-order
  contribution to the self-energy due to composite scattering
  is   given by diagram
    $a$
    in Fig.~\ref{fig:se}. In terms of
  the
  original
  interaction (wavy line),
 this diagram
 is equivalent to diagram
  $b$.
 Explicitly,
    \beq
  \Sigma_{\mathrm{comp}_1}(
 \bk_F, \Omega) =
 \int \frac{d^3Q}{(2\pi )^3}G(K_F+Q)\Gamma(K_F,K_F+Q;Q).
 \label{comp1}
    \eeq
      The intermediate fermion's momentum  is
    $P_F = K_F +Q$, i.e., if $Q$ is small, $P_F$ should be close to $K_F$.
    Integrating over the component of $\bq$ transverse to the FS and
    then
    over $\Omega_q$,
    using an explicit form of $\Gamma$ from Eq.~(\ref{1.7}),
    and
    continuing
    to real frequencies,  we obtain
          \bea
  \Sigma''_{\mathrm{comp}_1}(\bk_F, \Omega) &
  \sim
 & \frac{\bar g}{(v_F \delta k)^2}
 \mathrm{Im}\left[i\Omega
  \int^\infty_0
  dx
 \ln{\left(1
-i\Omega
\frac{{\bar g}}{
x^2
}\right)}\right]
\nn\\
&&
\sim
\frac{
 (\bar g)^{3/2}
\Omega^{3/2}
}
{(v_F\delta k)^2}, \label{1.8}
  \eea
 where
 $x = v_F \delta q$ and $\delta q$ is
 a component of $\bq$
 tangential to the FS.

A non-analytic, $\Omega^{3/2}$ dependence of $\Sigma''_{\mathrm{comp}_1}(\bk_F, \Omega)$
 from one-loop composite scattering was
obtained
 by HHMS
  along with a
nontrivial  logarithmic
 prefactor.
  Observe that,
   for the lowest $\Omega$, this form of $\Sigma''_{\mathrm{comp}_1}(\bk_F, \Omega)$
  holds for  all $\delta k$
  outside
  the hot region.
    The only condition of validity
    of Eq.~(\ref{1.8}) is the smallness of $q^2 \sim \gamma \Omega_q \sim \gamma \Omega$ compared with
  $(\delta k)^2$, i.e., $\Omega$ must be smaller than $\Omega_b$,
  where $\Omega_b$ is defined by Eq.~(\ref{omegab}).
  This is the same condition which separates lukewarm (or cold) fermions from hot fermions for ${\bf q}_\pi$ scattering.

Comparing Eq.~(\ref{1.8})
to
 the $\Omega^2$ term in
 $\Sigma''_{\bq_\pi} (\bk_F, \Omega)$
 due to
 $\bq_\pi$
  scattering [Eq.~(\ref{1.4})], we see that one-loop composite scattering
 gives a larger contribution to
  the imaginary part of the self-energy
  at
   frequencies
  $ \Omega < \Omega_b$, i.e., fermions
  outside the hot regions
  are damped stronger
  by composite scattering
 than by $\bq_\pi$  scattering.
 At $\Omega > \Omega_b
 $,
Eq. (\ref{1.8}) is no longer valid
 because  typical $q
  \sim
  \sqrt{
  \gamma \Omega}$
 become comparable to $\delta k$,
  and the logarithmic singularity in the composite vertex disappears.
 At
 these
  frequencies,
 $\bq_\pi$ scattering yields
 $\Sigma'' (\Omega) \propto \sqrt{\Omega}$ and one-loop composite scattering  adds only additional logarithmic factors to this dependence.\cite{acs,ms}

For comparison with other contributions, it is convenient to
re-write Eq.~(\ref{1.8})
in terms of
 the
 dimensionless variables from Eqs.~(\ref{ch_6}) and (\ref{ch_7}),
which yields
       \beq
  \Sigma''_{\mathrm{comp}_1}(\tk, {\bar \Omega})
  \sim \bar g  \frac{{\bar \Omega}^{3/2}}{\tk^2}
 \label{1.81}
  \eeq
valid for
${\bar \Omega} < {\bar \Omega}_b \sim \tk^2$.

 \subsubsection{
 Two-loop
 self-energy
 due to
 composite scattering}
 \label{sec:3b}
 \paragraph{{\bf Main features of two-loop composite scattering}.}
Another route to obtain a large imaginary part of the self-energy  is to
make use
 of the
singularities in
 the dynamic part of the particle-hole polarization bubble
 both
 at small and
  $2k_F$  momentum transfers.
\cite{millis_chitov,chubukov:03,chubukov:05,chubukov:06,aleiner:06,rosch:07,meier:11}
The polarization
 bubble
 behaves as $\Omega/q$
for
 $\Omega/v_F<q<k_F$,
 and as
 $\Omega\theta(2k_F-q)/\sqrt{2k_F-q}$
 for $q$ near $2k_F$.
  In a generic 2D FL liquid,
   both types of
   singularities
   give rise to
   a non-analytic
   form
    of the self-energy,
     $\Sigma'' (\bk_F, \Omega) \propto \Omega^2
      \ln\Omega$,
     which differs from the canonical FL form, $\Omega^2$, by a \lq\lq kinematic\rq\rq\/ logarithmic factor.
     \begin{figure}[h]
\begin{centering}
\includegraphics[scale=0.8]{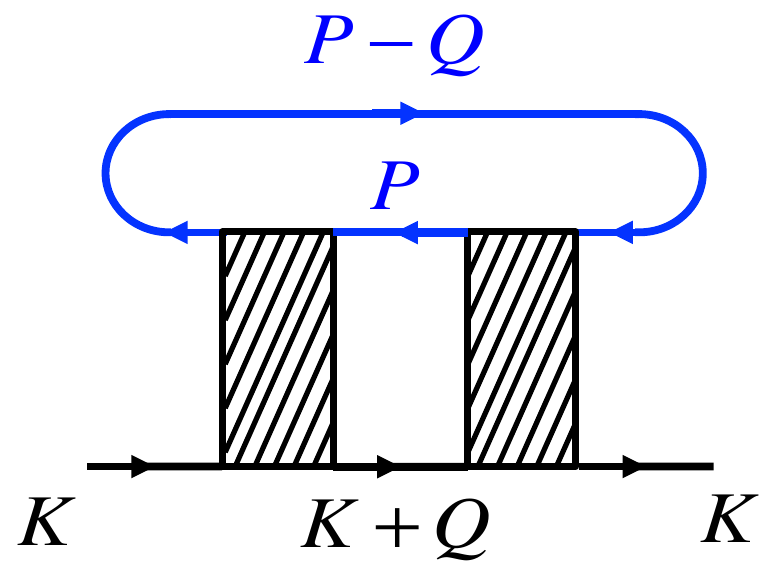}
\end{centering}
\caption{
Two-loop self-energy due to composite scattering.The hatched box is the composite vertex in Fig.~\ref{fig:vertex}$C$.}
\label{fig:se_2loop}
\end{figure}

     Similar singularities occur also in the two-loop self-energy from composite scattering, shown in Fig.~\ref{fig:se_2loop}.
     In  case of forward scattering, all the three internal fermions (with $2$-momenta
     $\bp-\bq$, $\bp$, and $\bk+\bq$) are near the same point on the FS as the initial one (with $2$-momentum $\bk$). In case of  $2k_F$ scattering, one of the internal momenta is near $\bk$ while the remaining two are near the diametrically opposite point, $-\bk$.
   In terms of the composite vertex,
 $\Gamma
 (P, K; Q)$ with 
 $2$-momentum transfer $\bq$,
 both processes correspond to small $q$,
  with typical $v_F q \sim \Omega_q \sim \Omega$,
  while
  $\bk
  $ is either near
  $\bp
  $ (forward scattering) or
  near
  $-\bp
  $
 ($2k_F$ scattering).

  A special feature of
  composite scattering
 is an additional logarithmic
 singularity of the
  composite
  vertex [cf. Eq.~(\ref{1.7}).]
For both
 forward and $2k_F$ scattering,
 the vertex
 can be approximated
by
\beq
\Gamma\equiv \Gamma
(P_F, K_F;Q) \sim \frac{\bg}{(\delta k)^2} \ln \frac{Z_{\bk_F} v_F |\delta k|}{\Omega}.
 \label{gamma_const}
\eeq

 Although the $
 i
 \Omega^2
 \ln \Omega$ term in the
 self-energy of
 a 2D FL comes from
 processes  in which
 all fermionic momenta are either  parallel or antiparallel to each other,
 it would be incorrect to think that these processes occur as if the system were one-dimensional (1D).
 Indeed,
  the information about 2D geometry of the FS
  in encoded in
  the  prefactor of the $
 \Omega^2
  \ln \Omega$
 term, which contains
 the
 local
 curvature
 of the FS.
 Namely, if
  the single-particle dispersion is parameterized as
   \beq
  \ve_{{\bf k}_F +{\bf q}} = v_F q_\perp + (\delta q)^2
  /2m^*,
  \label{disp}
  \eeq
   where $q_\perp$ and $\delta q$ are the components of $\bq$ along the normal and tangent to the FS, correspondingly,
  the prefactor
  of  the $
  i
  \Omega^2
  \ln \Omega$ term
  is proportional to
  $m^*/v^2_F$
  and thus diverges in the 1D limit, which corresponds to $m^*\to\infty$ at $v_F=\mathrm{const}$.
  (Although $m^*$ does vary along the FS, we will not display this dependence explicitly.)

    In a generic FL, the 1D regime, in which the curvature can be neglected, sets in only at energies above $\sim k_F^2/m^*$, which is comparable to the bandwidth, and is thus of little interest unless the FS has nested parts with large $m^*$.
   In the model considered in this paper, however, the 1D regime is realized even in the absence of nesting, and sets in at energies above
   the characteristic scale
   which is smaller than the bandwidth by the small parameter of the model, $\bar g/E_F$.
  In the following two sections, we consider the 2D and 1D regimes of composite scattering.

\paragraph{{\bf Two-dimensional regime.}}

We begin with the 2D regime, in which the FS curvature cannot be neglected.
The diagram for
the
two-loop self-energy
in Fig.~\ref{fig:se_2loop}
 reads
\bwt
\bea
\Sigma_{\mathrm{comp}_2} (\bk_F, \Omega)&=&-\int \frac{d\Omega_q}{2\pi}\int \frac{d\delta q}{2\pi } \int \frac{dq_\perp}{2\pi}
\int \frac{d\Omega_p}{2\pi}\int \frac{d\delta p}{2\pi}\int \frac{dp_\perp}{2\pi} \frac{1}
{\frac{{i\Omega_p}}{Z_\bp}-v_Fp_\perp-\frac{(\delta p)^2}{2m^*}}
\frac{1}{\frac{i(\Omega_p-\Omega_q)}{Z_{\bp-\bq}}-v_F(p_\perp-q_\perp)-\frac{(\delta p-\delta q)^2}{2m^*}}\nn\\
&&\times\frac{1}{\frac{i(\Omega+\Omega_q)}{Z_{\bk+\bq}}-v_Fq_\perp-\frac{(\delta k+\delta q)^2}{2m^*}}\Gamma^2(K,P;Q),
\label{se2}\eea
\ewt
where it is understood that all the $Z$ factors are evaluated on the FS.
First, we integrate the product of two Green's function in the first line over $p_\perp$ and re-define $\delta p$ by absorbing the $v_Fq_\perp$ term, and then integrate the Green's function in the second line over $q_\perp $. These two steps give
\bwt
\bea
\Sigma_{\mathrm{comp}_2} (\bk_F, \Omega)&=&\frac{1}{2v_F}\int \frac{d\Omega_q}{2\pi}\int \frac{d\delta q}{2\pi } \int \frac{d\Omega_p}{2\pi}\int \frac{d\delta p}{2\pi} \frac{\mathrm{sgn}(\Omega_{p}-\Omega_q)-\mathrm{sgn}(\Omega_p)}
{\frac{{i\Omega_p}}{Z_\bp}-\frac{i(\Omega_p-\Omega_q)}{Z_{\bp-\bq}}-\frac{\delta p \delta q}{m^*}+\frac{(\delta q)^2}{2m^*}} \mathrm{sgn}(\Omega+\Omega_q)\Gamma^2(K,P;Q).
\label{se3}\eea
\ewt
  Next, we assume that relevant $\delta q$ are much smaller than $\delta p\sim \delta k$,
    such that the $(\delta q)^2/(2m^*)$ in the denominator of (\ref{se3}) can be
    neglected
    and $Z_{\bp-\bq}$ can be approximated by $Z_{\bp}$.
    Integrating now over $\delta p$ and $\Omega_p$ in (\ref{se3}),
      we obtain  the usual Landau-damping form of the dynamic particle-hole bubble
$\sim m^*|\Omega_q|/v_F|\delta q|$.
   The kinematic logarithm is produced by integrating the $1/|\delta q|$ singularity of the particle-hole bubble over $\delta q$:
   $\int^{|\delta k}_{\Omega/(v_F Z_{\bk})} d\delta q/|\delta q|=\ln(Z_{\bk} v_F|\delta k|/\Omega)$. Finally, the integral over
 $\Omega_q$
 gives a FL-like factor of $\Omega^2$. While performing all the integrations indicated above, the factor of  $
\Gamma^2$ can be taken outside the integral.
Collecting all the factors together and performing analytic continuation, we obtain for the imaginary part of the self-energy
 \bea
 \Sigma''_{\mathrm{comp}_2} (\bk_F, \Omega)  &\sim&
 \frac{m^* \Omega^2
 }{v^2_F} \ln{\frac{Z_{\bk} v_F |\delta k|}{\Omega}}\Gamma^2 \label{2.1}
 \\
 & \sim & \Omega^2
 \frac{{\bar g}^2}{(v_F \delta k)^3} \frac{E^*_F}{v_F \delta k}  \ln^3{\frac{Z_{\bk} v_F |\delta k|}{\Omega}},\nn
  \eea
 where $\Gamma$ is given by Eq.~(\ref{gamma_const})
and
 the effective Fermi energy is defined as
 \beq
 E^*_F \equiv \frac 12m^* v^2_F.
 \eeq
  In dimensionless variables
of Eqs.~(\ref{ch_6}) and (\ref{ch_7}), Eq.~(\ref{2.1}) is expressed as
  \beq
 \Sigma''_{\mathrm{comp}_2} (\tk, {\bar \Omega}) \sim {\bar g}
 \frac{
 \bo^2}{\tk^4} \frac{E_F^*}{\bar g}
  \ln^3{\frac{Z_{\bk} |\tk|}{\bo}}.
  \label{2.1.1}
  \eeq
 Equations (\ref{2.1}) and (\ref{2.1.1}) are valid as long as
 the logarithmic factor is parametrically large, i.e., as long as $\bo/Z_{\bk} < |\tk|$.
  For $|\tk| <1$, $Z_{\bf k}
  \sim
   |\tk|$
 and hence the condition
 above reduces to
 $\bo < {\bar \Omega}_b = \tk^2$, which is the same as the
 condition
  to be
  outside the hot region. For $|\tk| >1$, $Z_{\bk} \approx 1$ and the condition is $\bo < |\tk| = ({\bar \Omega}_b)^{1/2}$.

  Note that Eq.~(\ref{2.1}) describes both
 the
  forward-
  and $2k_F$-scattering contributions; indeed,
  the result is the same
  regardless of whether one considers the case of $\bp\approx \bk$ or $\bp\approx -\bk$.
  In this regard, the case of an anisotropic FS with the $Z$ factor varying rapidly around the hot spots, considered in this paper, differs from that of an isotropic FS with $Z=\mathrm{const}$, considered in previous studies of forward- and $2k_F$ contributions to the self-energy.
   \cite{chubukov:03,chubukov:05}
   In the latter case, the forward-scattering part of the self-energy has a singularity on the mass-shell, which is regularized by resumming the perturbation theory and taking into account the curvature of the fermionic dispersion, whereas the $2k_F$ part is regular on the mass-shell.

   In the case considered here, even the forward-scattering part is regular on the mass-shell. This is so because the mass-shell of the external fermion $\Omega=Z_\bk\left(v_Fk_\perp+(\delta k)^2/2m^*\right)$ 
   contains 
  a local value of the $Z$ factor, corresponding to
  a
   point $\bk$ on the FS. On the other hand, the mass-shell of the internal fermion contains the $Z$ factor at 
   the
   point $\bk+\bq$,
      where $\bq$ is the running variable in the integral for the self-energy.
      As a result, the external and internal mass-shells do not coincide
         and the \lq\lq resonance\rq\rq\/, which leads to the mass-shell singularity in the isotropic case, is absent.

Two-loop composite scattering
was
considered by HHMS
for the case  of
$|\tk| <1$.
 Equation (\ref{2.1}) is reproduced if one inserts
 finite curvature into
 Eq. (5.18)
of  Ref.~\onlinecite{max_last}.
 However, the form of
 $\Sigma''_{\mathrm{comp}_2} (\bk_F, \Omega) \propto \Omega^2$
 in Eqs.~(5.36) and (5.37) of
  Ref.~\onlinecite{max_last} has an extra small factor
  of
  $|\delta k|/k_F\ll 1$ compared with
  Eq.~(\ref{2.1}). The reason for the
   discrepancy is that HHMS
  considered the case when
  the
  FS curvature is
  absent at the bare level
   but
   generated
   dynamically
   by the
    interaction.\cite{max_thanks} In this
   case,
   $E^*_F/{\bar g}$ by itself scales as
  $\tk$
  and $ \Sigma''_{\mathrm{comp}_2} (\bk_F, {\bar \Omega})$ in Eq.~(\ref{2.1}) scales as
$1/|\tk|^3$.\\

\paragraph{{\bf One-dimensional regime.}}

Equation (\ref{2.1})  [or (\ref{2.1.1})] is not the full story, however. Indeed, our reasoning leading to Eq.~(\ref{2.1}) is valid provided that one can integrate
over $\delta p$ in Eq.~(\ref{se3}) in infinite limits.  In reality, internal $|\delta p|$ and $|\delta q|$ are
bounded
 from above by external $\delta k$.
 At larger $\delta p$
and $\delta q$, the composite vertex falls off quickly.
The largest value of the $\delta p \delta q/m^*$ and $(\delta q)^2/2m^*$ terms in Eq.~(\ref{se3}) is then of order $(\delta k)^2/m^*$.
 On the other hand,
the internal frequencies, $\Omega_p$ and $\Omega_q$, are
 on the order of the external one, $\Omega$.
 Integration over $\delta p$ in infinite limits
 can then be justified
 if
 $\Omega < Z_{\bk} \delta k^2 /m^*$ or ${\bar \Omega} < {\bar \Omega}_{b_1}$,
 where
 $\bar\Omega_{b_1}\equiv  \tk^2 Z_{\tk} ({\bar g}/E^*_F)$.  For $|\tk|
 <1$,
 $Z_{\tk}\sim \tk$
  and thus
 ${\bar \Omega}_{b_1}
 \sim
 |\tk|^3 ({\bar g}/E_F)$;
 for $|\tk| >1$, $Z_{\tk} \approx 1$ and thus ${\bar \Omega}_{b_1} = |\tk|^2 ({\bar g}/E_F)$.
 In both cases, ${\bar \Omega}_{b_1} <
  {\bar \Omega}_b = \tk^2$, i.e., the condition  $\bo < {\bar \Omega}_{b_1}$ is valid only for a subset of
  fermions outside
  the
  hot regions.

 For the remaining
fermions with frequencies in the interval
${\bar \Omega}_{b_1} < {\bar \Omega} < {\bar \Omega}_b$,
the energy associated with the FS curvature is the smallest
energy scale in the problem, and we are thus in the effectively one-dimensional 
regime.
Had we been considering a real 1D system, the self-energy would have exhibited two characteristic features. First,
the self-energy due to scattering of fermions from the same hot spot (forward scattering) would have
had a pole on the mass shell, indicating the \lq\lq infrared catastrophe\rq\rq\/.
\cite{bychkov,maslov_review}
Second, the imaginary part of the self-energy due to scattering of fermions from the opposite hot spots ($2k_F$ scattering) would have vanished on the mass shell, indicating the absence of relaxational processes in a 1D system with linearized dispersion. (To obtain finite relaxation rate in 1D, one needs to
include the curvature of the dispersion.~\cite{glazman}) What makes our system
 different from a real 1D one
  is again
   the variation of the $Z$ factor along the FS surface. Even if we neglect (as we will) the curvature term in the fermionic dispersion, the variation of the $Z$ factor prevents either of the two characteristic 1D features described above to develop. The resulting self-energy is finite on the mass-shell both for forward- and $2k_F$ cases
    and,
    at fixed position on the FS, scales with frequency in a MFL way: $\Sigma\propto \Omega\ln\Omega$.

To see this explicitly, we
neglect the curvature terms in Eq.~(\ref{se2}) and 
take
 into account that the velocities corresponding to the momenta $\bk$ and $\bp$ are near each other for the forward-scattering case and opposite to each other for the $2k_F$-scattering case.
 Then the self-energy in the 1D regime can be written as \bwt
\bea
\Sigma^{\pm}_{\mathrm{comp}_2} (\delta k,k_\perp, \Omega)&=&-\int \frac{d\Omega_q}{2\pi}\int \frac{d\delta q}{2\pi } \int \frac{dq_\perp}{2\pi}
\int \frac{d\Omega_p}{2\pi}\int \frac{d\delta p}{2\pi}\int \frac{dp_\perp}{2\pi} \frac{1}
{\frac{{i\Omega_p}}{Z_\bp}\mp v_Fp_\perp}
\frac{1}{\frac{i(\Omega_p-\Omega_q)}{Z_{\bp-\bq}}\mp v_F(p_\perp-q_\perp)}\nn\\
&&\times\frac{1}{\frac{i(\Omega+\Omega_q)}{Z_{\bk+\bq}}-v_F(k_\perp+q_\perp)}\Gamma^2(K,P;Q),
\label{se2_1Da}\eea
\ewt
where $\pm$ corresponds to 
forward/$2k_F$ scattering. In contrast to 
 the regimes considered in the previous sections,
  the self-energy in the 1D regime
depends on the momentum across the FS ($k_\perp$), and we made this dependence explicit in Eq.~(\ref{se2_1Da}).
Integrating the product of two Green's functions in the first line of Eq.~(\ref{se2_1Da}) first over $p_\perp$ and then over
$\Omega_p$, we obtain objects which play the role of the (dynamic) polarization bubbles of 1D fermions
\bea
\Pi^{\pm}_{1D}=\frac{1}{2\pi v_F}\frac{1}{Z^{-1}_{\delta p}-Z^{-1}_{\delta p-\delta q}}\ln\frac{\frac{i\Omega_q}{Z_{\delta p}}\pm v_Fq_\perp}{\frac{i\Omega_q}{Z_{\delta p-\delta q}}\pm v_Fq_\perp}.\label{pi1D}
\eea
For a momentum-independent $Z$ factor, Eqs.~(\ref{pi1D}) reduce to familiar expressions for the polarization bubbles of fermions of the same ($+$) and opposite ($-$) chiralities:
\bea
\Pi^{\pm}_{1D}=\frac{1}{2\pi v_F}\frac{i\Omega_q}{\frac{i\Omega_q}{Z}\pm v_Fq_\perp}.
\eea
Both characteristic features of the self-energy in 1D are related to the fact that the imaginary part of the 1D bubble is a $\delta$ function centered on the bosonic mass-shell: subsequent convolution of $\I \Pi^{\pm}_{1D}$ with the remaining fermionic spectral function produces either a $\delta$ function singularity or zero in the imaginary part of mass-shell self-energy for forward- and $2k_F$ cases, correspondingly.  We will see later on, however, that $\delta p\sim\delta q\sim \delta k$ in our case,
which implies that $Z_{\delta p}\sim Z_{\delta p-\delta q}$ but $Z_{\delta p}\neq Z_{\delta p-\delta q}$.
In 
our
case, we have for the imaginary part of the bubble on the real frequency axis
\bea
\I \Pi^{\pm}&=&\frac{1}{2v_F}\frac{1}{Z^{-1}_{\delta p}-Z^{-1}_{\delta p-\delta q}}\\
&\times&\left[ \theta\left(\mp v_Fq_\perp-\frac{\Omega_q}{Z_{\delta p}}\right)-\theta\left(\mp v_Fq_\perp- \frac{\Omega_q}{Z_{\delta p-\delta q}}\right)\right].\nn
\eea
Again, a purely 1D case is recovered in the limit $Z_{\delta p}\to Z_{\delta p-\delta q}$ by using an identity $\lim_{\ve\to 0}\theta(x+\ve)=\theta(x)+\ve\delta(x)$.

 We continue
 Eq.~(\ref{se2_1Da}) to real frequencies, project the self-energy onto the 1D-like mass shell ($v_Fk_\perp=\Omega/Z_{\delta k}$), and integrate over $q_\perp$. These steps yield for the imaginary part of the self-energy on the real frequency axis
\bwt
\bea
\I \Sigma^{\pm,R}_{\mathrm{comp}_2} (\delta k,k_\perp=\Omega/v_FZ_{\delta k}, \Omega)&=&\frac{\Omega}{16 v_F^2\pi^2}\int \int d\delta q  d\delta p \frac{ \Gamma^2(K,P;Q)}{Z^{-1}_{\delta p}-Z^{-1}_{\delta p-\delta q}}
\int^0_{-1} dx\left\{ \theta\ls\pm  \lr\frac{1}{Z_{\delta k+\delta q}}-\frac{1}{Z_{\delta k}}\rr\pm x \lr\frac{1}{Z_{\delta k}}\mp \frac{1}{Z_{\delta p}}\rr\rs\right.\nn\\
&-&\left.\theta\ls\pm  \lr\frac{1}{Z_{\delta k+\delta q}}-\frac{1}{Z_{\delta k}}\rr\pm x \lr\frac{1}{Z_{\delta k}}\mp \frac{1}{Z_{\delta p-\delta q}}\rr\rs\right\},\label{se2_1Db}
\eea
\ewt
where $x=\Omega_q/\Omega$. In deriving Eq.~(\ref{se2_1Db}), we neglected the dependence of $\Gamma(K,P;Q)$ on $\Omega_q$ which is permissible in the leading logarithmic approximation.  The integrals over the tangential components of the momentum ($\delta p$ and $\delta q$) are effectively cut at $\delta p\sim\delta q\sim \delta k$, because the vertex decreases at larger $\delta p$ and $\delta q$. This implies that all the $Z$ factors in Eq.~(\ref{se2_1Db}) are of order of $Z_{\delta k}$. The range of integration over $x$ is $\max\{-1,x_1\}\leq x \leq \min\{0,x_2\}$, where $x_{1,2}$ are constraints imposed by the $\theta$ functions. Since all the $Z$ factors inside the $\theta$ functions are of the same order, $|x_{1,2}|\sim 1$, and the integral over $x$ produces a number of order one.
 The vertex $\Gamma(K,P;Q)$ can then be approximated by its value at $\delta p\sim\delta q\sim\delta k$, i.e., by $\bar g/(\delta k)^2$, and taken out of the integral (note that the logarithmic factor in Eq.~(\ref{1.7}) is of order one to this accuracy). The remaining integrals over $\delta p$ and $\delta k$ give a factor of $(\delta k)^2$. Collecting all the approximations mentioned above, we obtain an order-of-magnitude estimate for the imaginary part of the self-energy
  \beq
\Sigma''_{\mathrm{comp}_2}(\delta k, \Omega)\sim \lr \frac{\bar g }{v_F\delta k}\rr^2Z_{\delta k}\Omega.
\label{se2_11}\eeq
Since the forward- and $2k_F$-contributions to the self-energy happen to be of the same order, we suppress the superscript $\pm$ from now on.

 Restoring the real part of the self-energy via the Kramers-Kronig relation, we obtain
 \beq
 \Sigma_{\mathrm{comp}_2}(\delta k, \Omega)\sim i \lr \frac{\bar g }{v_F\delta k}\rr^2Z_{\delta k}\Omega
\ln\frac{v_F|\delta k| Z_{\delta k}}{\Omega}\label{se2_10}
 \eeq
or, in dimensionless variables,
 \beq
\Sigma_{\mathrm{comp}_2} (\tk, {\bar \Omega}) \sim i {\bar g} \frac{{\bar \Omega} Z_\tk}{\tk^2} \ln{\frac{|\tk|Z_\tk}{\bo}}.
\label{se2_5}
\eeq
Equation (\ref{se2_5}) is valid for $\bo < |\tk|Z_\tk$. At larger ${\bar \Omega}$,
 the logarithmic factor in Eq.~(\ref{se2_5}) disappears, and the self-energy becomes regular and small.

 Equations (\ref{se2_10}) and (\ref{se2_5}) imply that, in a certain range of frequencies,
  the
  self-energy
 near
  an SDW
  instability
  in 2D
  is of a MFL form. This is a
  much
  desired result because
  the
  phenomenological assumption about
  the
  MFL behavior~\cite{mfl} allows one
  to
  explain
the
  key
  experimental
  results
   for the cuprates.
  We emphasize, however, that the prefactor
  of
  the
  $\Omega \ln \Omega$ term  depends strongly on $\delta k$ and,
   in this respect,
    the result of
    the microscopic
    theory,
   Eq.~(\ref{se2_10}),
   differs from
   the
   MFL phenomenology,\cite{mfl}
   which assumes that the
   self-energy does not vary along the FS.

Equation (\ref{se2_5})
along with
the crossover scale ${\bar \Omega}_{b_1}$
were obtained by HHMS for $|\tk|<1$, when $Z_{\tk} \approx |\tk|$.
We found that $\Omega \ln \Omega$ form also holds for $|\tk|>1$,
where $Z_\tk
\sim 1$.
Explicitly, we have
\bea
&&\Sigma''_{\mathrm{comp}_2} (\tk, {\bar \Omega}) \sim\left\{
\begin{array}{l}
  {\bar g} \frac{{\bar \Omega}}{|\tk|},   ~~|\tk| <1; \label{se2_7} \\
{\bar g}
\frac{{\bar \Omega}}{|\tk|^2},   ~~|\tk| >1. \label{se2_8}
\end{array}
\right.
\eea
As we see, the prefactor in the $|\tk|>1$ region falls off rapidly (as $1/\tk^2$) with $\tk$.
This will be important for
the analysis of the
optical conductivity in Sec.~\ref{sec:3}.

Comparing the imaginary parts of the two-loop self-energies
in the 2D and 1D regimes
[Eqs.~(\ref{2.1}) and (\ref{se2_8}), correspondingly], we see that they match at ${\bar \Omega} \sim {\bar \Omega}_{b1} = \tk^2 Z_
{\tk}({\bar g}/E_F)$ (modulo a logarithm). At
${\bar \Omega}<\bar\Omega_{b1}$, the curvature plays the dominant role and scattering is of the 2D type; at
${\bar \Omega}>\bar\Omega_{b1}$, the curvature can be neglected and the self-energy
is of the 1D type.
The upper
boundary
of the
1D regime depends on
the position on the FS
relative to the hot spot, specifically, on
whether $|\tk|$ is larger or smaller than unity.

\subsection{
Fermionic self-energy: Summary of the results}
\label{sec:se_summary}
We now collect the contributions to
the self-energy from
all the scattering mechanisms considered so far:
$\bq_{\pi}$ scattering,
and one-
and
 two-loop composite scattering. Each of the three
 forms
  represents
  a
  different physical process,
  e.g.,
  one-loop scattering captures
  physics associated with the logarithmic singularity of
the composite vertex at small momentum transfers,
while
the two-loop composite contribution
represents
physics associated with forward-
and $2k_F$ processes,
 and also with 1D scattering in the regime when the curvature of the FS can be neglected.

In dimensionless units,
the
imaginary part of the
self-energy from $\bq_\pi$ scattering is
\bea
\Sigma''_{\bq_{\pi}} (\tk, {\bar\Omega})
\sim
\left\{
\begin{array}{l}
{\bar g} \frac{{\bar\Omega}^2}{|\tk|^3},  ~~\mathrm{for}\; {\bar\Omega} < \tk^2\\
{\bar g} \sqrt{\bar\Omega}, ~~\mathrm{for}\; {\bar\Omega} > \tk^2.
\end{array}
\right.
\label{sa_1}
\eea

The self-energy from one-loop composite scattering is
\beq
 \Sigma''_{\mathrm{comp}_1}(\tk, {\bar \Omega})
  \sim \bar g  \frac{
\bo^{3/2}}{\tk^2}, ~~\mathrm{for}\;{\bar\Omega} < \tk^2. \label{sa_2}
  \eeq
The
form of
self-energy from two-loop composite scattering
depends on whether
$|\tk| <1$
or $|\tk| >1$,
because the quasiparticle residue behaves as $Z_\tk\sim \min\{\tk,1\}$.
For $|\tk| <1$,
\bea
\Sigma''_{\mathrm{comp}_2} (\tk, {\bar\Omega}) \sim
\left\{
\begin{array}{l}
{\bar g} \frac{{\bar\Omega}^2}{|\tk|^4} \frac{E^*_F}{\bar g}  \ln^3{\frac{\tk^2}{\bo}},~~\mathrm{for}\; {\bar\Omega} < |\tk|^3
 \frac{\bg}{E_F^*};\\
{\bar g} \frac{\bar\Omega}{|\tk|},~~\mathrm{for}\; |\tk|^3
\frac{\bg}{E_F^*}
 < {\bar\Omega} < \tk^2.\\
\end{array}
\right.\nn\\
\label{sa_3}
\eea
while for $|\tk| >1$
\bea
\Sigma''_{\mathrm{comp}_2} (\tk, {\bar\Omega})
\sim
\left\{
\begin{array}{l}
 {\bar g} \frac{\bar\Omega}{\tk^2},~~\mathrm{for}\; \tk^2
 \frac{\bg}{E_F^*}
  <{\bar\Omega} < \tk^2;
   \\
{\bar g} \frac{{\bar\Omega}^2}{|\tk|^4} \frac{E^*_F}{\bar g}  \ln^3{\frac{
\tk}{\bo}},~~ \mathrm{for}\;{\bar\Omega} < \tk^2
 \frac{\bg}{E_F^*}.
\end{array}
\right.
\label{sa_4}
\eea

Each of the
asymptotic forms in Eqs.~(\ref{sa_1}-\ref{sa_4})
represents the dominant contribution to
 $\Sigma'' (\tk, {\bar\Omega})$ in some range of $\tk$ and ${\bar \Omega}$.  Comparing Eqs.~(\ref{sa_1}-\ref{sa_4}) and selecting the largest contribution,
 we obtain the imaginary part of the full fermionic self-energy,
 shown
 schematically
 as a function of $\bo$ at fixed $\tk$ in Fig.~\ref{fig:sigma_vs_omega},
 and as a function of $\tk$ at fixed $\bo$ in Figs.~\ref{fig:sigma_vs_k_a} and \ref{fig:sigma_vs_k_b}.
 \bwt
 \begin{figure}[!]
 \begin{centering}
\includegraphics[scale=0.15]
{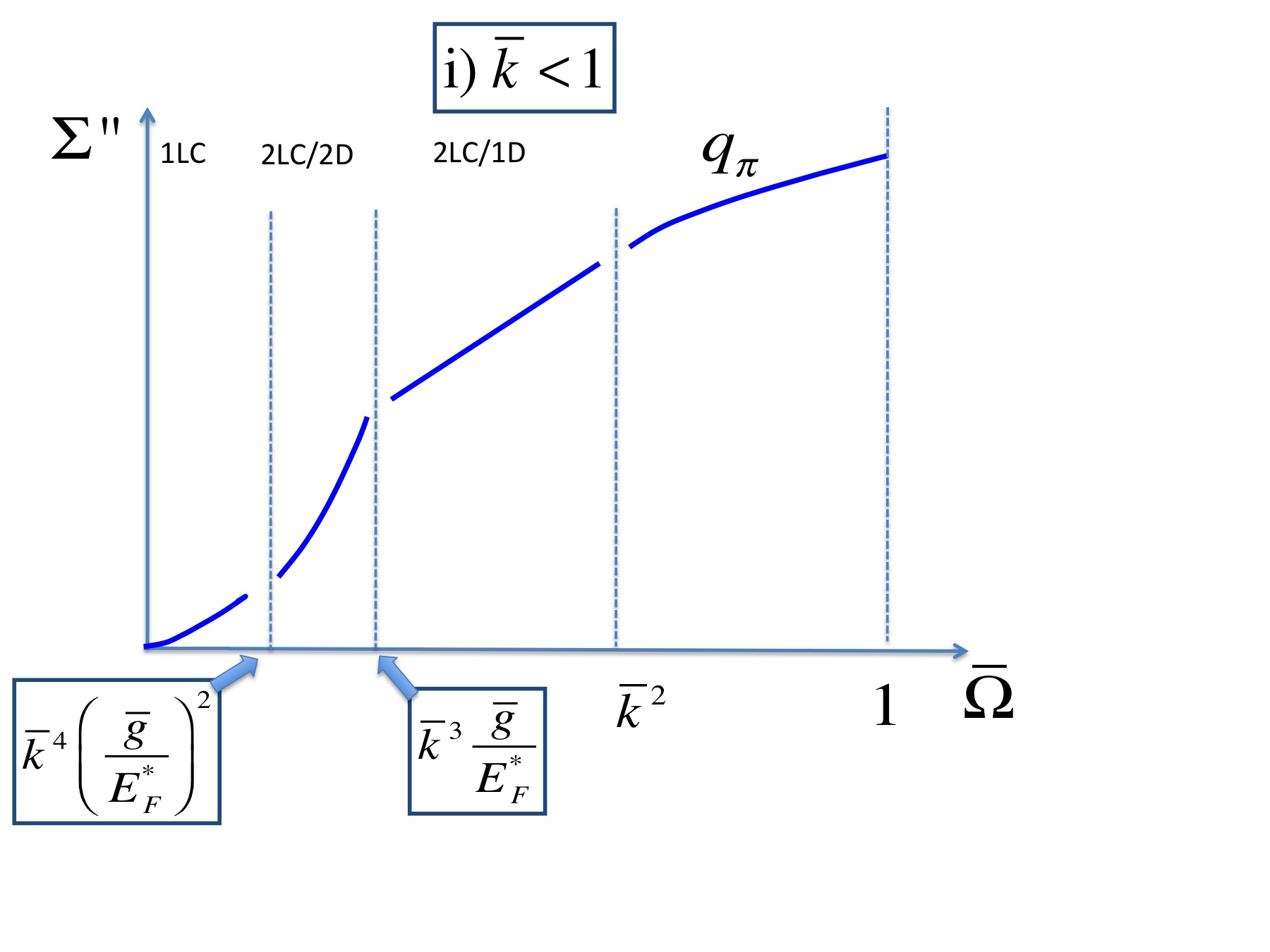}
\includegraphics[scale=0.15]
{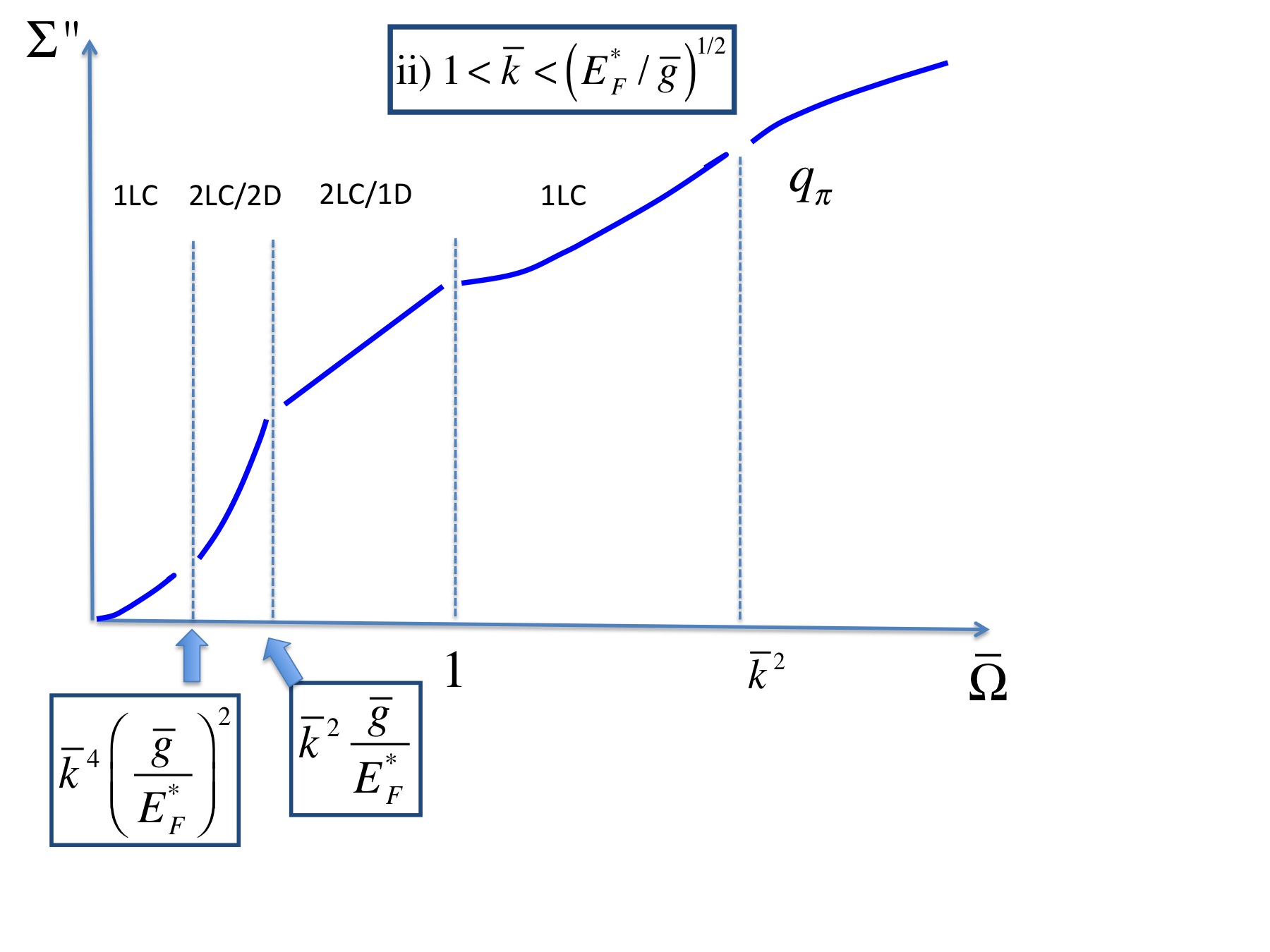}
\includegraphics[scale=0.15]
{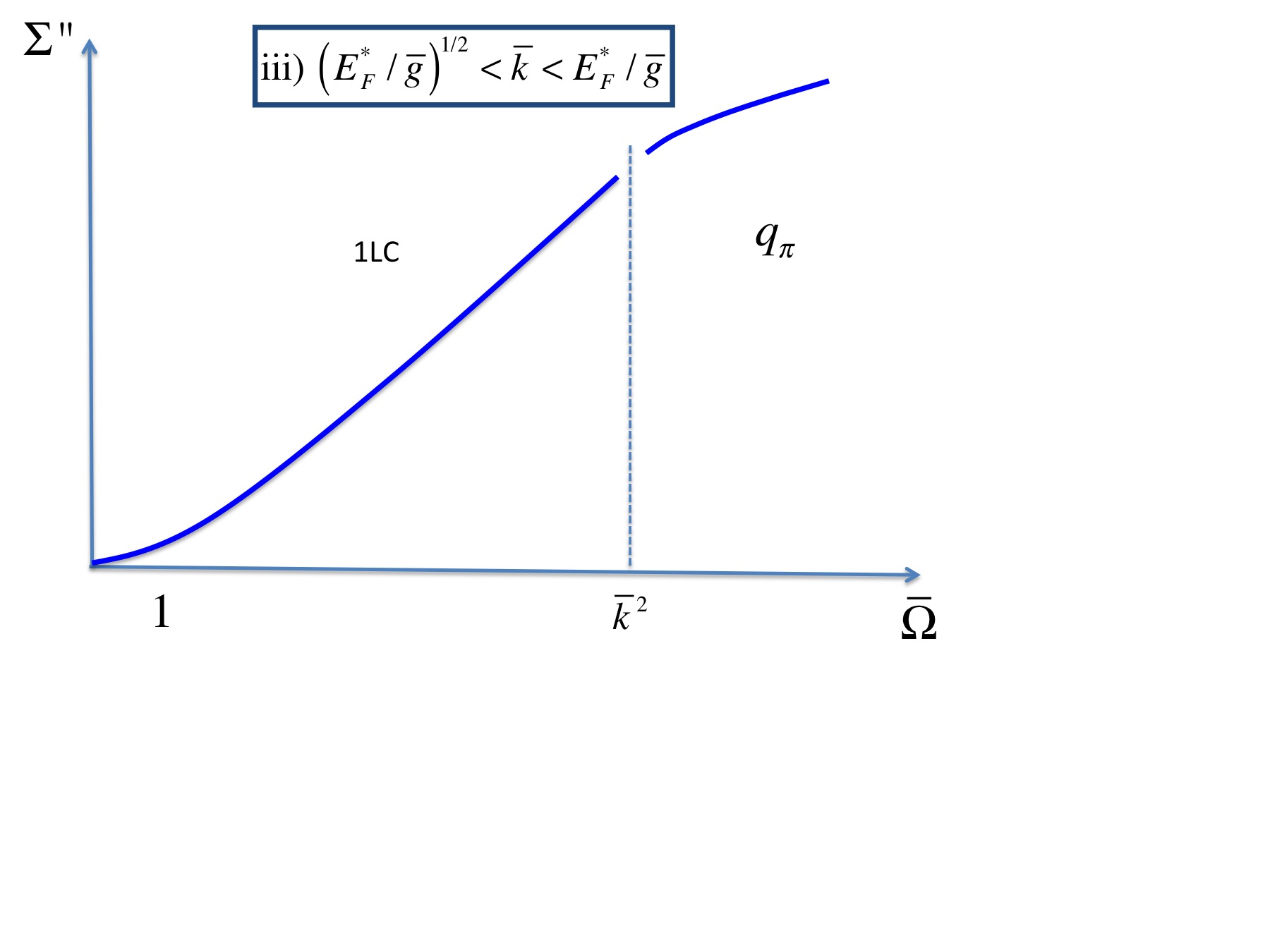}
\end{centering}
\vspace{-3 cm}
\caption{A sketch of the imaginary part of the self-energy, $\Sigma''(\tk,\bo)$, as a function of the dimensionless frequency, $\bo=\Omega/\bg$, at fixed (dimensionless) distance from the hot spot, $\tk=\delta k v_F/\bg$. Abbreviations of the asymptotic regimes and asymptotic forms
 of $\Sigma''$ are given in Table \ref{table:crossovers}.}
\label{fig:sigma_vs_omega}
\end{figure}
\ewt
\vspace{1 cm}
  In each case, there is a sequence of crossovers
  around which the functional form of $\Sigma'' (\tk, {\bar\Omega})$  changes.
At fixed $\tk$,
 the sequence of crossovers of $\Sigma'' ({\bar\Omega})$ as a function of $\bo$
 is different in the following three
 regions of $\tk$:
 \begin{itemize}
 \item[i)] $|\tk| <1$, \item[ii)]
 $1<|\tk|<(E^*_F/{\bar g})^{1/2} $,
 and \item[iii)]
 $(E_F^*/{\bar g})^{1/2}<|\tk|<
 E_F^*/{\bar g}
 $.
 \end{itemize}
The behavior of $\Sigma''$ as a function of $\bo$ is sketched in the three panels of Fig.~\ref{fig:sigma_vs_omega}.
Abbreviations of the asymptotic regimes along with the corresponding forms of $\Sigma''$ are given in Table \ref{table:crossovers}.
 At
 $|\tk|>E_F^*/\bg$, the
 entire
 FS becomes hot, and our
 model
  is no longer applicable.

  Similarly, the sequence of crossovers
  in
  $\Sigma''
  $ at
  fixed ${\bar \Omega}$
  depends on whether $\bo$ is in one
 of the following four
 regions:
 \begin{itemize}
 \item[i)] $\bo <({\bar g}/E^*_F)^2$,
 \item[ii)] $({\bar g}/E^*_F)^{2} < \bo < {\bar g}/E^*_F$,
 \item[iii)] ${\bar g}/E^*_F < \bo <1$, and
 \item[iv)] $1<\bo < (E^*_F/{\bar g})^{2}$.
 \end{itemize}
 The behavior of $\Sigma''$ as a function of $\tk$ is sketched in Figs.~\ref{fig:sigma_vs_k_a} [for $\bo$ in regions i) and ii)] and \ref{fig:sigma_vs_k_b}  [for $\bo$ in regions iii) and iv)].
 At
 $\bo>(E^*_F/{\bar g})^{2}$ the
 entire
 FS becomes hot.
 \begin{figure}[!]
 \begin{centering}
\includegraphics[scale=0.18]
{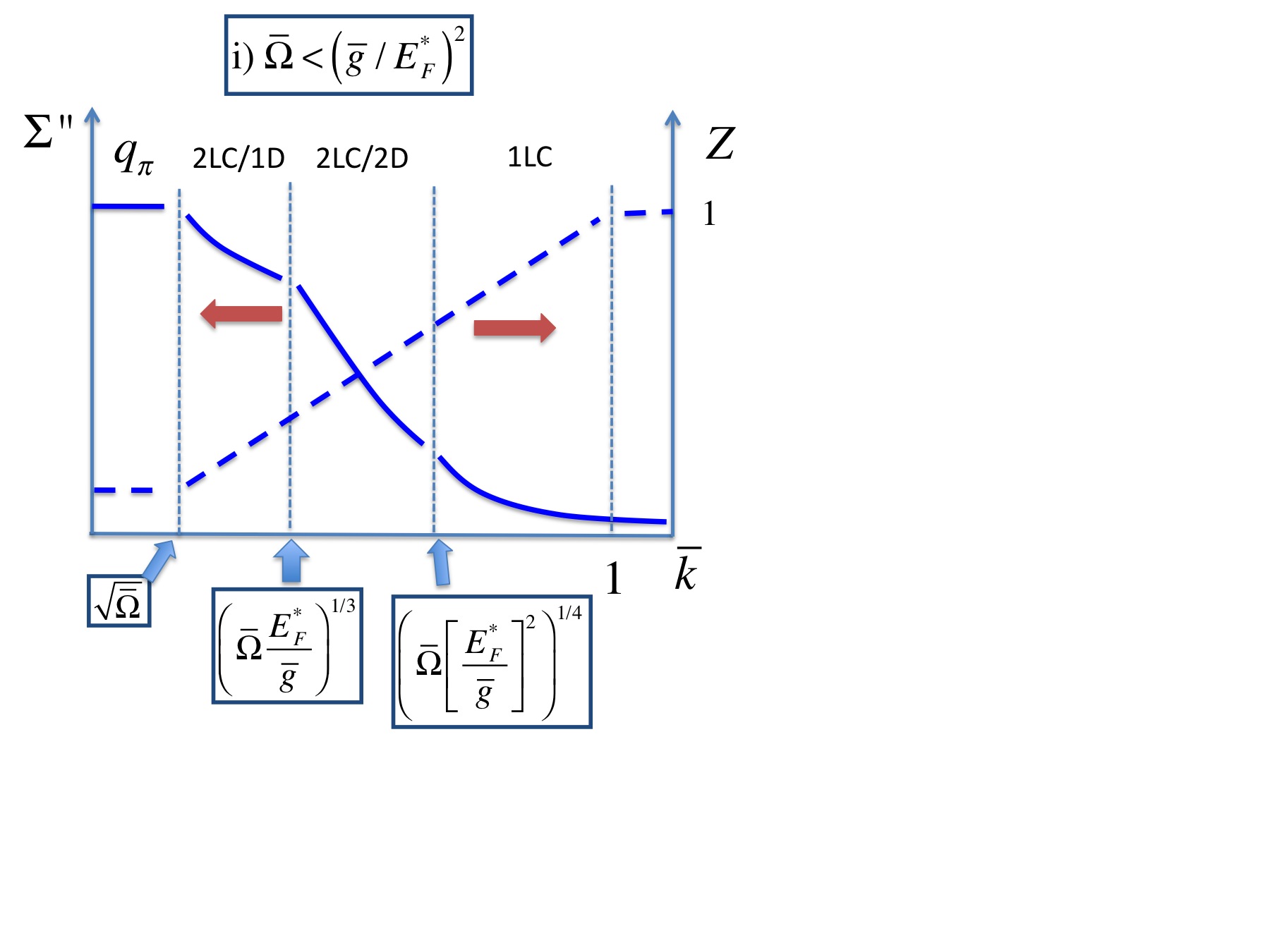}
\includegraphics[scale=0.18]
{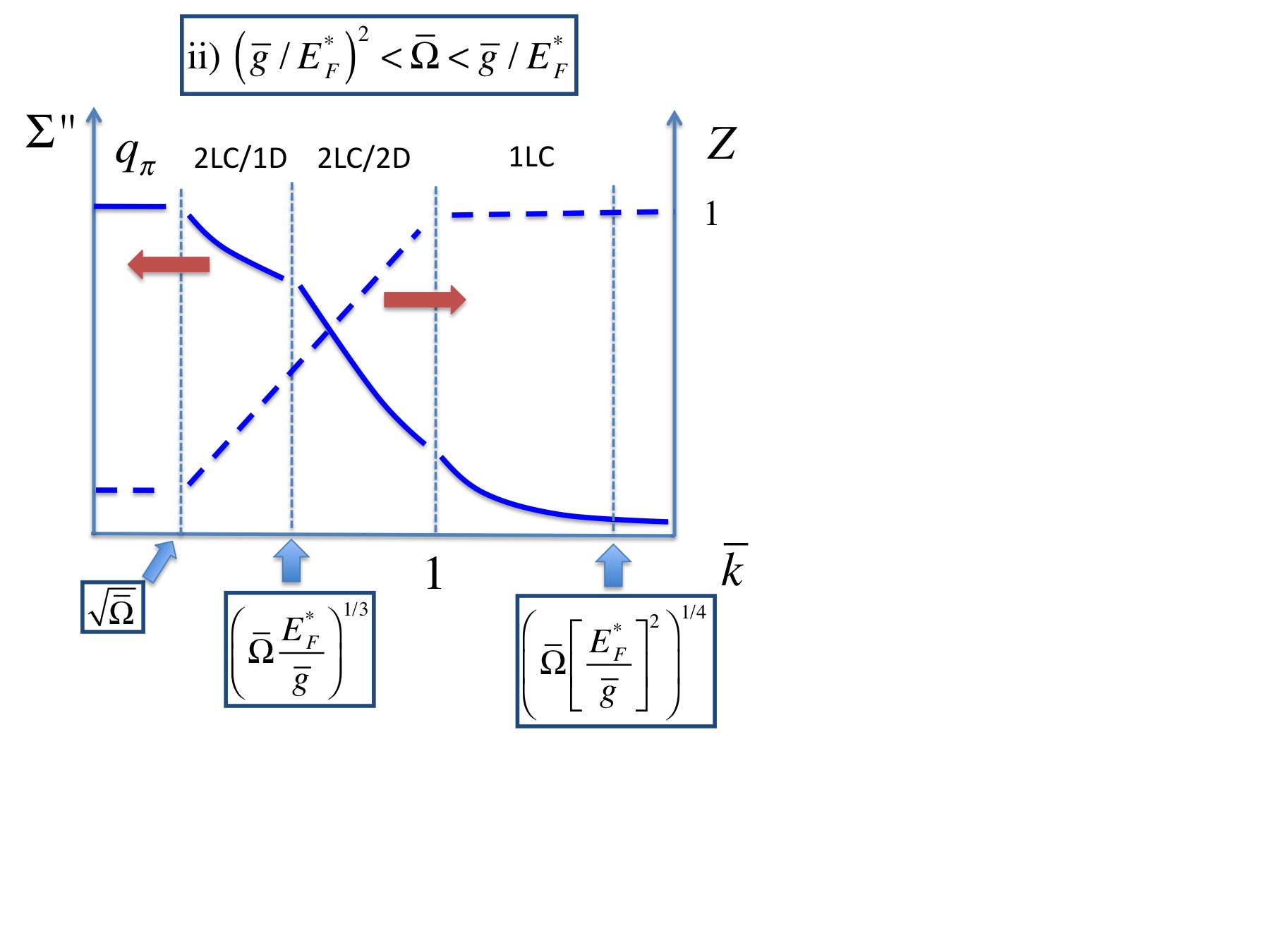}
\vspace{-2 cm}
\caption{
 A sketch of the imaginary part of the self-energy, $\Sigma''(\tk,\bo)$ (solid), and quasiparticle residue, $Z$ (dashed), as functions of the dimensionless distance from the hot spot, $\tk=\delta k v_F/\bg$, at fixed (dimensionless) frequency, $\bo=\Omega/\bg$. Abbreviations of the crossover regimes and asymptotic forms
 of $\Sigma''$ are given in Table \ref{table:crossovers}.
\label{fig:sigma_vs_k_a}}
\end{centering}
\end{figure}
 \begin{figure}[!]
 \begin{centering}
 \includegraphics[scale=0.18]
{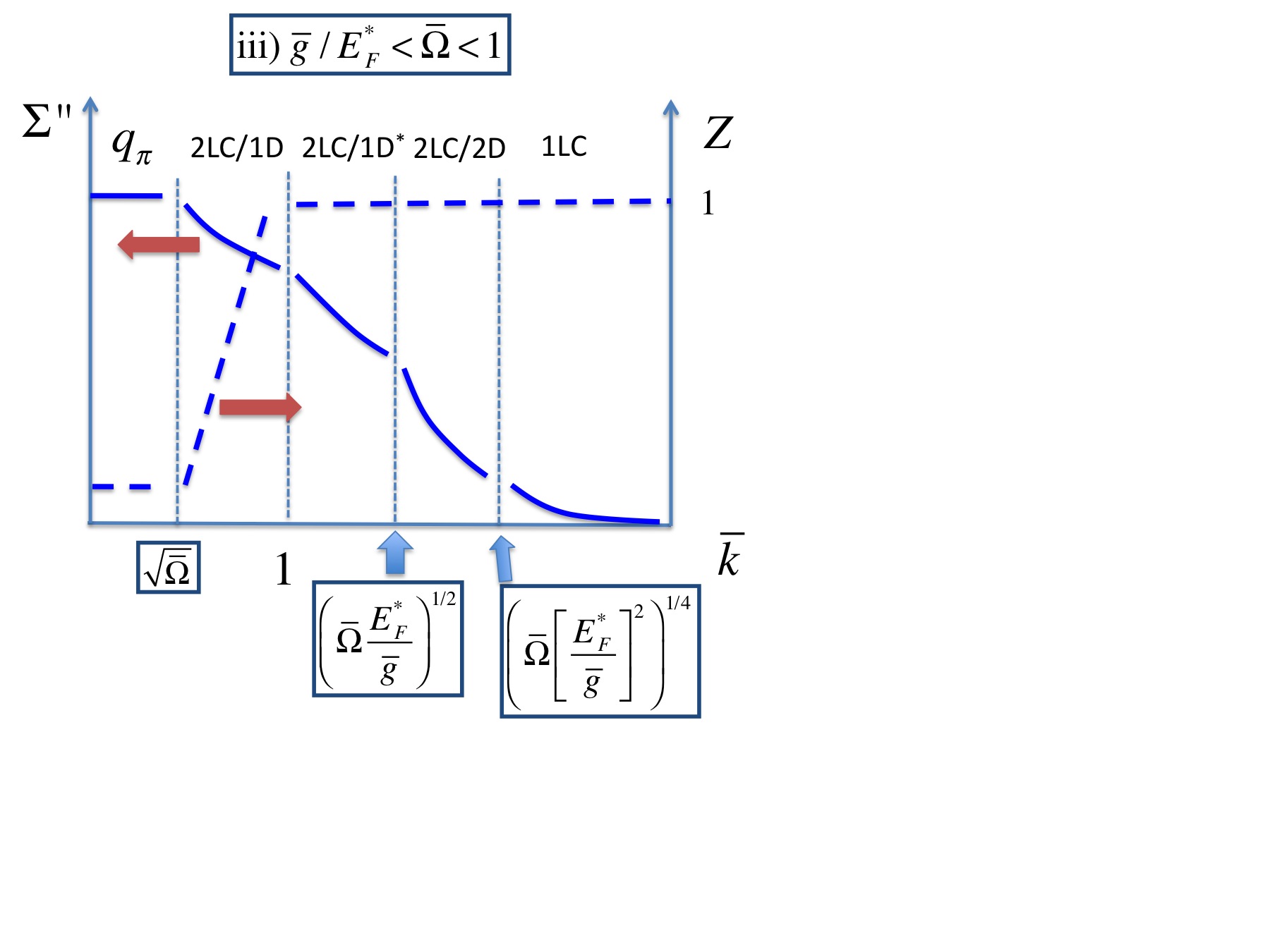}
\includegraphics[scale=0.18]
{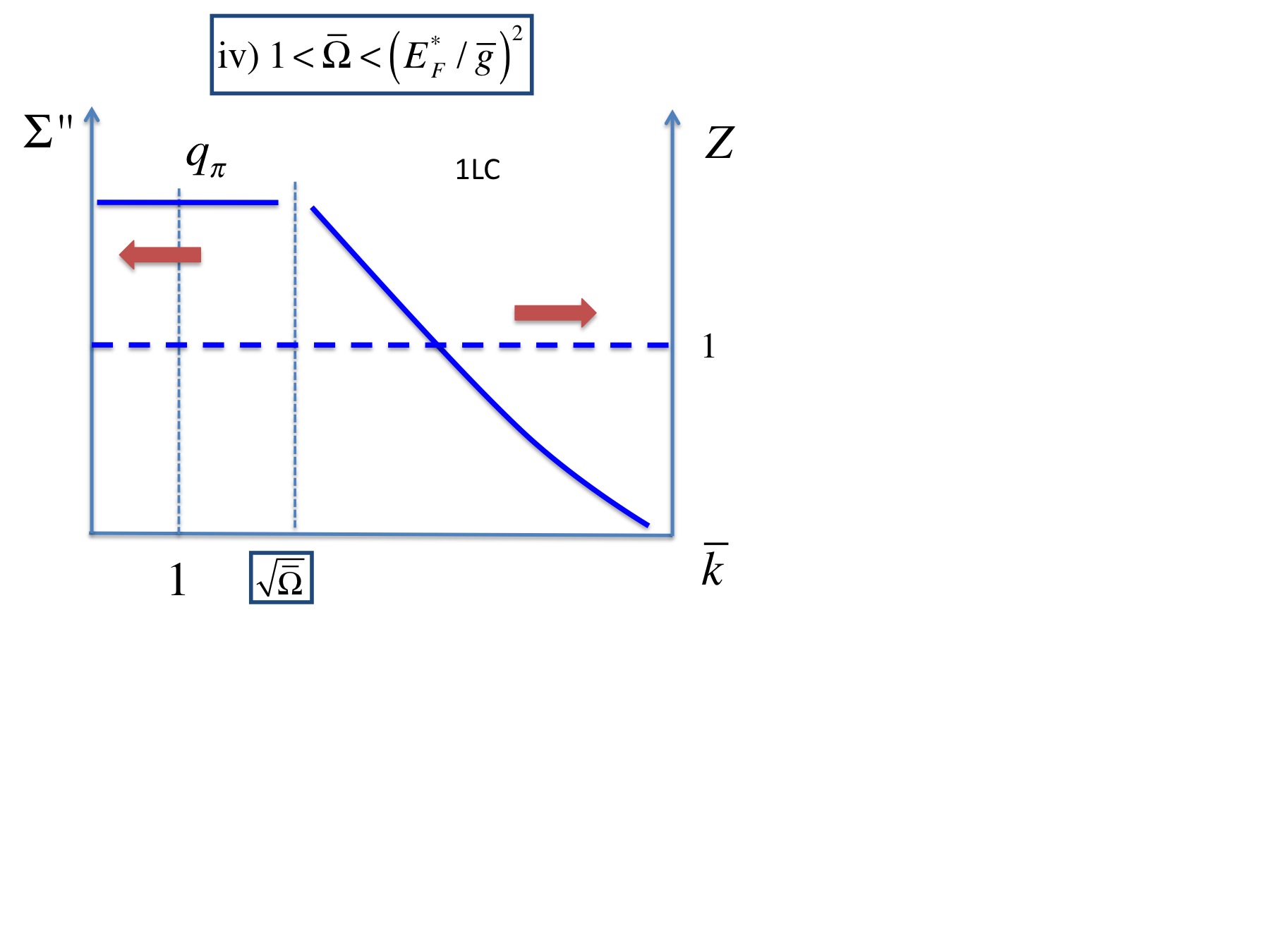}
\end{centering}
\vspace{-3 cm}
\caption{
A sketch of the imaginary part of the self-energy, $\Sigma''(\tk,\bo)$ (solid), and quasiparticle residue, $Z$ (dashed), as functions of the dimensionless distance from the hot spot, $\tk=\delta k v_F/\bg$, at fixed (dimensionless) frequency, $\bo=\Omega/\bg$. Abbreviations of the crossover regimes and asymptotic forms
 of $\Sigma''$ are given in Table \ref{table:crossovers}.}
\label{fig:sigma_vs_k_b}
\end{figure}
   \bwt
   \begin{table*}
    [!]
\caption{Asymptotic forms of $\Sigma''$.} 
\centering 
\begin{tabular}{|c| c| c|} 
\hline 
Abbreviation& Dominant scattering process &$\Sigma''/\bg$ \\ [1ex] 
\hline 
$q_\pi$ & $\bq_\pi$-scattering &$\sqrt{\bo}$  \\ 
\hline
1LC & 1-loop composite scattering &\raisebox{-1ex}{$\bo^{3/2}/\tk^2$}\\
\hline
2LC/1D & 2-loop composite scattering/1D regime for $\tk<1$&\raisebox{-1ex}{$\bo/\tk$}\\
\hline
2LC/1D$^*$ & 2-loop composite scattering/1D regime for $\tk>1$&\raisebox{-1ex}{$\bo/\tk^2$}\\
\hline
2LC/2D & 2-loop composite scattering/2D regime &\raisebox{-1ex}{$(E_F^*/\bg)(\bo^2/\tk^4)\ln^3(\min\{\tk^2,1\}/\bo)$}\\
[1ex] 
\hline 
\end{tabular}
\label{table:crossovers} 
\end{table*}
\ewt
The dominant contribution to the real part of the self-energy
in all the regimes comes from $\bq_\pi$ scattering:
\bea
\Sigma'_{\bq_{\pi}} (\tk, {\bar\Omega})
\sim
\left\{
\begin{array}{l}
 {\bar g} \sqrt{{\bar\Omega}}, ~~ \mathrm{for}\;
 |\tk|<\sqrt{\bo},
 \\
{\bar g} \frac{{\bar\Omega}}{|\tk|},~~ \mathrm{for}\;
|\tk|>\sqrt{\bo}.
\end{array}
\right.
\label{sa_5}
\eea
The quasiparticle residue $Z_{\tk} = (1 +
 \bg^{-1}\partial \Sigma'(\tk, {\bar\Omega})/\partial {\bar \Omega})^{-1}$ as a function of $\tk$
 is sketched  in Figs.~\ref{fig:sigma_vs_k_a} and \ref{fig:sigma_vs_k_b} (dashed lines).

\subsection{
Classification of fermions as \lq\lq cold\rq\rq\/, \lq\lq lukewarm\rq\rq\/, and \lq\lq hot\rq\rq\/
in the presence of composite scattering}
\label{sec:hclw}

The classification of fermions as \lq\lq hot\rq\rq\/, \lq\lq cold\rq\rq\/, and \lq\lq lukewarm\rq\rq\/ in Sec.~\ref{sec:3aa} was based on the behavior
of the self-energy with only $\bq_\pi$ scattering taken into account.
In particular,
fermions were classified as \lq\lq hot\rq\rq\/,
if their $\Sigma_{\bq_\pi}$
scales as $\sqrt{\Omega}$ (and is independent
of $\delta k$);
as \lq\lq cold\rq\rq\/,
if their $\Sigma_{\bq_\pi}$
 has
  a FL form and is small compared to bare $\Omega$;
and, finally,  as \lq\lq lukewarm\rq\rq\/, if their $\Sigma_{\bq_\pi}$
 had a FL form but the quasiparticle residue was smaller than unity.
 In this classification scheme,
 the boundary between the
 hot
  and lukewarm
 regimes
 is at
 ${\bar \Omega} \sim \tk^2$  (with the hot behavior corresponding to higher ${\bar \Omega}$).
  With composite scattering taken into account, this classification scheme
   still holds
   for $\bo > \tk^2$. However, the behavior of $\Sigma''$
   for
   $\bo < \tk^2$
   becomes more complex.
   First, we see from the top panel of Fig.~\ref{fig:sigma_vs_k_a} that, for $\tk <1$, the
   region of $\bo<\tk^2$ which
   was identified
   before
   as \lq\lq lukewarm\rq\rq\/,
  now contains subregions
  of a conventional FL ( $\Sigma'' (\Omega) \propto \Omega^2$), unconventional FL  ( $\Sigma'' (\Omega) \propto \Omega^{3/2}$), and MFL behavior ( $\Sigma'' (\Omega) \propto \Omega$).

 Second, for $|\tk| >1$, the
  region of ${\bar \Omega} < \tk^2$ was
  earlier
  classified as \lq\lq cold\rq\rq\/,
 because
 $\Sigma''_{\bq_\pi}\propto\Omega^2$ and
  and $Z_{\tk} \approx 1$ in this region.
  However, we see from the middle panel of Fig.~\ref{fig:sigma_vs_k_b} that,
   with composite scattering taken into account, the region ${\bar \Omega} < \tk^2$
   also contains subregions of both conventional and unconventional FL behaviors,
   as well as
   a
  MFL
  subregion.
  ted.

  To streamline the terminology, we will still
  be classifying
   fermions in the region ${\bar \Omega} < \tk^2$ as \lq\lq lukewarm\rq\rq\/ for $|\tk| <1$ and as \lq\lq cold\rq\rq\/ for $|\tk| >1$, because in all the three cases the FL
  criterion that $\Omega + \Sigma' (\Omega)$ must be larger than
   $\Sigma'' (\Omega)$ is satisfied.
   Nevertheless, there are clear differences between the
   behaviors obtained with only ${\bf q}_\pi$
 scattering and
 both ${\bf q}_\pi$ and composite scattering
 taken into account.
\
\subsection{Higher-loop orders in composite scattering}
\label{sec:higher_loops}
\begin{figure}
\begin{centering}
\includegraphics[scale=0.6]{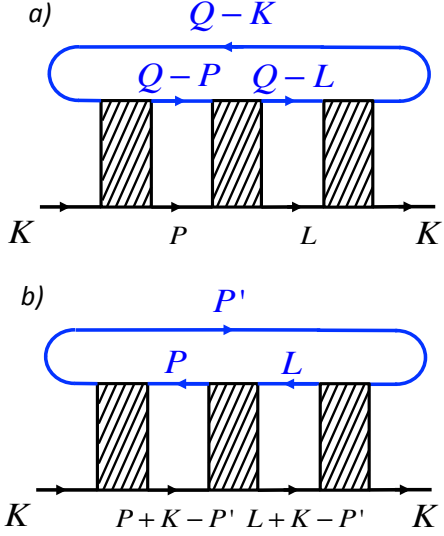}
\end{centering}
\vspace{-4.5 cm}
\caption{
Examples of the three-loop self-energy diagrams.  $a$) Particle-particle (Cooper) channel. $b$) Particle-hole channel. The hatched box is the composite vertex in Fig.~\ref{fig:vertex}.}
\label{fig:se_3loop}
\end{figure}

A natural question is whether higher-loop orders in composite scattering modify the results of the previous section.
We begin with the
regime
of the smallest $\Omega$,
when
 $\Sigma''_{\mathrm{comp}_2}
 (\delta k, {\bar\Omega})\propto\Gamma^2 \Omega^2 \ln\left(
\tilde\Lambda/\Omega\right)
$,
where
$\Gamma \propto \left[{\bar g}/(
\delta k)^2\right] \ln 
\left(
\tilde \Lambda/
\Omega
\right)
$  is the  composite vertex and $\tilde\Lambda
 \sim (v_F
 \delta k )^2/{\bar g}$.

 In an ordinary 2D FL, the prefactor
 of
 the
 $\Omega^2 \ln {
 \Omega}$ term in the imaginary part of the self-energy is the sum of
  the fully renormalized backscattering and forward scattering amplitudes.~\cite{chubukov:05} The forward scattering amplitude
   approaches
   a constant value at zero frequency, hence the corrections from higher orders do not change the second order result, at least qualitatively.
   The backscattering amplitude contains the series of logarithms from the Cooper channel.~\cite{aleiner:06,aleiner_efetov}  In our case, the situation with higher-order
    corrections from Cooper channel is
    somewhat different:
    integration over the internal momentum eliminates the logarithm in the vertex entering the three-loop Cooper diagram
 (Fig.~\ref{fig:se_3loop}$a$) but brings in an additional Cooper logarithm, so that the renormalized vertex has the same logarithmic factor as the original one.

  To see this,
   we recall that the
   argument of the
   logarithmic factor in $\Gamma$
   is
   actually
   $
   (\delta k)^2/(q^2 + \gamma |\Omega_q|)
   $,
  where $q$ is the transferred
  momentum, see Eq.~(\ref{1.7}).
 Suppose now that we consider the three-loop composite self-energy as the two-loop self-energy with one-loop vertex correction.  The vertex correction part involves two
 vertices and two fermionic Green's functions.
  Integrating
  the product of the two Green's functions over $
  q_\perp$,
  we obtain the vertex correction as
  \beq
   \tilde\Gamma \sim \int d
   \delta q \int_\Omega d \Omega_q \frac{\Gamma^2 (\delta q, \Omega_q) Z_{\delta q}}{|\Omega_q|}
   \eeq
    where 
    integration over $\delta q$ is restricted to $|\delta q| < |\delta k|$.
   Substituting $Z_{\delta q} = v_F |\delta q|/{\bar g}$ and $\Gamma
   \sim
    ({\bar g}/(
   \delta k)^2 \ln
   (\delta k)^2/\ls(\delta q)^2 + \gamma |\Omega_q|\rs$,
   we
    find that the integral over $\delta q$
    comes from the region $|\delta q| \sim |\delta k|$, and the renormalized vertex is
\beq
   \tilde\Gamma \sim \frac{\bar g}{
   (\delta k)^2} \int^\Lambda_\Omega \frac{d \Omega_q}{|\Omega_q|} \sim \Gamma.
   \eeq
   We see that the renormalized vertex is of the same order as the bare one, hence
   the three-loop self-energy,
   $\Sigma
   _{\mathrm{comp}_3} (
   \delta k,
   \Omega)$,  is of the same order as
   $\Sigma
   _{\mathrm{comp}_2} (
   \delta k,
   \Omega)$.

 A somewhat different result
 is obtained for
 particle-hole
 three-loop diagram
 in Fig.~\ref{fig:se_3loop}$b$.  The contribution to this diagram
  from
  a
  $2k_F$ process,
  in which
  the momenta on the closed fermionic loop are almost opposite to the external momentum,
  has a higher power of frequency ($\Omega^{5/2}$, see below) but, at the same time, more singular dependence on $
  \delta k$.
  As result, the three-loop diagram, evaluated for $\Omega$ and $\delta k$ relevant for the conductivity.
happens to be of the same order as the two-loop one.

 For an estimate of
    the diagram in  Fig.~\ref{fig:se_3loop}$b$,
    we
     replace
    the actual vertices by a constant [$=\Gamma$ from Eq.~(\ref{gamma_const})] and take them out of the integral.
In addition, we
 replace the actual $Z$ factors entering the diagram by some average value, $\langle Z\rangle$.
Integrating over the $(2+1)$ momenta $P$ and $L$, we then obtain
\bea
\Sigma_{\mathrm{comp},3}(\delta k,\Omega)&\sim& \Gamma^3\int _{P'} G(P')\Pi_{2k_F}^2(P'-K)\label{3rd_1}
\\&&=\Gamma^3\int_QG(K+Q)\Pi^2_{2k_F}(Q),\nn\eea
where $\Pi_{2k_F}(Q)$ is the $2k_F$ part of the polarization bubble. Unlike the two-loop self-energy, the three-loop one cannot be re-written in terms of the $q=0$ part of the bubble, and we need to use an explicit form of  $\Pi_{2k_F}(Q)$. The singular part of $\Pi_{2k_F}(Q)$ is given by
\bea
\Pi_{2k_F}(Q)=\frac{m^*\langle Z\rangle}{4\pi}\ls\frac{-\tilde q v_{F}+\sqrt{(\tilde q v_{F})^2+(\Omega_q/\langle Z\rangle)^2}}{E^*_{F}}\rs^{1/2},\nn\\
\eea
where $\bq$ almost coincides
 with the chord of length $2k_F$, which connects two diametrically opposite hot spots, and $\tilde q\equiv 2k_F-q$.
A singular, $\Omega^2\ln\Omega$ contribution  from $2k_F$ scattering to the two-loop self-energy of a 2D FL comes from the region of $\tilde q>|\Omega_q|/v_F\langle Z\rangle>0$ (see, e.g.,  Appendix A in Ref.~\onlinecite{chubukov:05}), where $\Pi_{2k_F}(Q)$ can be approximated by
\beq
\Pi_{2k_F}(Q)=\frac{m^*}{4\pi}\frac{|\Omega_q|}{\sqrt{2E_F^*\tilde q}}.\label{pi2kF}
\eeq
Assuming that the singular part of the three-loop self-energy comes from the same region, we substitute Eq.~(\ref{pi2kF}) into the last line of Eq.~(\ref{3rd_1}) and write the internal Green's function as $G(K+Q)=\left[i(\Omega+\Omega_q)/\langle Z\rangle+v_F\tilde q-4E_F^*\theta^2\right]^{-1}$, where $\theta$ is a (small) angle between $\bq$ and the chord. For $\tilde q$ in the interval specified above, we have
\beq
\int d\theta G(K+Q)=i\pi \frac{\mathrm{sgn}(\Omega+\Omega_q)}{\sqrt{v_F\tilde q}}.
\eeq
The factor of $\mathrm{sgn}(\Omega+\Omega_q)$ confines the integral over $\Omega_q$ to the interval $(0,\Omega$), and we obtain for the Matsubara self-energy
\bwt
\bea
\Sigma_{\mathrm{comp},3}(\delta k,\Omega)\sim i\Gamma^3 \frac{k_F(m^*)^2}{(v_FE_F^*)^{3/2}}\int^{\Omega}_0d\Omega_q\Omega_q^2\int_{\frac{|\Omega_q|}{\langle Z\rangle v_F}}^\infty \frac{d\tilde q}{\tilde q^{3/2}}\sim i\frac{(m^*\Gamma)^3\langle Z\rangle ^{1/2}}{(E_F^*)^{3/2}}\Omega^{5/2}.
\eea
\ewt
A non-analytic, $\Omega^{5/2}$ scaling of the Matsubara self-energy implies that, on the real frequency axis, $\Sigma'_{\mathrm{comp},3}\sim
\Sigma''_{\mathrm{comp},3}\propto \Omega^{5/2}$. In dimensionless variables and on using Eq.~(\ref{gamma_const}) for $\Gamma$, we find
\beq
\Sigma''_{\mathrm{comp},3}(\tk,\bar\Omega)\sim \bar g\left(\frac{E_F^*}{\bar g}\right)^{3/2}\frac{\langle Z\rangle^{1/2}\bar\Omega^{5/2}}{\tk^6}\ln^3\frac{\langle Z\rangle\tk}{|\bar\Omega|}.
\label{3loop}
\eeq
Although the $\Omega^{5/2}$ dependence of $\Sigma''_{\mathrm{comp},3}$ is subleading to the $\Omega^2$ dependence of $\Sigma''_{\mathrm{comp},2}$ 
in Eq.~(\ref{2.1}),
the three-loop self-energy
 in Eq.~(\ref{3loop})
 has a more singular
  dependence on
  the distance to the hot spot ($\delta k$), and can thus compete with the two-loop one. Using Eq.~(\ref{2.1.1}) for  $\Sigma''_{\mathrm{comp},2}$, we find
for the ratio
\beq
\frac{\Sigma''_{\mathrm{comp},3}(\tk,\bar\Omega)}{\Sigma''_{\mathrm{comp},2}(\tk,\bar\Omega)}\sim\left(\frac{E_F^*}{\bar g}\frac{\bar\Omega \langle Z\rangle}{\tk^4}\right)^{1/2}.\label{23ratio}
\eeq

In Sec.~\ref{sec:3}, we will see that the two-loop self-energy gives the dominant contribution to the conductivity ($\sigma'\propto \bar\Omega^{-1/3}$) if $\bar\Omega<\bar g/E_F^*$, and that the relevant values of $\tk$ in this regime are $\tk^*\sim \left(\bo E_F^*/\bg\right)^{1/3}<1$.  Recalling that $\langle Z\rangle \sim \tk$ for $\tk<1$ and substituting $\tk^*$ for $\tk$ into Eq.~(\ref{23ratio}),
we find that,
  for $\Omega$ and $\delta k$ relevant for the conductivity,
the $2k_F$ three-loop composite self-energy is of the same order
 the
 two-loop
  self-energy.
   Combining this
   result
   with
   that for the three-loop self-energy
   in the Cooper
   channel,
    we conclude that,
  as far as the conductivity is concerned,
  $\Sigma''_{\mathrm{comp},3}\sim  \Sigma''_{\mathrm{comp},2}$.

 It can be readily checked that the same is true also for higher ($n\geq 4$) orders, and also for the forward-scattering case.
  Therefore, an expansion in powers of the composite vertex is not, strictly speaking, controlled,
   but
    it also does not generate stronger singularities.
In reality, convergence of the series is determined by the numerical prefactors which we do not attempt to compute here.

We now turn to the 1D regime, where $\Sigma_{\mathrm{comp},2}$ scales as $\bg\Omega\ln \Omega$ [Eq.~(\ref{se2_5})]. In true 1D,
higher-order diagrams produce terms of the type $\lambda^n\Omega \ln^n (\Omega/\Lambda)$, where $\lambda$ is the dimensionless coupling constant and $\Lambda$ is the ultraviolet cutoff of the theory. The perturbation theory breaks down at the energy scale  $\Omega_{\mathrm{LL}}\sim \Lambda \exp(-1/\lambda)$, below which the Luttinger-liquid behavior
 emerges.
Computing the three-loop self-energy in the 1D regime,
 we obtain
\beq
\Sigma''_{\mathrm{comp},3}=\bg\frac{\langle Z\rangle^2}{\tk^3}\bo\ln\frac{\langle Z\rangle \tk}{\bo}.
\eeq
 In our case, the 1D regime exists only at sufficiently high energies, namely, for $\bo>(\bg/E_F^*)\min\{\tk^2,\tk^3\}$. As will see in Sec.~\ref{sec:3}, the conductivity in this regime is controlled by the region
$\tk\sim 1$.
At $\tk\sim 1$, the effective coupling constants in both two-loop and three-loop self-energies is of order one, and their ratio contains only a logarithmic factor:
\beq
\frac{\Sigma''_{\mathrm{comp},3}(\tk\sim 1,\bo)}{\Sigma''_{\mathrm{comp},2}(\tk\sim 1,\bo)}\sim \ln \frac{1}{\bo}.
\label{231D}
\eeq
At the lowest frequency marking the beginning of the 1D regime ($\bo\sim \bg/E_F^*<1$), the logarithm in Eq.~(\ref{231D}) is large,
 indicating that MFL form 
exists only
 at the
 two-loop order,
 while the actual form of $\Sigma''$ contains an anomalous dimension:
  $\Sigma''\propto \bo^
  {-(1+\alpha)}$ with $\alpha\neq 0$.
  A computation of $\alpha$ requires non-perturbative methods, e.g.,  multi-dimensional bosonization,
  and is beyond the scope of this paper.

 \section{Optical conductivity at a quantum critical point}
\label{sec:3}
In this section, we discuss the optical conductivity. Our analysis is presented in the following order. First, in Sec.~\ref{sec:sigma_se}, we discuss only the self-energy contribution to the conductivity in the various frequency regimes, while neglecting entirely the vertex corrections. In Sec.~\ref{sec:qual}, we present qualitative arguments, based on the Boltzmann equation, which explain why vertex corrections play a relatively
insignificant role in our problem. This conclusion is confirmed in Sec.~\ref{sec:cond}, where
 we compute vertex corrections diagrammatically and show that
   they
   change at most the logarithmic factors in the results of Sec.~\ref{sec:sigma_se}, while the power-law scaling forms of the conductivity remain intact.
\subsection{Self-energy contribution to the optical conductivity\label{sec:sigma_se}}
In this section, we calculate only the self-energy contribution to the real part of the optical conductivity, $\sigma'_\Sigma(\Omega)$,
   while neglecting the vertex part.
  The conductivity $\sigma'_\Sigma(\Omega)$ is obtained by convoluting two Green's functions in the current-current 
  correlator. For a quasi-2D system with lattice spacing $c$ in the $z$ direction and in-plane tetragonal symmetry,
  the in-plane conductivity is given by
\bea
\sigma'_\Sigma(\Omega)&=&\frac{e^2}{\Omega c}\int^0_{-\Omega}\frac{d\omega} {\pi}\oint \frac{d\bk_F}{(2\pi)^2}\int dk_\perp v^2_\bk \I G^R(\bk,\omega+\Omega)\nn\\
&&\times\I G^R(\bk,\omega),
\label{sse1}
\eea
where
 $d\bk_F$ is an element of the FS contour
 and
$G^R(\bk,\omega)$ is the retarded Green's function. Except for the regime of 1D-like two-loop composite scattering, which will be discussed separately, the self-energy of our problem depends very weakly on $k_\perp$. If this dependence is neglected, one can integrate Eq.~(\ref{sse1}) over $k_\perp$. In addition, we make use of the fact that $\sse$ is controlled by the narrow regions near the hot spots, where the bare Fermi velocity, $v_F$, varies
 slowly, and thus can be taken out of the integral. Integral over $\bk_F$ can then be replaced by that over $\delta k$ around each of the $N_{\mathrm{h.s.}}$ hot spots. 
 ($N_{\mathrm{h.s.}}=8$ for the FS in Fig.~\ref{fig:hotspot}).
   With these simplifications, $\sse$ is cast into the following form
\bea
\sigma'_\Sigma(\Omega)&=&\frac{e^2v_FN_{\mathrm{h.s}}}{4\pi^2 c}\int^\Omega_{0} \frac{d\omega}{\Omega}\int d\delta k\nn\\ &&\times\frac{\Sigma''(\delta k,\Omega-\omega)+\Sigma''(\delta k,\omega)}{\lr\frac{\Omega}{Z_{\delta k}}\rr^2+\ls\Sigma''(\delta k,\Omega-\omega)+\Sigma''(\delta k,\omega)\rs^2}.\nn\\
\label{sse2}
\eea
For an order-of-magnitude estimate, one can replace $\int^\Omega
_0
 d\omega\left[\Sigma''(\delta k,\Omega-\omega)+\Sigma''(\delta k,\omega)\right]$ by $\Sigma''(\delta k,\Omega)$ and neglect $\Sigma''$ in the denominator of Eq.~(\ref{sse2}). Introducing the nominal conductivity
\beq
\sigma_0\equiv \frac{e^2 N_{\mathrm{h.s}}}{4\pi^2 c}
\label{sse3}
\eeq
and using the dimensionless variables defined by Eq.~(\ref{ch_6}),
we obtain
\beq
\sse\sim\frac{\sigma_0}{\bo^2}\int d\tk Z_\tk^2\frac{\Sigma''(\tk,\bo)}{\bg}.
\label{sse4}
\eeq
Now, we substitute $\Sigma''(\tk,\bo)$ and $Z_\tk$ found in the previous section into Eq.~(\ref{sse4}) and select the largest contribution to the integral.

In the frequency interval $0<\bo<\bg/E_F^*$, which includes both the top and bottom panels of Fig.~\ref{fig:sigma_vs_k_a}, the largest contribution to $\sigma'_\Sigma(\bo)$ comes from the region 2LC/2D  (two-loop composite scattering in the 2D regime), where $\Sigma''\sim \bg(E_F^*/\bg)\lr\bo^2/\tk^4\rr\ln^3(\tk^2/\bo)$ and $Z_\tk\sim \tk$. Because the integrand falls off rapidly (as $\tk^{-2}$) with $\tk$ in this regime, the upper limit of integration can be extended to infinity, while the lower limit coincides with the lower boundary of the 2LC/2D regime, i.e., $\tk\sim (\bo E_F^*/\bg)^{1/3}$. Substituting  expressions for $\Sigma''$ and $Z$ into Eq.~(\ref{sse4}), we obtain
\bea
\sse&\sim&\frac{\sigma_0}{\bo^2}\frac{E_F^*}{\bg}\int^\infty_{\lr\bo E_F^*/\bg\rr^{1/3}} d\tk \frac{\bo^2}{\tk^2}\ln^3\frac{\tk^2}{\bo}\nn\\
&\sim& \sigma_0\lr\frac{E_F^*}{\bg}\rr^{2/3}\frac{1}{\bo^{1/3}}
\ln^3\ls\left(\frac{E_F^*}{\bg}\right)^2\frac{1}{\bo}\rs\nn\\
&=&\sigma_0\ls\frac{(E_F^*)^2}{\bg\Omega}\rs^{1/3}
\ln^3\frac{(E_F^*)^2}{\bg\Omega}.
\label{sse5}
\eea
 As we see, $\sigma'_\Sigma(\Omega)$ in this regime exhibits a NFL behavior, i.e., an $\Omega^{-1/3}$ divergence at $\bo\to 0$
  (modulo a logarithmic factor).

For $\bg/E_F^*<\bo<1$ (Fig.~\ref{fig:sigma_vs_k_b}, top panel), the dominant contribution comes from the regions 2LC/1D and 2LC/1D$^*$ (two-loop composite scattering in the 1D regime for $\tk<1$ and $\tk>1$, correspondingly). As we said at the beginning of this section, the self-energy in this regime depends
both
on
 $\Omega$ and $v_F k_\perp$;
thus Eq.~(\ref{sse4}), derived from the Kubo formula for the case of $k_\perp$ independent self-energy, is not, strictly speaking,
  applicable. However, following the same steps that lead us to Eq.~(\ref{se2_1Db}), it can be readily shown that the mass-shell and FS values of the self-energy are of the same order and given by Eq.~(\ref{se2_8}). It is thus permissible to use Eq.~(\ref{se2_8}) for an estimate of the conductivity. We recall that $Z_\tk\sim \tk$ in the 2LC/1D region and $Z_\tk\approx 1$ in the 2LC/1D$^*$ region. Since the integral over $\tk$ in the 2LC/1D region converges at $\tk\to 0$, its lower limit ($\sqrt{\bo}$) can be set equal to zero. Likewise, the integral over $\tk$ in the 2LC/1D$^*$ region converges at $\tk\to\infty$ so that its upper limit [$\lr\bo E_F^*/\bg\rr^{1/2}$] can be extended to infinity. Combining these two contributions, we find
\bea
\sse&\sim&\frac{\sigma_0}{\bo^2}\left( \int^1_{0}d\tk \tk^2 \frac{\bo}{\tk}+ \int _1^{\infty}d\tk \frac{\bo}{\tk^2}\right)\nn\\
&\sim& \frac{\sigma_0}{\bo}=\sigma_0\frac{\bg}{\Omega}.
\label{sse6}
\eea
The integrals in both terms in the first line of Eq.~(\ref{sse6}) come from the region $\tk\sim 1$, which separates the 2LC/1D and 2LC/1D$^*$ regimes.

Finally, we come to the interval $1<\bo<(E_F^*/\bg)^2$ (Fig.~\ref{fig:sigma_vs_k_b}, bottom panel). The dominant contribution to
 conductivity in this case
  comes from
  the hot region ($0<\tk<\sqrt{\bo}$), where $\Sigma''(\tk,\bo)\sim \bg\sqrt{\bo}$. At lower frequencies, the hot-region contribution to the conductivity is reduced due to a small value of the $Z$ factor. At $\bo>1$, however, the $Z$ factor is almost equal to unity and does not affect the conductivity, which is given by
\bea
\sse\sim\frac{\sigma_0}{\bo^2} \int^{\sqrt{\bo}} d\tk \sqrt{\bo}\sim \frac{\sigma_0}{\bo}=\sigma_0\frac{\bg}{\Omega},
\label{sse7}
\eea
which is the same scaling as in Eq.~(\ref{sse6}). Therefore, the MFL, $1/\Omega$ scaling of $\sigma'_\Sigma$ spans over a wide frequency region: $\bg/E_F^*<\bo<(E_F^*/\bg)^2$ (or $\bg^2/E_F^*<\Omega<(E_F^*)^2/\bg$),
 although the prefactor changes
  between
 the regions of
  $\bg^2/E_F^*<\Omega < {\bar g}$ and ${\bar g} < \Omega < (E_F^*)^2/\bg$.

Summarizing, $\sigma'_\Sigma(\Omega)$ is given by
\bea
\sigma'(\Omega)\sim \sigma_0\times\left\{
\begin{array}{l}
\ls\frac{(E_F^*)^2}{\bg\Omega}\rs^{1/3}
\ln^3\frac{(E_F^*)^2}{\bg\Omega},~~\mathrm{for}\;0<\Omega<\bg^2/E_F^*\\
\bg/\Omega,~~\mathrm{for}\; \bg^2/E_F<\Omega<(E_F^*)^2/\bg.\\
\label{sigma_Sigma}
\end{array}
\right.
\eea

\subsection{
Vertex corrections: Boltzmann equation
}
\label{sec:qual}
    The estimates for the conductivity in the previous section [Eqs.~(\ref{sse5}-\ref{sse7})] were obtained by taking into account only the self-energy contribution to the current-current correlation function while neglecting the vertex corrections. In certain cases, the vertex corrections reduce the self-energy contribution significantly, and even cancel it out entirely (for the case of a Galilean-invariant system). At first glance, one may expect a strong cancelation between the self-energy and vertex contributions to occur in our case as well.  Indeed, all the relevant processes, considered in Sec.~\ref{sec:sfm}, involve fermions with either almost parallel or almost antiparallel momenta. Had we been dealing with a generic FL, a contribution of such processes to the transport relaxation rate would have been much smaller than that to the self-energy.
 We will show,  however, that the cancelation between the self-energy and vertex-correction contributions for our case -- which is a strongly anisotropic and strongly correlated FL/NFL -- turns out to be much less dramatic:  the self-energy result
 overestimates the actual conductivity  by at most a logarithmic factor, while a power-law singularity
 of $\sigma'$ remains intact.

    To see this result qualitatively, we recall that, within the Boltzmann-equation approach, a contribution to the $jj$ component of
    the conductivity tensor from a
    four-fermion interaction process contains a \lq\lq current-imbalance factor\rq\rq\/ \cite{rosch:2005,maslov:2011}
    \beq
    \Delta\equiv \ls \bv_j(\bk)+\bv_j(\bp)-\bv_j(\bkp)-\bv_j(\bpp)\rs^2,
    \label{Delta}
    \eeq
   averaged with the scattering probability over the FS. It is the presence of $\Delta$ that makes the transport scattering rate to be, in general, different
   from the quasiparticle decay rate.
   The role of $\Delta$ is to ensure gauge-invariance and
    time-reversal symmetry. Gauge-invariance implies that there is no contribution to the conductivity from strictly forward scattering, when
    $\bkp=\bk$ and $\bpp=\bp$ (or $\bkp=\bp$ and $\bpp=\bk$)
     in which case $\Delta=0$.
     Time-reversal symmetry guarantees that there is also no contribution from scattering in the Cooper channel,  when $\bp+\bk=\bm{0}=\bkp+\bpp$ and hence the total currents carried by the incoming and outgoing fermions are equal to zero.
    A $2k_F$ scattering process, 
    as defined
    in this paper,  is a subcase of 
    the
     Cooper process with additional constraints
    $\bkp\approx \bk$ and $\bpp\approx \bp$, and hence $\Delta=0$ in this case as well.
    The question now is how strongly do these constraints reduce the transport scattering rate of lukewarm and hot  fermions compared to the quasiparticle
    decay rate.

   For a forward-scattering process, all four lukewarm fermions are near the same hot spot, i.e., $\bk=\bk_{\mathrm{h.s.}}+\delta\bk$,
   $\bp=\bk_{\mathrm{h.s.}}+\delta\bp$, $\bkp=\bk_{\mathrm{h.s.}}+\delta\bk+\delta\bq$, and $\bpp=\bk_{\mathrm{h.s.}}+\delta\bp-\delta\bq$,
   where all the \lq\lq$\delta$ vectors\rq\rq\/ are tangential to the FS.
   For a $2k_F$ scattering process, two out of the four fermions are near the same the hot spot, while the other two are near the opposite spot, e.g.,
  $\bk=\bk_{\mathrm{h.s.}}+\delta\bk$,
   $\bp=-\bk_{\mathrm{h.s.}}+\delta\bp$, $\bkp=\bk_{\mathrm{h.s.}}+\delta\bk+\delta\bq$, and
   $\bpp=-\bk_{\mathrm{h.s.}}+\delta\bp-\delta\bq$.
   Obviously, $\Delta$  vanishes in the limit of $\delta q\to 0$ for both types of scattering.

  If the quasiparticle velocity varies smoothly along the FS,
   the velocities entering Eq.~(\ref{Delta}) can be expanded near the corresponding hot spots as
   $\bv(\bk_{\mathrm{h.s.}}+\delta \bk)\approx \bv(\bk_{\mathrm{h.s.}})+(\delta\bk\cdot\nabla)\bv(\bk_{\mathrm{h.s.}})+\mathcal{O}(\delta k^2)$,
   and similarly for other terms in $\Delta$. The linear terms of the expansion then cancel out, and the
   contribution to the conductivity is reduced by a factor of $\Delta\propto \delta k^4$.
   Such a situation would be encountered in a generic FL
   (in which case $\bk_{\mathrm{h.s.}}$ is to be understood as just an arbitrary
    point on the FS rather than a hot spot).
   However,  the situation is very much different for a FL near SDW criticality, in which case the (renormalized) quasiparticle velocity varies rapidly around the hot spot.
   Using the definition of the $Z$ factor from Eq.~(\ref{1.4}) and assuming that the {\em bare} velocity, $\bv^0(\bk)$,
   varies smoothly along the FS, we can re-write the velocities in Eq.~(\ref{Delta}) as
   $\bv(\bk_{\mathrm{h.s.}}+\delta\bk)=\bv^0(\bk_{\mathrm{h.s.}})Z_{\delta k}$, etc.~\cite{comment_sigmak}
   Consequently, the current-imbalance factor is reduced to
   \bea
   \Delta=\ls\bv^0_j(\bk_{\mathrm{h.s.}})\rs^2\ls Z_{\delta k}\pm Z_{\delta p}\mp Z_{\delta k+\delta q}\pm Z_{\delta p-\delta q}\rs^2,\nn\\
   \label{DeltaZ}\eea
   where $\pm$ corresponds to forward/$2k_F$ scattering.
   The combination of the four $Z$ factors form a scaling function of $\delta k$, $\delta p$, and $\delta q$.
   In  the lukewarm regime, for example, this function is obtained by substituting Eq.~(\ref{sa2}) into Eq.~(\ref{DeltaZ}).
  While the deviations from the hot spots, $\delta k$ and $\delta p$, as well as the momentum transfer,
   $\delta q$, are small compared with $k_F$, the momentum transfer is not small compared with $\delta k$ and $\delta p$; instead,
   $\delta q\sim \delta k\sim \delta p$. Therefore,  one cannot expand the combination of the four $Z$ factors any further, and $\Delta$ is small only as the square of the $Z$ factor itself,
    e.g., only as $\delta k^2$ in the lukewarm regime. This smallness has already been taken into account in the \lq\lq naive\rq\rq\/ estimate for the conductivity; indeed,
     the $Z$ factor
     in the denominator of Eq.~(\ref{cond}) accounts  for velocity renormalization.\cite{kotliar}

     In the 2LC/2D regime, the transport rate is smaller than the quasiparticle decay rate only by a logarithmic factor present in the latter [cf. Eq.~(\ref{2.1})].  Indeed, a cube of the logarithm in Eq.~(\ref{2.1}) comes from two sources. Two out of three logarithms come from the logarithmic singularity of the composite vertex in thee regime of $\delta q<\min\{\delta k,\delta p\}$. However,
in processes relevant for the conductivity $\delta q\sim\delta k\sim\delta p$, and thus the logarithmic singularity of the vertex is replaced by a number of order one. The third logarithm comes from the $1/\delta q$ singularity of the integrand in the self-energy but this
 singularity is  canceled by the vanishing of $\Delta$ at $\delta q=0$.
  The remainder of the self-energy comes from the region
  $\delta q\sim\delta k\sim\delta p$ and has no logarithms.

     The power-law singularity of the conductivity, $\sigma'(\Omega)\propto 1/\Omega^{1/3}$, comes from
   the $1/(\delta k)^4$ singularity of the self-energy, which is not affected by the factors described above.  One should then expect the actual low-frequency form of the conductivity to be
   \beq
   \sigma'(\Omega)\sim \sigma_0\ls\frac{(E_F^*)^2}{\bg\Omega}\rs^{1/3}. \label{cond13}
   \eeq
   for $0<\Omega<\bg^2/E_F^*$.

   At higher frequencies ($\bg^2/E_F^*<\Omega<(E_F^*)^2/\bg$), the self-energy contribution to the conductivity contains no logarithmic factors [cf.~Eqs.~(\ref{sse6}) and (\ref{sse7})], and thus $\sigma'(\Omega)$ differs from $\sigma'_{\Sigma}(\Omega)$ 
   by
     at most a number of order one, i.e.,
   \beq
   \sigma'(\Omega)\sim \sigma_0\frac{\bg}{\Omega}.\label{cond14}
   \eeq
     The conductivity as a function of $\Omega$ is sketched in Fig.~\ref{fig:sigma_omega}.
   \begin{figure}[h]
\includegraphics[scale=0.5]{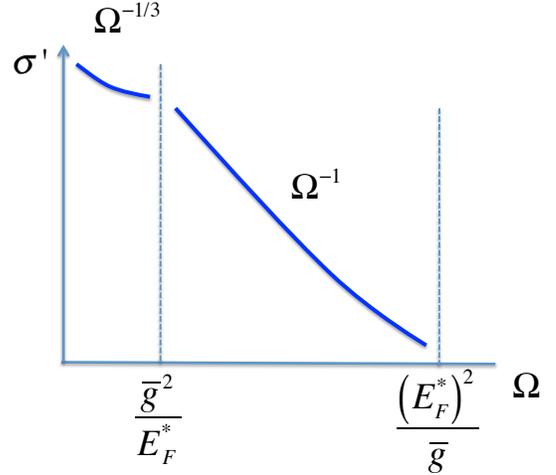}
\vspace{-1.5 in}
\caption{The real part of the conductivity as a function of frequency.}
\label{fig:sigma_omega}
\end{figure}

   To two-loop order,
   Eq.~(\ref{cond13}) was obtained by HHMS
   who argued, however, that a singular behavior of the conductivity
   comes only from $2k_F$ scattering, while the forward-scattering contribution is canceled by vertex corrections. Our analysis does not reveal major
   differences between forward- and $2k_F$ scattering
   to two-loop order.

   We should point out, however, that the reasoning based on the Boltzmann equation is not precise. While the canonical form of the Boltzmann equation is valid only to second order in a {\em static} interaction (or else for an effective interaction obtained in the Random Phase Approximation),\cite{keldysh} scattering at composite bosons corresponds to {\em fourth} order in the {\em dynamic} interaction--the staggered spin susceptibility.
  Our situation, however, is simplified by the fact that the intermediate fermions are far off their mass shells.  As a result, the four-leg vertex, which should {\em a priori} depend on all three fermionic frequencies (the fourth one is fixed by energy conservation),
    actually depends
    only on the frequency transfer. For such a vertex, cancelations between the diagrams occur in the same way as predicted by the Boltzmann equation.
    In the next section, we will present a detailed analysis of the diagrams for the conductivity which
    confirms the qualitative arguments given in this section.

 \subsection{
 Diagrams for the conductivity}
  \label{sec:cond}
  \subsubsection{Terminology and notations}
  We use the Kubo formula for
 the
 conductivity at
 finite
  frequency
 $\sigma'_{jj}(\Omega) \propto
  \mathrm{Im }{\cal P}_{jj} (\Omega)/\Omega$.
 The current-current correlator
 ${\cal P}_{jj} (\Omega) \equiv {\cal P} (\Omega)$
 is given by a particle-hole bubble with zero momentum transfer and  frequency transfer $\Omega$,
  and with velocities of internal fermions ${\bf v} ({\bf p})$ at the vertices.

The two diagrams for the current-current correlator
${\cal P}
(\Omega)$ with
 self-energy insertions are shown in Fig.~\ref{fig:AB}~A.
Other contributions to
${\cal P}  (\Omega)$ are the vertex-correction diagram
 (Fig.\ref{fig:AB}~B) and two Aslamazov-Larkin diagrams (Fig.~\ref{fig:CD})
 (see, e.g., Ref.~\onlinecite{larkin}).
Depending on whether the momenta on the solid and dashed are near the same or opposite hot spot,
we are dealing with a forward or $2k_F$ scattering process, correspondingly.

Before we proceed further,
  a
  brief remark on
  terminology is in order.
  We believe that the diagram
  identified by
  HHMS
  as
  a
  \lq\lq vertex correction\rq\rq\/
 is actually
 the first of the
 two Aslamazov-Larkin
   diagrams (Fig.~\ref{fig:CD}~C), while the
   actual
   vertex-correction diagram (Fig.~\ref{fig:AB}
   B~)
   was not considered by HHMS.
  We will use
  this terminology
  throughout the rest of this section.

Our analysis proceeds in two steps. First, in Sec.~\ref{sec:cancel}, we show that diagrams $A$ and $B$ in Fig.~\ref{fig:AB}, as well as diagrams $C$ and $D$ in Fig.~\ref{fig:CD},  cancel each other {\em if} one neglects the variations of the $Z$ factor around the FS.
Next, in Sec.~\ref{sec:no cancel},
 we show that allowing for the variation of the $Z$ factor prevents complete cancelation and
does lead to power-law singularities in the conductivity, as announced in Eqs.~(\ref{cond13} and (\ref{cond14}).

\subsubsection{Cancellation of diagrams under the conditions of strict forward- and $2k_F$-scattering}
\label{sec:cancel}
 It is convenient to consider mutual cancelations
 between
  the diagrams in Fig.~\ref{fig:AB} and Fig.~\ref{fig:CD} separately.
\begin{figure}
\includegraphics[scale=0.35]{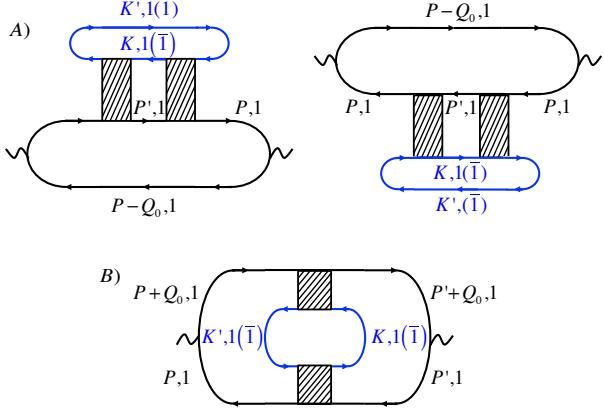}
\caption{Diagrams for the conductivity.
  Labels $1$ and $\bar 1$ corresponds to hot spots in Fig.~\ref{fig:hotspot}. $Q_0=({\bf 0},\Omega)$ is the $(2+1)$ momentum of the external electric field. A)  Self-energy diagrams. B) Vertex-correction diagram.}
\label{fig:AB}
\end{figure}
\begin{figure}
\includegraphics[scale=0.35]{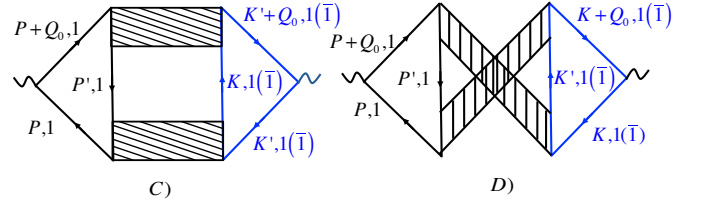}
\caption{Aslamazov-Larkin diagrams for the conductivity.
Notations are the same as in Fig.~\ref{fig:AB}.}
\label{fig:CD}
\end{figure}
As before, we use $(2+1)$ notations for the energy and momentum, such that $P=(\bp,
\Omega_p)$, etc. The external $(2+1)$ momentum has only the
 frequency  component:
  $Q_0\equiv ({\bf 0},\Omega)$, where $\Omega$  is the frequency of the external electric field
 (chosen to be positive
  for convenience.)\\

\paragraph{{\bf Self-energy and vertex-correction diagrams.}}

First, we discuss the diagrams $A$ and $B$, whose contributions to the current-current correlator are given by
\bse
\bea
{\cal P}_A&=&
-
\int_P\bv_j^2(\bp)\left[G(P-Q_0)G(P)+G(P)G(P+Q_0)\right]\notag\\
&&\times G(P)\Sigma(P)\label{A}\\
{\cal P}_B&=&\int_{P\dots K'} \bv_j(\bp)\cdot \bv_j(\bpp) G(P)G(P+Q_0)\notag\\
&&\times G(P')G(P'+Q_0)G(K)G(K')\Gamma\Gamma_{Q_0,1},\label{B}
\eea
\ese
where
\bse
\bea
\Gamma&\equiv& \Gamma\lr P,K;P',K'\rr,
\label{gamma}\\
\Gamma_{\Omega,1}&\equiv&\Gamma\left(P+Q_0,K;P'+Q_0,K'\right),
\label{gamma1}
\eea
\ese
 while the self-energy reads
\beq\Sigma(P)=-\int_{P',K,K'}\Gamma^2 G(P')G(K)G(K').\label{se}\eeq

For the time being, we are not specifying a particular form of the interaction vertex.
 The only requirement we impose is that the  vertex satisfies the microscopic reversibility condition:
$\Gamma(P,K;P',K')=\Gamma(P',K';P,K)$,
 which we have already used in Eq.~(\ref{se}).
 Note that the velocities $\bv(\bp)$ and $\bv(\bpp)$ in Eqs.~(\ref{A}) and (\ref{B}), as well as all velocities in the formulas below, are the {\em bare} ones.~\cite{comment_bare} Velocity renormalization
 by the interaction is accounted for by the $Z$ factors which occur explicitly in the Kubo formalism.

The Green's functions in the diagrams $A-D$ are renormalized by $\bq_{\pi}$ scattering, which determines the $Z$ factor. Therefore, the \lq\lq bare\rq\rq\/ Green's functions in the diagrams $A-D$ are of the form
\beq
G(P)=\lr\frac{i\Omega_p}{Z_{\bp}}-\ve_{\bp}\rr^{-1},\label{Gf}
\eeq
with $Z_\bp$ given by Eq.~(\ref{sa2}).
 Green's functions of the form (\ref{Gf}) satisfy the following identity
\beq
G(P)G(P+Q_0)=\frac{Z_{\bp}}{i\Omega}\ls G(P)-G(P+Q_0)\rs.\label{id}
\eeq
Splitting the products of the Green's functions in Eq.~(\ref{A}) with the help of
this identity,
 we  re-write ${\cal P}_A$ as
\beq
{\cal P}_A=
\frac{1}{i\Omega}\int_P \bv_j^2(\bp)Z_{\bp}\ls\Sigma(P)-\Sigma(P+Q_0)\rs
G(P)G(P+Q_0).
\label{A1}
\eeq
As we saw in Sec.~\ref{sec:sigma_se}, the diagram $A$ by itself produces singular terms in the conductivity,
 given by Eqs.~(\ref{sse5}-\ref{sse6}).
By construction,
 the momenta along
 both
the
top
and bottom
lines of the composite vertex $\Gamma(P,K; P',K')$
 are close to each other, i.e., $\bk' \approx \bk$ and ${\bf p}' \approx {\bf p}$,
and so are the velocities in diagram $B$: $\bv(\bp)\approx \bv(\bpp)$.
To see if the singular contributions from diagrams $A$ and $B$ cancel each other,
   we first neglect the differences between $\bv(\bp)$ and $\bv(\bpp)$,
   and also between $Z_\bp$ and $Z_{\bpp}$.
   The first constraint corresponds to either strict forward scattering, when the momenta on the solid and dashed lines are near the same hot spot,
   or to strict $2k_F$ scattering, when these momenta are near the opposite hot spots. The constraint $Z_\bp=Z_{\bpp}$ will be relaxed in the
   Sec.~\ref{sec:no cancel}.
   Imposing these constraints and
   applying
    identity (\ref{id}) to the product $G(P')G(P'+Q_0)$ in diagram $B$, we
    re-write  ${\cal P}_B$ as
   \bwt
\bea
{\cal P}_B&=&\frac{1}{i\Omega}\int_{P\dots K'} \bv_j^2(\bp)Z_\bp \ls G(P)G(P+Q_0)G(P')\Gamma\Gamma_{\Omega,1}- G(P)G(P+Q_0)G(P'+Q_0)\Gamma\Gamma_{\Omega,1}\rs
G(K)G(K').
\label{3_4}
\eea
\ewt
Next, we
 re-write $\Gamma\Gamma_{\Omega,1}$ entering the first and second terms in the square brackets of Eq.~(\ref{3_4}) as
 $\Gamma\Gamma_{\Omega,1} + \Gamma^2 - \Gamma^2$
 and $\Gamma\Gamma_{\Omega,1} + \Gamma_{\Omega,1}^2 - \Gamma_{\Omega,1}^2$,
correspondingly.
Then ${\cal P}_B$
 can be represented as a sum of three terms: ${\cal P}_B={\cal P}_B^1+{\cal P}_B^2+{\cal P}_B^3$, where
\bse
\bea
{\cal P}^{1}_B&=&\frac{1}{i\Omega}\int_{P\dots K'} \bv_j^2(\bp)Z_{\bp} \ls \Gamma^2 G(P')-\Gamma_{\Omega,1}^2 G(P'+Q_0)\rs \notag\\
&&\times G(P)G(P+Q_0)G(K)G(K'),\label{B1}\\
{\cal P}^{2}_B&=&\frac{1}{i\Omega}\int_{P\dots K'}  \bv_j^2(\bp)Z_{\bp}\Gamma \ls\Gamma_{\Omega,1}-\Gamma \rs \notag\\
&&\times G(P')G(P)G(P+Q_0)G(K)G(K'),\label{B2}\\
{\cal P}^{3}_B&=&\frac{1}{i\Omega}\int_{P\dots K'}  \bv_j^2(\bp)Z_{\bp}\Gamma_{\Omega,1} \ls\Gamma_{\Omega,1}-\Gamma \rs \notag\\
&&\times G(P'+Q_0)G(P)G(P+Q_0)G(K)G(K').\notag\\
\label{B3}\eea
\ese
 Using the self-energy from Eq.~(\ref{se}), we re-write ${\cal P}^{1}_B$ as
\beq
{\cal P}^{1}_B=\frac{1}{i\Omega}\int_P  \bv_j^2(\bp)Z_{\bp} \ls \Sigma(P+Q_0)-\Sigma(P)\rs G(P)G(P+Q_0).
\eeq
Comparing this result with $P_{A}$ in Eq.~(\ref{A1}) we see that this part of the diagram $B$ cancels out the entire diagram $A$:
${\cal P}_A+{\cal P}_B^1=0$.

If $\Gamma$
 were
 an arbitrary dynamic vertex,
 each of the two remaining  terms, ${\cal P}^2_B$ and ${\cal P}_B^3$,
 would, in general, be of the same order as
${\cal P}_A$.
 Our case, however, is special in
 the sense that, within the approximation adopted for the composite vertex
 in Sec. \ref{sec:vertex},
   the
    frequency dependence of $\Gamma(P,K;P',K')$
    involves only one variable --
 the difference of the frequencies of the initial and final states:
\bea
\Gamma(P,K;P',K')=
 F (|\Omega_p-\Omega_{p'}|; \bp-\bp',\bp,\bk'),\nn\\
 \label{3_3}
\eea
 where an explicit form of the function $F$
can be read off from Eq.~(\ref{1.7}).
 Since  $\Gamma_{\Omega1}$ differs from $\Gamma$ only by a shift of the initial and final frequencies by $\Omega$ [see Eqs.~(\ref{gamma}) and (\ref{gamma1})], it follows from Eq.~(\ref{3_3}) that
$\Gamma_{\Omega,1}=\Gamma$,
 and
  thus ${\cal P}^2_B={\cal P}^3_B=0$. Therefore, the sum of diagrams
$A$ and $B$ is equal to zero.\\

\paragraph{{\bf Aslamazov-Larkin diagrams.}}

We now turn to Aslamazov-Larkin diagrams
 $C$ and $D$ in Fig.~\ref{fig:CD}.
The corresponding contributions to the current-current correlator
 are:
\bse
\bea
{\cal P}_C=
\int_{P\dots K'}&&\bv_j(\bp)\cdot\bv_j({\bkp}) G(P)G(P+Q_0)G(K')\notag\\
&&\times G(K'+Q_0)G(P')G(K)\Gamma_{\Omega,2}\Gamma,\label{C}\\
{\cal P}_D=
\int_{P\dots K'}&&\bv_j(\bp)\cdot \bv_j({\bk})G(P)G(P+Q_0) G(K) \notag\\
&&\times G(K+Q_0)G(P')G(K')\Gamma_{\Omega,3}\Gamma_{\Omega,4}\;,\nn\\\label{D}
\eea
\ese
where $\Gamma$ is given by Eq.~(\ref{gamma}), and
\bse
\bea
\Gamma_{\Omega,2}&\equiv&\Gamma(P+Q_0,K;P',K'+Q_0),\label{gamma2}\\
\Gamma_{\Omega,3} &\equiv&\Gamma(P,K+Q_0;P',K'),\label{gamma3}\\
\Gamma_{\Omega,4}&\equiv&\Gamma(P+Q_0,K;P',K').\label{gamma4}
\eea
\ese
In each vertex, the sum of the incoming momenta/frequencies is equal to the sum of the outgoing
ones.
For a forward-scattering process, all momenta in the diagrams $C$ and $D$ are close to each other, i.e., $\bp\approx \bpp\approx\bk\approx\bkp$.
 For a $2k_F$ process,
 the momenta are related to each other as $\bp\approx \bpp\approx -\bk\approx -\bkp$.
Accordingly, the current vertices can be simplified as
 \beq \bv_j(\bp)\cdot \bv_j({\bk})=\bv_j(\bp)\cdot\bv_j({\bkp})=
 \pm
 \bv^2_j(\bp),\label{current}\eeq where the $+$($-$) sign corresponds to forward ($2k_F$) scattering.
 As it was done for the diagrams $A$ and $B$, we also
 set all the $Z$ factors to be equal to $Z_\bp$ for the time being.
 Applying
  identity (\ref{id}) to the products of the first four Green's functions in Eqs.~(\ref{C}) and (\ref{D}), we represent both ${\cal P}_C$ and ${\cal P}_D$ as a sum of four terms:
\bse
\bwt
\bea
{\cal P}_C&=&
\pm
\frac{1}{\Omega^2}\int_{P,P',K,K'}\left\{\bv^2_j(\bp)
Z^2_\bp\ls {G(P)G(K')}+G(P+Q_0)G(K'+Q_0)-{G(P)G(K'+Q_0)}-{G(P+Q_0)G(K')}\rs\right. \notag\\
&&\times G(P')G(K)\Gamma\Gamma_{\Omega,2}\Big\}\label{C1}\\
{\cal P}_D&=&
\pm
\frac{1}{\Omega^2}\int_{P,P',K,K'}\left\{\bv^2_j(\bp)
 Z^2_\bp\ls {G(P)G(K)}+{G(P+Q_0)G(K+Q_0)}-{G(P)G(K+Q_0)}-{G(P+Q_0)G(K)}\rs\right. \notag\\
&&\times G(P')G(K')\Gamma_{\Omega,3}\Gamma_{\Omega,4}\Big\}.\label{D1}
\eea
\ewt
\ese
 Shifting the
   momenta by the external
  momentum $Q_0$,
  we reduce the sum of
Eqs.~(\ref{C1}) and (\ref{D1}) to the following form
\bea
{\cal P}_C+{\cal P}_D=
\pm
\frac{1}{\Omega^2}\int_{P,P',K,K'}&&\bv^2(\bp)Z_\bp^2 G(P)G(K')G(P')G(K)\notag\\
&&\times{\cal G}(P,K,P',K',Q_0),
\eea
where ${\cal G}(P,K',P',K,Q_0)$ is a bilinear combination of the vertices given by
\bwt
\bea
{\cal G}(P,K,P',K',Q_0)&=&\Gamma(P,K;P',K')\ls \Gamma(P+Q_0,K;P',K'+Q_0)-\Gamma(P+Q_0,K-Q_0;P',K')
\rs\notag\\
&&
+
\Gamma(P,K;P',K')\ls
\Gamma(P-Q_0,K;P',K'-Q_0) -\Gamma(P-Q_0,K+Q_0;P',K')\rs\notag\\
&&+\Gamma(P-Q_0,K;P',K')\ls \Gamma(P,K-Q_0;P',K')-\Gamma(P,K;P',K'+Q_0)\rs\notag\\
&&+\Gamma(P+Q_0,K;P',K')\ls \Gamma(P,K+Q_0;P',K')-\Gamma(P,K;P',K'-Q_0)\rs .
\label{CD}
\eea
\ewt
 Again,
  if $\Gamma$
  were
   an arbitrary vertex,
    ${\cal G}$
      would
       be
      non-zero.
     However, for
       our form
       composite vertex, the vertices in each of the four square brackets in Eq.~(\ref{CD}) cancel each other. For example,
       in the first line
        of
         Eq.~(\ref{CD})
        we have $\Gamma(P+Q_0,K;P',K'+Q_0) =
        F(|\Omega_p+\Omega - \Omega_{p'}|, \bp-\bp^{'},\bp,\bk)$
       and $\Gamma(P+Q_0,K+Q_0;P',K')=
       F(|\Omega_p+\Omega - \Omega_{p'}|, \bp-\bp^{'},\bp,\bk)$, i.e., the two
       vertices
        are equal.
       Likewise,  the remaining three lines in Eq.~(\ref{CD}) also vanish. Therefore, ${\cal G}=0$ and ${\cal P}_C+{\cal P}_D=0$.

 Therefore, if one focuses on strict forward and $2k_F$ scattering
 and neglects the variation of the  $Z$ factor along the FS,
 the contributions to the conductivity from  all the diagrams cancel each other.

\subsubsection{Absence of cancelation of the power-law singularity in the conductivity}
\label{sec:no cancel}
We are now relaxing the constraints of strict forward- and $2k_F$-scattering by taking into account that the $Z$ factors of fermions with different,
albeit close,
 momenta are different. The bare fermionic velocities will be still  taken at either the same or opposite; however,
 as explained in Sec.~\ref{sec:qual}, the renormalized velocities which, in our model, differ from the bare ones by the $Z$ factors, vary rapidly near the hot spots.
Since allowing for such a variation will be already sufficient for eliminating the cancelation of the diagrams even for the special form
  of the composite vertex in Eq.~(\ref{3_3}), we initially restrict our analysis to that form of the vertex.
  Consequently, the vertices entering the diagrams $A$-$D$  are related to each other as
  \beq
  \Gamma_{\Omega,1}
  =\Gamma_{\Omega,3}=\Gamma;\;\Gamma_{\Omega,4}=\Gamma_{\Omega,2}.\label{gamma_con}\eeq

   With these constraints on the vertices and also with $\bv_j(\bp)=\bv_j(\bpp)$, the sum of the diagrams $A$ and $B$ is reduced to
  \bea
  {\cal P}_A+{\cal P}_B&=&\frac{1}{i\Omega}\int_{P\dots K'}\bv^2_j(\bp)G(P)G(P+Q_0)G(K)G(K')\nn\\
 &&\times\Gamma^2 \lr Z_{\bp'}-Z_{\bp}\rr\ls G(P')-G(P'+Q_0)\rs.\label{AB}
  \eea
We define the \lq\lq auxiliary self-energy\rq\rq\/
  as
  \beq
  \Sigma_Z(P)\equiv-\int_{P',K,K'} \Gamma^2 G(P')G(K)G(K')\frac{Z_{\bpp}}{Z_{\bp}},
  \eeq
  which differs from the usual self-energy [Eq.~(\ref{se})] by the ratio of the $Z$ factors under the integral. Defining also the difference of the usual and auxiliary self-energies,
  $\Delta\Sigma(P)\equiv \Sigma(P)-\Sigma_Z(P)$, we re-write Eq.~(\ref{AB})  as
  \bea
  {\cal P}_A+{\cal P}_B&=&\frac{1}{i\Omega}\int_{P}\bv^2_j(\bp)
  Z_\bp\ls \Delta\Sigma(P)-\Delta\Sigma(P+Q_0)\rs\nn\\
 &&\times  G(P)G(P+Q_0).\label{AB1}
  \eea
  Now the sum of the diagrams $A$ and $B$ has a form similar to that of the diagram $A$ itself [Eq.~(\ref{A1})], except for the usual self-energy in Eq.~(\ref{A1}) is replaced by $\Delta\Sigma$ in Eq.~(\ref{AB1}).
  Therefore, to compare Eqs.~(\ref{A1}) and (\ref{AB1}), we only need to compare the two-loop self-energy  with
  \bea
  \Delta\Sigma(K)=
 -\int_{P,P',K'}\Gamma^2 G(P)G(P')G(K')\frac{Z_{\bk}-Z_{\bkp}}{Z_{\bk}}.\nn\\\label{dse}
  \eea
  In what follows, we consider explicitly only the 2D regime of two-loop composite scattering with the self-energy given by
  Eq.~(\ref{2.1}).
 When evaluating the usual self-energy in Eq.~(\ref{se2}), we integrated over $p_\perp$ and then over $q_\perp$, which led us to Eq.~(\ref{se3}). Performing the same integrations in Eq.~(\ref{dse}) and using an explicit form of the $Z$-factor from Eq.~(\ref{sa2}), we arrive at
\bwt
\bea
\Delta\Sigma_{\mathrm{comp}_2} (\delta k, \Omega_k)&=&\frac{1}{2v_F}\int \frac{d\Omega_q}{2\pi}\int \frac{d\delta q}{2\pi } \int \frac{d\Omega_p}{2\pi}\int \frac{d\delta p}{2\pi} \frac{\mathrm{sgn}(\Omega_{p}-\Omega_q)-\mathrm{sgn}(\Omega_p)}
{\frac{{i\Omega_p}}{Z_\bp}-\frac{i(\Omega_p-\Omega_q)}{Z_{\bp-\bq}}-\frac{\delta p \delta q}{m^*}+\frac{(\delta q)^2}{2m^*}}
{\cal Z}_{\delta k,\delta q}
\nn\\ &&\times\mathrm{sgn}(\Omega_k+\Omega_q)\Gamma^2(K,P;Q).
\label{ser}\eea
\ewt
where ${\cal Z}_{\delta k,\delta q}\equiv 1-|\delta k+\delta q|/|\delta k|$.
Since ${\cal Z}_{\delta k,\delta q}$
vanishes as $\delta q$ at $\delta q\to 0$, the $1/|\delta q|$ singularity of the particle-hole bubble is eliminated. In the absence of the $1/\delta q$ singularity, the internal momenta are of order of the internal one: $\delta q\sim \delta p\sim \delta k$. Therefore, the logarithmic factor in the vertex [Eq.~(\ref{1.7})] is replaced by a number of order one, whereas the third (kinematic) logarithm simply does not occur. As a result, $\Delta\Sigma$ contains no logarithmic factors.  However, ${\cal Z}_{\delta k,\delta q}\sim 1$ at relevant $\delta q\sim\delta k$
and thus does not affect power-counting of the rest of the result, which reads
\beq\Delta\Sigma''\sim  \Omega^2\frac{{\bar g}^2}{(v_F \delta k)^3} \frac{E^*_F}{v_F \delta k}.
\eeq
Therefore, the combined contribution of the diagrams $A$ and $B$ differs only by a logarithmic factor from the self-energy contribution (the diagram $A$).

The imaginary parts of the self-energies two-loop composite scattering in the 1D regime [Eq.~(\ref{se2_10})]
and from $\bq_\pi$ scattering of hot fermions [Eq.~(\ref{1.3}) with $\delta k=0$] contain no logarithmic factors.  Since ${\cal Z}_{\delta k,\delta q}\sim 1$ in these cases as well, the combined contribution of the diagrams $A$ and $B$ differs from that of the diagram $A$ only by a number of order one.

 We now turn to the diagrams $C$ and $D$. Using constraints (\ref{gamma_con}) for  the interaction vertices and (\ref{current})
 for the current vertices but keeping the momentum dependence of the $Z$-factors, we obtain for the sum of the diagrams $C$ and $D$
 \bwt
 \bea
 {\cal P}_C+{\cal P}_D=\pm \frac{1}{i\Omega}\int_{P\dots K'}\bv^2_j(\bp)\Gamma\Gamma_{\Omega,2}\lr Z_{\bk}-Z_{\bkp}\rr
\ls G(K)G(K'+Q_0)G(P')-G(K)G(K')G(P')\rs G(P)G(P+Q_0). \nn\\\label{CD1}
 \eea
 \ewt
 [In deriving this result, we also used properties (\ref{id}) and (\ref{3_3}).] In general, Eq.~(\ref{CD1}) cannot be expressed via the self-energy because it contains a product of different interaction vertices, $\Gamma$ and $\Gamma_{\Omega,2}$, whereas the
self-energy contains $\Gamma^2$, and also because the external frequency enters the first term in the square brackets in a different way as compared to the self-energy diagram. In our case, however, these differences are immaterial.
Indeed, Eq.~(\ref{gamma2}) shows that $\Gamma_{\Omega,2}$ differs from $\Gamma$ only in that the first and last fermionic frequencies are shifted by the external frequency $\Omega$. Since the composite vertex in Eq.~(\ref{1.7}) depends on the frequency only logarithmically, the difference between $\Gamma$ and $\Gamma_{\Omega_2}$ is not important to logarithmic accuracy. If we identify $\Gamma$ with $\Gamma_{\Omega,2}$, the second term in the square brackets, taken without the $(Z_\bk-Z_{\bkp})$ factor, reduces to $\Sigma(P)$. As it was the case for the sum of diagrams $A$ and $B$, the role of the $(Z_\bk-Z_{\bkp})$ factor is to regularize the $1/\delta q$ singularity of the particle-hole bubble. After this regularization, the second term in the square brackets gives the same contribution to the conductivity as the self-energy diagram without an extra logarithm.

In the first term, the frequency of the fermion $K'$ is shifted by the external frequency. Denoting again $K'=K+Q$ and $P'=P-Q$, it is easy to see that this shift changes
  the
  frequency of the particle-hole bubble formed by fermions $K$ and $K+Q+Q_0$, such that instead of $|\Omega_q|/|\delta q|$ we now have $|\Omega_q+\Omega|/|\delta q|$. The change has the same effect as shifting the frequency of the incoming fermion from $\Omega_p$ to $\Omega_p+\Omega$:
       the $(Z_\bk-Z_{\bkp})$ factor again removes one of the logarithms.

We thus see that the combined contribution of diagrams $C$ and $D$ is of the same order as that of diagrams $A$ and $B$. The two groups of diagrams cancel each other
to leading logarithmic order.
Beyond
 this order,
 however, the vertices in diagrams $C$ and $D$ differ from those in $A$ and $B$, and thus a cancelation cannot happen. We therefore conclude, that the sum of the four diagrams differs from the self-energy diagram by at most a logarithmic factor, and the conductivity does indeed scale as announced in Eqs.~(\ref{cond13}) and (\ref{cond14}).

\subsubsection{ Subleading non-singular terms in the optical conductivity}

 For completeness, we also analyze the form of the subleading terms in the optical conductivity, which are present even
    under the
    assumptions
    adopted in Sec.~\ref{sec:cancel},
i.e.,
strict forward- and $2k_F$-scattering and constant $Z$ factor. These subleading terms
appear because the diagrams for the conductivity do not cancel
each other
if
the  frequency dependence of the composite vertex
is taken into account.
  Indeed,
    when
   deriving Eq.~(\ref{1.7})
   we approximated the fermionic propagator
      $G(K+Q_\pi)$
   by
  its
    static form ($-1/v_F \delta p$),
     and similarly for the second propagator,   $G(P+Q_\pi)$.
      The full fermionic propagator
      depends on the frequency via
    the
    $\Omega_k/Z_{\bk}$ term.  All internal frequencies in the diagrams for ${\cal P}$ are of order $\Omega$,
    hence the extra terms which distinguish between, e.g.,
    $\Gamma_{\Omega,1}$ and $\Gamma$, come in powers of
    $\Omega/Z_\bk v_F \delta k \propto \Omega/\delta k |\delta k|$,
    where we used that
    $Z_\bk\propto |\delta k|$ for lukewarm fermions.  The first-order term again vanishes by parity,
    and the leading term in $\Gamma_{\Omega,1} -\Gamma$ scales as
    $\Gamma \Omega^2/(\delta k)^4$.
    In the 2D regime of composite scattering,
    typical $|\delta k|
    \propto\Omega^{1/3}$, hence the extra term is of order $\Omega^{2/3}$,
    and the corresponding contribution to conductivity
    scales as $\Omega^{1/3}$, i.e., $\sigma'(\Omega)\sim\mathcal{O}(\Omega^{-1/3})+\mathcal{O}(\Omega^{1/3})$.
This dependence is non-analytic yet
 subleading to  a constant, FL  term in the conductivity. In the 1D regime, typical $\delta k$ are frequency independent, hence
the correction to the conductivity scales as $\Omega^2$, i.e., $\sigma'(\Omega)\sim \mathcal{O}(\Omega^{-1})+\mathcal{O}(\Omega^{2})$.

 \section{Conclusions}
\label{sec:4}
In this paper, we considered
the $T=0$ optical conductivity
of a clean
two-dimensional metal near a
spin-density-wave instability with momentum $\bq_\pi = (\pi,\pi)$.
It is well established by now
that critical magnetic fluctuations
  destroy fermionic coherence in hot regions,
 but
     coherent quasiparticles survive
    on
      the  rest of the FS.
   Recent analysis by HHMS
   (Ref. ~\onlinecite{max_last})
   has demonstrated that the contribution to
   the
   conductivity from hot fermions
    is reduced by vertex corrections, and is subleading to a constant, Fermi-liquid contribution from cold fermions.  These authors
    also argued that
  composite scattering between lukewarm fermions
     (which
     behave
     as
     Fermi-liquid albeit strongly renormalized quasiparticles)
   gives  a singular contribution to the conductivity because
  the
   diagrams
   with self-energy and vertex-correction insertions
    do not cancel each other.

 We found
 that
 the imaginary part of the fermionic self-energy from
 two-loop composite
  scattering
 scales
 as  $\Sigma'' (\bk_F, \Omega) \propto
 \Omega^2/\delta k
 ^4
 \ln^3{|v_F \delta k/\Omega|}$,
 for $\Omega$ below some characteristic scale, and as $\Omega \min\{v_F\delta k/\bg,1\} /\delta k^2$, above that scale.
  The
  conductivity
  obtained by
   inserting such
  a
  self-energy into the current-current correlator exhibits a NFL,
  singular dependence on $\Omega$:
$\sigma'_\Sigma(\Omega)\propto \ln^3\Omega/\Omega^{1/3}$  and $\sigma'_\Sigma(\Omega)\propto 1/\Omega$ for $\Omega$ below and above $\Omega_{\min}=\bg^2/E_F$, correspondingly. At the high-frequency end, the $1/\Omega$ scaling of $\sigma'_\Sigma(\Omega)$ extends all the way up to the bandwidth, above which the low-energy theory becomes inapplicable.

We showed that the
 vertex-correction and Aslamazov-Larkin diagrams
  cancel
  out a part of but not all the self-energy contribution. Namely, the low-frequency form of the full conductivity loses the logarithmic prefactor
but retains a power-law, $\Omega^{-1/3}$ singularity, whereas the high-frequency, $1/\Omega$ form remains intact (up to a number). The full conductivity behaves as specified by Eq.~(\ref{sigma_final}).

As a word of caution, Eq.~(\ref{sigma_final}) is only a two-loop result. As shown in Sec.~\ref{sec:higher_loops}, corrections to the self-energy from higher loops are of the same order as the two-loop result at lower frequencies and are formally larger than the two-loop result by a logarithmic factor at higher frequencies.
This means, in particular, that the scaling form of the conductivity in the high-frequency regime should acquire an anomalous exponent: $\sigma'(\Omega)\propto 1/\Omega\to \sigma'(\Omega)\propto 1/\Omega^{1+\beta}$.
A calculation of $\beta$
requires non-perturbative methods and
is beyond the scope of this paper.

  We emphasize that
   non-analytic terms in the conductivity,
   considered in this paper,
    are different from the ones in the presence of impurities.\cite{imp}
  In the latter case, non-analytic terms appear as corrections to a constant Drude term due to impurity scattering and predominantly come from hot fermions. We caution, however, that
  a
  computation of the conductivity in near-critical dirty systems requires a special care.~\cite{subir_matthias}

    Strictly speaking, the range for the $1/\Omega$ scaling of $\sigma'(\Omega)$ is well-defined only under the assumption that the spin-fermion coupling is weak, i.e., $\bar g < E_F$.  The actual behavior of $\sigma'(\Omega)$
    is
    determined by the numerical coefficients which are hard to calculate in a consistent way.
 It is still encouraging, however, to see that
 a microscopic model predicts
a $1/\Omega$ scaling in
a  (at lest formally) wide frequency range,
which
 is consistent
 with the behavior observed
 in the cuprates.\cite{basov}
    The scale $\Omega_{\min} \sim {\bar g}^2/E_F$ is parametrically smaller than the scale of the superconducting $T_c  \sim {\bar g}$, hence $1/\Omega^{1/3}$ behavior is likely to be  masked by superconductivity (or finite $T > T_c$).

An interesting question
to be addressed
elsewhere is whether there is $\Omega/T$ scaling of the conductivity and,
 in particular,
whether the $1/\Omega$  behavior of the conductivity at $T=0$ is
paralleled by
a
the linear-in-$T$ behavior of the resistivity in
a similarly wide temperature range.

\label{sec:concl}
\acknowledgments
We thank J. Betouras,
S. Hartnoll,
S. Kivelson, G. Kotliar,
S. Maiti, M. Metlitski, I. Paul, S. Sachdev, and P. W{\"o}lfle for fruitful discussions.
 We are particularly thankful to M. Metlitski for critical comments on the first version of the manuscript.
This work was supported by the
Department of Energy via grant No. DE-FG02-ER46900 (A.V.C.), National Science Foundation via grant No. DMR-1308972 (D.L.M.),  and Russian Foundation for Basic Research via grant No. 12-02-00100 (V.I.Y.). D.L.M. and A.V.C. thank the Aspen Center for Physics
 and MPIPKS (Dresden) for hospitality
  during the work on this project.
   The Aspen Center for Physics is supported by the National Science Foundation via grant PHYS-1066293.


\begin{thebibliography}{99}
\bibitem{comm_a} A similar behavior with $\rho (T) \propto T^{3/2}$
observed
 in the helimagnet MnSi
 under pressure
 ~\cite{rosch:07}
 is unlikely
 to be
 associated with quantum criticality, as
 the
 $T^{3/2}$ behavior
 is seen in
 a wide range of pressures above the critical one.
\bibitem{comment_2kF} To be more precise, the initial and the final states belong to the opposite hots spots.
 Once the fermions at hot spots $2$ and $\bar 2$ are integrated out, the effective scattering process can be labeled as $(1,\bar 1;1,\bar 1)$, i.e.,
 the fermions  located initially at spots $1$ and $\bar 1$ remain in the vicinity of their respective hot spots.  In the terminology of
 Ref.~\onlinecite{chubukov:05}, such a process can be viewed either as a $q=0$ part of the \lq\lq backscattering process\rq\rq\/, with an amplitude given by the difference of the charge and spin components of the scattering amplitude, or as a special case of
 a Cooper process, in which the initial fermions with the total momentum close to zero retain their initial momenta.
\bibitem{comment_sigmak} We neglect the dependence of the self-energy on $k_\perp$,  so that the velocity renormalization factor
 $(1+\frac{\partial\Sigma'}{v_F\partial k_\perp})(1+\frac{\partial\Sigma'}{\partial\Omega})^{-1}\approx (1+\frac{\partial\Sigma'}{\partial\Omega})^{-1}$ coincides with the $Z$ factor.
\bibitem{comment_bare} In general, the current vertex contains not the bare
velocity but rather the \lq\lq charge\rq\rq\/ one,
which differs from the bare velocity by a factor of  $1+\frac{\partial\Sigma'}{v_F\partial k_\perp}$.~\cite{eliash:62,finn,mats2}
In our case, however, this derivative can be neglected  (see comment \onlinecite{comment_sigmak}), and thus the charge velocity is almost equal to the bare one.
\end{thebibliography}
\end{document}